\documentclass[apj]{emulateapj}

\slugcomment{published in The Astrophysical Journal}

\usepackage{epsfig}
\usepackage{epstopdf}
\usepackage{amsmath}
\usepackage{color}
\usepackage{natbib}
\usepackage{graphicx}
\usepackage{verbatim}
\usepackage{booktabs} 
\usepackage{multirow}
\usepackage{ulem}
\usepackage[caption=false]{subfig}
\usepackage{morefloats}

\newcommand{\ra}[1]{\renewcommand{\arraystretch}{#1}}

\newcommand{\atlas}{$\rm{ATLAS^{3D}}$}
\newcommand{\Reff}{$R_{\rm e}$}
\newcommand{\kms}{km~s$^{-1}$}

\begin{document}

	\title{The SLUGGS Survey: Wide-Field Stellar Kinematics of Early-Type Galaxies}

	\author{Jacob~A.~Arnold\altaffilmark{1}}
	\author{Aaron~J.~Romanowsky\altaffilmark{2,3}${}^\star$}
	\author{Jean~P.~Brodie\altaffilmark{2}}
	\author{Duncan~A.~Forbes\altaffilmark{4}}
	\author{Jay~Strader\altaffilmark{5}}
	\author{Lee~R.~Spitler\altaffilmark{6}}
	\author{Caroline~Foster\altaffilmark{7}}
        \author{Christina~Blom\altaffilmark{4}}
	\author{Sreeja~S.~Kartha\altaffilmark{4}}
	\author{Nicola~Pastorello\altaffilmark{4}}
	\author{Vincenzo~Pota\altaffilmark{4}}
	\author{Christopher~Usher\altaffilmark{4}}
	\author{Kristin~A.~Woodley\altaffilmark{2}}

	\altaffiltext{1}{Department of Astronomy and Astrophysics, University of California, Santa Cruz, CA 95064, USA}
	\altaffiltext{2}{University of California Observatories, 1156 High Street, Santa Cruz, CA 95064, USA}
	\altaffiltext{3}{Department of Physics and Astronomy, San Jos\'e State University, One Washington Square, San Jose, CA 95192, USA}
	\altaffiltext{4}{Centre for Astrophysics \& Supercomputing, Swinburne University, Hawthorn, VIC 3122, Australia}
	\altaffiltext{5}{Department of Physics and Astronomy, Michigan State University, East Lansing, MI 48824, USA}
	\altaffiltext{6}{Department of Physics and Astronomy, Faculty of Sciences, Macquarie University, Sydney, NSW 2109, Australia}
	\altaffiltext{7}{Australian Astronomical Observatory,  P.O. Box 915, North Ryde, NSW, Australia\\ \qquad \qquad ${}^\star$ romanow@ucolick.org}

	\date{\today}

	\begin{abstract}
	
We present stellar kinematics of 22 nearby early-type galaxies (ETGs), based on two-dimensional (2D) absorption line stellar
spectroscopy out to $\sim$\,2--4\,\Reff\ (effective radii), as part of the ongoing SLUGGS Survey.
The galaxies span a factor of 20 in intrinsic luminosity, as well as a full range of environment and ETG morphology.
Our data consist of good velocity resolution ($\sigma_{\rm inst} \sim 25$\,$\mathrm{km\;s}^{-1}$)
integrated stellar-light spectra extracted from the individual slitlets of custom made Keck/DEIMOS slitmasks.
We extract stellar kinematics measurements ($V$, $\sigma$, $h_3$, and $h_4$) for each galaxy. 
Combining with literature values from smaller radii, we present 2D  spatially resolved maps of the large-scale kinematic
structure in each galaxy.  We find that the kinematic homogeneity found inside 1~\Reff\ often breaks down at larger radii, where
a variety of kinematic behaviors are observed. While central slow rotators remain slowly rotating in their halos,
central fast rotators show more diversity, ranging from rapidly increasing
to rapidly declining specific angular momentum profiles in the outer regions.
There are indications that the outer trends depend on morphological type,
raising questions about the proposed unification 
of the elliptical and lenticular (S0) galaxy families in the \atlas\ survey.
Several galaxies in our sample show multiple lines of evidence for
distinct disk components embedded in more slowly rotating spheroids, and
we suggest a joint photometric--kinematic approach for robust bulge--disk decomposition.
Our observational results appear generally consistent with a picture of two-phase (in-situ plus accretion) galaxy formation. 

	\end{abstract}


	\maketitle

	\section{Introduction}\label{sec:intro}

The relict signatures of the mass assembly and growth of galaxies
are evident in their present-day internal dynamical structure. 
Within the $\Lambda$CDM cosmological framework, galaxies
grow through gas infall and in-situ star formation,
along with hierarchical merging and accretion of smaller galaxies
\citep[e.g.,][]{White:1978uk}. 
The relative dominance of these modes is conventionally thought to 
connect to morphological type, with late-type (spiral) galaxies
dominated by the in-situ mode, and early-type galaxies
(ETGs; ellipticals and lenticulars) by the merging mode.
		
Merging naturally produces stellar spheroids through violent relaxation, but
four decades of detailed observations have shown that ETGs
are more than simple featureless spheroids.
While remarkably uniform in some respects, 
e.g., following tight scaling relations \citep{Djorgovski:1987ij,Dressler:1987ds}, 
ETGs also exhibit a diversity of structural features and dynamical behaviors. 
These include wide variations in central rotation \citep{Bertola:1975,Illingworth:1977er,Davies:1983bc}, isophote shape \citep{Bender:1989wm}, 
surface brightness profiles \citep[e.g.,][]{Lauer:1995dm,Kormendy:2009iw}, and radio and X-ray luminosity \citep[e.g.,][]{Bender:1987ta,Bender:1989wm}.	
Because many of these properties are correlated \citep{Bender:1989wm,Nieto:1991tj}, 
the ellipticals can be sub-divided into two fairly distinct varieties:  boxy-nonrotating and disky-rotating,
where the latter are also closely connected to the lenticulars \citep{Kormendy:1996fm}.

This bimodal ETG classification scheme has recently been refined through
the landmark \atlas\ survey \citep{Cappellari:2011ej},
which used data from the SAURON integral field spectrograph 
for detailed, homogeneous analyses of kinematics, dynamics,
and stellar populations.
It was determined that $\sim$\,80\%--90\% of ETGs belonged to
the family of centrally fast/regular rotators,
characterized by rapid rotation and oblate axisymmetric shapes
that reflect underlying disk-like components
(\citealt{Emsellem:2011br,Krajnovic:2011jj,Krajnovic:2013ho};
see also \citealt{Rix:1990kp,Kuntschner:2010io}).
The remaining $\sim$\,10\%--20\% of ETGs were classified 
as centrally slow/non-regular rotators, which
show a variety of complex features such as kinematically distinct cores.

The \atlas\ findings motivated a proposed paradigm shift away
from traditional morphological classifications and toward a
kinematically motivated grouping of S0s with fast-rotator ellipticals
\citep{Emsellem:2011br,Cappellari:2011kh}.
These galaxies are in turn thought to have close structural and
evolutionary connections with spiral galaxies, with remnant
disks formed through dissipative processes,
while the slow rotators have lost their disks through more active
merger histories \citep{Khochfar:2011gh}.

\begin{figure}
\epsfxsize=9.2cm
\epsfbox{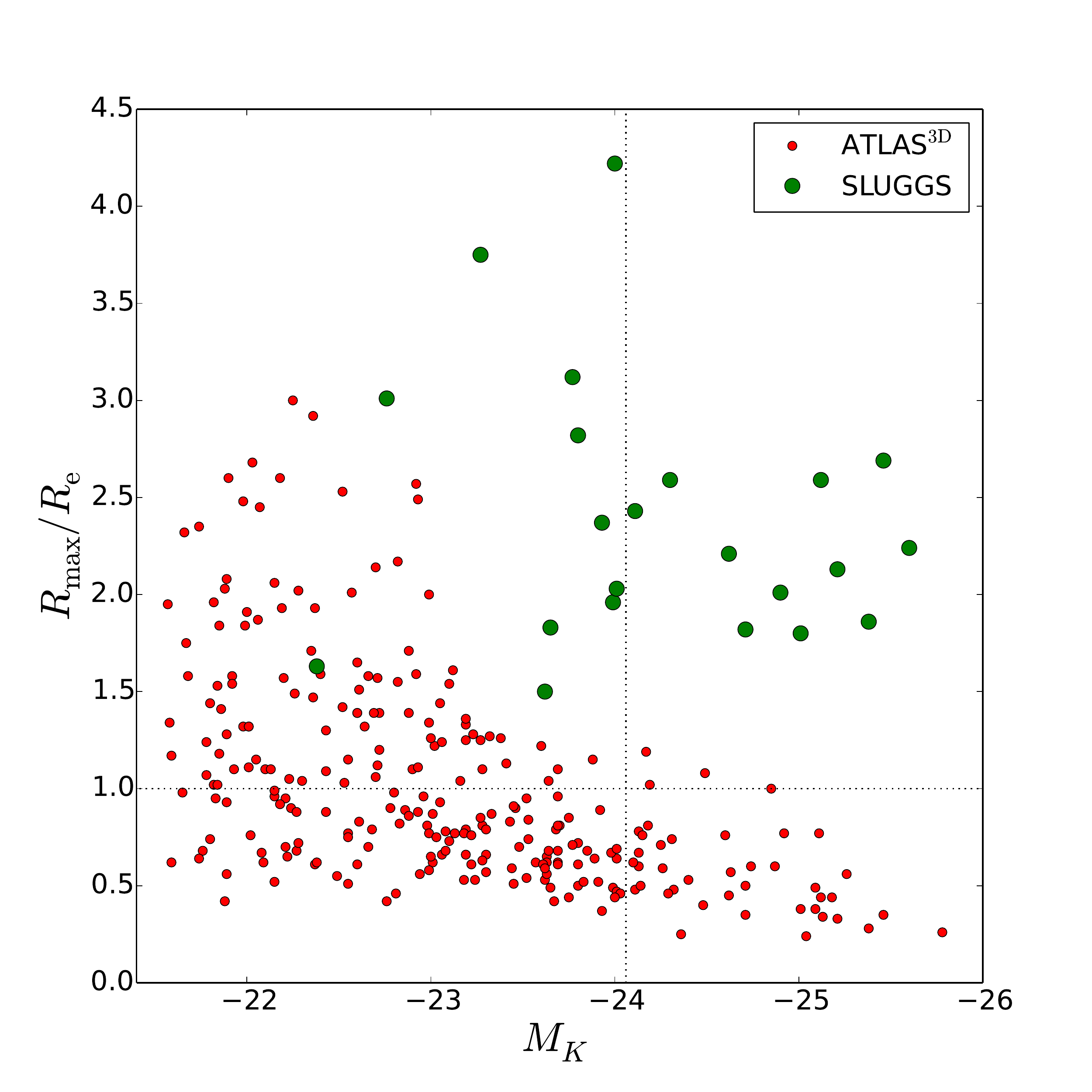}
\caption{Radial extent of stellar-spectroscopy data, in units of the effective radius (\Reff),
vs.\ galaxy $K$-band absolute magnitude.
Horizontal and vertical dotted lines mark 1~\Reff\ and the
characteristic luminosity \citep{Bell:2003}, respectively.
A given instrument with a fixed field-of-view observes a relatively smaller region
for the more luminous galaxies.
At a fixed luminosity, the SLUGGS observations extend to around four times
larger radii than \atlas\ (see Section~\ref{sec:finalmaps} for the $R_{\rm max}$ derivation).
}
\label{fig:Rmax}
\end{figure}

Although the \atlas\ results are remarkable, the conclusions about galaxy types are not unequivocal.
It is well established from simulations that fast rotators are a very generic
category that may include a variety of formational pathways (e.g., \citealt{Burkert08,Naab14,Moody14}),
while the finer details of the kinematic data are suggestive of distinct sub-types \citep{Krajnovic:2008fi,Hoffman:2009ju}.
Furthermore,  the SAURON data were spatially limited.
The survey extent in galactocentric radius was to $\sim$\,0.9~\Reff\ typically,
and to only $\sim$\,0.6~\Reff\ for the more luminous galaxies ($M_K < -23.5$),
where \Reff\ is the effective radius enclosing half of the projected light (see Figure~\ref{fig:Rmax}).

This limitation leads to several concerns and questions. First, how certain is it that
the defining characteristics of ETGs are found on these scales,
and can be used to transform our understanding of galaxies?
Second, given the critical importance of kinematics for recognizing
the pervasive {\it presence} of disk-like structures in ETGs, how well can
the {\it properties} of the disk and spheroid subcomponents be determined with the available data? 
Third, how much of the evolutionary story of ETGs is missed without
observations of their outer regions, which may undergo different assembly processes than their inner parts?

The last point is particularly relevant in light of the emerging
picture of two-phase assembly for ETGs, which is 
motivated both by observational evidence of dramatic size-growth
\citep[e.g.,][]{Daddi:2005ek,Cimatti:2008kj,Dokkum:2008fv}
and by cosmologically based simulations.
The first phase involves the formation of a compact ($\sim$\,1~kpc) in-situ bulge at redshifts $z > 2$ through violent dissipative events such as gas rich major-mergers or a fragmenting turbulent disk fed by cold flows \citep[e.g.,][]{Noguchi:1999gm,Elmegreen:2008gv,Dekel:2009bn,Ceverino:2010eh}.
The second phase is slower, taking place from $z \sim 3$ down to $z=0$
\citep{Oser:2011bp}, and consists of mostly gas-poor minor mergers which drive an outward growth of the galaxy
\citep{Naab:2009di,Bezanson:2009fk,Oser:2010dr,vanDokkum:2010bn,Newman:2012hp}.

In this view, nearby ETGs are layered structures where
the central regions are the remnants of the primordial in-situ bulges, 
and the outer extended halos consist of predominantly accreted material from disrupted satellites.
At intermediate radii there should be a transition zone comprised of both accreted material and stars scattered outward from the inner bulge \citep{Zolotov:2009dt}. Tantalizing observational evidence for this two-component structure
has been found through photometry of nearby ETGs
\citep{Forbes:2011jr,Huang:2013a,Huang:2013b,Petty:2013}, but
kinematical and dynamical information is needed for a clearer picture.

\begin{figure*}
\epsfxsize=18.0cm
\epsfbox{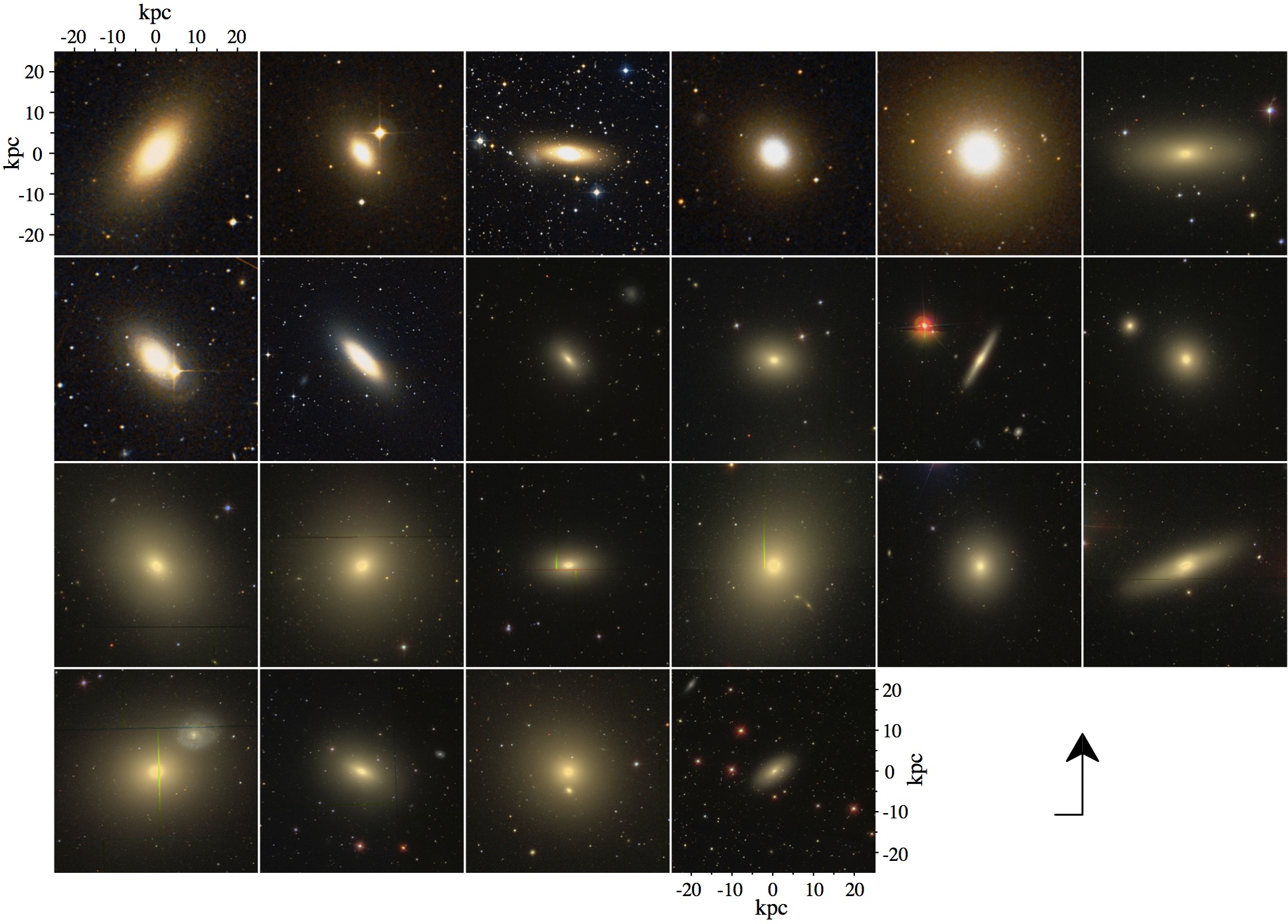}
\caption{Color images of each galaxy, displayed on the same physical scale. The scale is indicated in the top left and lower right. 
Images were taken from the Sloan Digital Sky Survey (SDSS) where available, and otherwise the Digital Sky Survey (DSS) using http://sky-map.org.
Galaxies are ordered left to right, top to bottom by ascending NGC number.
\textbf{Row~1}:~NGC~720, 821, 1023, 1400, 1407, 2768.
\textbf{Row~2}:~NGC~2974, 3115, 3377, 3608, 4111, 4278.
\textbf{Row~3}:~NGC~4365, 4374, 4473, 4486, 4494, 4526.
\textbf{Row~4}:~NGC~4649, 4697, 5846, 7457. The arrow shows that north is up and east is to the left.
}
\label{fig:galimage}
\end{figure*}

Wide-field kinematics observations have been obtained
in ETGs using integrated-light stellar kinematics (e.g., \citealt{Graham98,Rix99,Kronawitter00,Halliday01}),
as well as planetary nebulae (PNe) and globular clusters (GCs) as proxies
for the field stars \citep[e.g.,][]{Coccato:2009je,Arnold:2011kp,Pota:2013en}.
The latter methods have turned up kinematical transitions at large radii
that imply deviations from the central region trends, along
with hints that fast and slow rotators follow similar total
angular momentum scaling relations \citep{Romanowsky:2012}.
However, the results from discrete velocities
tend to be somewhat noisy, and there are uncertainties about how
these tracers connect to the underlying stellar populations.
The gold standard for galaxy kinematics is direct stellar-light spectroscopy,
including two-dimensional (2D) information both to map out any off-axis rotation
and to provide strong constraints on the internal dynamics \citep{Cappellari:2008}.
Occasional forays have been made in this direction but have
been too restricted either in radius or in 2D coverage
(e.g., \citealt{Statler:1999hq,Weijmans:2009hl,Murphy:2011hf}).

A remarkable new approach in this context is to use general-purpose wide-field multi-slit spectrographs to obtain
spatially well-sampled 2D spectra out to several \Reff. This ``Stellar Kinematics from Multiple Slits'' (SKiMS)
technique was pioneered by \citet{Norris:2008ea}
and \citet{Proctor:2008ce}, and reached fruition in \cite{Proctor:2009iy}
by using the Keck/DEIMOS spectrograph \citep{Faber:2003ev}.
DEIMOS SKiMS was applied to a total of eight ETGs through a subsequent
series of papers \citep{Arnold:2011kp,Foster:2011hh,Foster:2013}, and here is expanded to a sample of 22
(including reanalysis of the previously observed systems using new observations and an updated methodology).
The large spatial scales of these data, relative to \atlas, are illustrated by Figure~\ref{fig:Rmax},
with galaxy parameters reported in Table~\ref{info}.
This is the first systematic study of wide-field
stellar kinematics for a large sample of ETGs,
which represents a critical step forward in understanding the details of stellar halo assembly.  
		
These 22 galaxies were observed as part of the ongoing
SAGES Legacy Unifying Globulars and GalaxieS Survey
(SLUGGS\footnote{http://sluggs.ucolick.org}) of
25 ETGs in the local universe ($D<30$~Mpc; \citealt{Brodie14}).  The survey combines stellar and GC
system photometry, kinematics, and metallicities to explore the
halo regions of ETGs.  A key goal of the survey
is to target a representative sample of galaxies, spanning a range of luminosities,
morphologies, and environments.  The diversity in
shape and intrinsic brightness of our sample is
immediately obvious from Figure~\ref{fig:galimage}, which displays
color images of each galaxy on a common spatial scale.
The environment of our 22 galaxies includes 5 field galaxies ($23\%$), 12 in groups ($54\%$), and 5
in clusters ($23\%$). Further details of the SLUGGS survey are described in \citet{Brodie14}. Analyses
of GC metallicities for a subset of the SLUGGS sample have been presented in \citet{Usher:2012}, of GC
kinematics in \citet{Pota:2013en}, and of stellar metallicities in \citet{Pastorello14}.

This rich stellar kinematics data set will be analyzed in increasing depth
through a series of papers, and here we take a first
look at some critical questions motivated above. Do the kinematical trends established
with \atlas\ extend to larger radii, and support their revamped galaxy classification schemes?
Or do novel patterns emerge that require a reassessment of the situation?
Do the wide-field data provide better constraints on bulge--disk decomposition?
Are kinematical transitions detected that support the two-phase assembly paradigm?

We describe the observational methodology and data
reduction process in Section~\ref{sec:data}.  In Section~\ref{sec:kinmeas}, we
describe our kinematic measurements of $V$, $\sigma$,
$h_3$, and $h_4$, and perform repeatability tests and
literature comparisons.  In Section~\ref{sec:kinematic_maps}
we introduce a new 2D smoothing algorithm in order to produce
wide-field two-dimensional kinematic maps for our sample.
We explore some general kinematic trends of the sample in
Section~\ref{sec:disc}, and summarize in Section~\ref{sec:conclude}.

\begin{table*}
\ra{1.3}
\caption{Galaxy Information}\label{info}
\begin{tabular*}{18cm}{@{\extracolsep{\fill}}@{}ccccrccccccccc@{}}\toprule
	\multicolumn{0}{c}{Galaxy}  &  \multicolumn{0}{c}{R.A.} & \multicolumn{0}{c}{Decl.} & \multicolumn{0}{c}{\Reff} & \multicolumn{0}{c}{$R_{\rm max}$} & \multicolumn{0}{c}{$\lambda_{R_e}$} & \multicolumn{0}{c}{Rot$_{R_e}$} 
		& \multicolumn{0}{c}{$M_{K}$} & \multicolumn{0}{c}{Morph.} & \multicolumn{0}{c}{P.A.} & \multicolumn{0}{c}{$\epsilon$} 
		& \multicolumn{0}{c}{ $V_{\rm sys}$} & \multicolumn{0}{c}{Dist.} & \multicolumn{0}{c}{Env.}		\\
	
	\multicolumn{0}{c}{NGC}  & \multicolumn{0}{c}{(h m s)} & \multicolumn{0}{c}{(d m s)} & \multicolumn{0}{c}{(arcsec)} & \multicolumn{0}{c}{(arcsec)} & \multicolumn{0}{c}{} & \multicolumn{0}{c}{} 
		& \multicolumn{0}{c}{(mag)} & \multicolumn{0}{c}{} & \multicolumn{0}{c}{(deg)} & \multicolumn{0}{c}{}
		& \multicolumn{0}{c}{($\rm{km~s^{-1}}$)} & \multicolumn{0}{c}{(Mpc)} & \multicolumn{0}{c}{} 		\\
	
	\multicolumn{0}{c}{(1)} & \multicolumn{0}{c}{(2)} & \multicolumn{0}{c}{(3)} & \multicolumn{0}{c}{(4)} & \multicolumn{0}{c}{(5)} & \multicolumn{0}{c}{(6)} 
		& \multicolumn{0}{c}{(7)} & \multicolumn{0}{c}{(8)} & \multicolumn{0}{c}{(9)} & \multicolumn{0}{c}{(10)} & \multicolumn{0}{c}{(11)} & \multicolumn{0}{c}{(12)} 
		& \multicolumn{0}{c}{(13)} & \multicolumn{0}{c}{(14)}	\\			
	
	\midrule
	0720		& $01~53~00.5$	& $-13~44~19$	& 33.9	& 68 & 0.12		& S	& $-24.6$	& E5		& 142.3	& 0.43	& 1745	& 23.4 	& F		\\
	0821		& $02~08~21.1$	& $+10~59~42$	& 39.8	& 78 & 0.27		& F	& $-24.0$	& E6		& 31.2	& 0.35	& 1718	& 23.4	& F		\\	
	1023		& $02~40~24.0$	& $+39~03~48$	& 47.9	& 97  & 0.39		& F	& $-24.0$	& S0	& 83.3	& 0.63	& 602	& 11.1	& G		\\
	1400		& $03~39~30.8$	& $-18~41~17$	& 29.3	& 76 & 0.27        	& F	& $-24.3$	& E1/S0	& 36.1	& 0.11	& 558	& 26.8 	& G		\\
	1407		& $03~40~11.8$	& $-18~34~48$	& 63.4	& 142 & 0.08        	& S	& $-25.4$	& E0		& 58.3	& 0.05	& 1779	& 26.8 	& G		\\
	2768		& $09~11~37.5$	& $+60~02~14$	& 63.1	& 115 & 0.25		& F	& $-24.7$	& E6/S0		& 91.6	& 0.57	& 1353	& 21.8	& G		\\
	2974		& $09~42~33.3$	& $-03~41~57$	& 38.0	& 57 & 0.66		& F	& $-23.6$	& E4		& 44.2	& 0.37	& 1887	& 20.9	& F		\\
	3115		& $10~05~14.0$	& $-07~43~07$	& 34.4	& 145 & 0.58		& F	& $-24.0$	& S0		& 43.5	& 0.49	& 663	& 9.4		& F		\\
	3377		& $10~47~42.3$	& $+13~59~09$	& 35.5	& 107 & 0.52		& F	& $-22.8$	& E5--6	& 46.3	& 0.33	& 690	& 10.9	& G		\\
	3608		& $11~16~58.9$	& $+18~08~55$	& 29.5	& 54 & 0.04		& S	& $-23.7$	& E1--2 	& 82.0	& 0.20	& 1226	& 22.3	& G		\\
	4111		& $12~07~03.1$	& $+43~03~57$	& 12.0	& 45 & 0.62		& F	& $-23.3$	& S0 	& 150.3	& 0.79	& 792	& 14.6	& G		\\
	4278		& $12~20~06.8$	& $+29~16~51$	& 31.6	& 89 & 0.18		& F	& $-23.8$	& E1--2	& 39.5	& 0.09	& 620	& 15.6	& G		\\
	4365		& $12~24~28.3$	& $+07~19~04$	& 52.5	& 112 & 0.09		& S	& $-25.2$	& E3		& 40.9	& 0.24	& 1243	& 23.3	& G		\\
	4374		& $12~25~03.7$	& $+12~53~13$	& 52.5	& 136 & 0.02		& S	& $-25.1$	& E1 	& 128.8	& 0.05	& 1017	& 18.5	& C		\\
	4473		& $12~29~48.9$	& $+13~25~46$	& 26.9	& 84 & 0.23		& F	& $-23.8$	& E5 	& 92.2	& 0.43	& 2260	& 15.3	& C		\\
	4486		& $12~30~49.4$	& $+12~23~28$	& 81.3	& 151 & 0.02		& S	& $-25.4$	& E0/cD	& 151.3	& 0.16	& 1284	& 17.7	& C		\\
	4494		& $12~31~24.1$	& $+25~46~31$	& 49.0	& 119 & 0.21		& F	& $-24.1$	& E1--2	& 176.3	& 0.14	& 1342	& 16.6	& G		\\
	4526		& $12~34~03.1$	& $+07~41~58$	& 44.7	& 99 & 0.45		& F	& $-24.6$	& S0	& 113.7	& 0.76	& 617	& 16.4	& C		\\
	4649		& $12~43~40.0$	& $+11~33~10$	& 66.1	& 178 & 0.13		& F	& $-25.5$	& E2/S0		& 91.3	& 0.16	& 1110	& 17.3	& C		\\
	4697		& $12~48~35.9$	& $-05~48~03$	& 61.7	& 146 & 0.32		& F	& $-23.9$	& E6 	& 67.2	& 0.32	& 1252	& 11.4	& G		\\
	5846		& $15~06~29.3$	& $+01~36~20$	& 58.9	& 106 & 0.03		& S	& $-25.0$	& E0--1/S0 & 53.3	& 0.08	& 1712	& 24.2	& G		\\
	7457		& $23~00~59.9$	& $+30~08~42$	& 36.3	& 59 & 0.52		& F	& $-22.4$	& S0		& 124.8	& 0.47	& 844	& 12.9	& F		\\
	\bottomrule
\end{tabular*}
\\

			Notes: (1) Galaxy NGC number.  (2) Right
                        ascension and (3) declination in J2000 coordinates taken from the
                        NASA/IPAC Extragalactic database (\url{http://ned.ipac.caltech.edu/})
                        (NED).  (4) Effective (half-light) radius in units of arcseconds.  Effective radii for
                        \atlas\ galaxies were taken from  \citet{Cappellari:2011ej}, while the remaining
                        values (for NGC~720, NGC~1400, NGC~1407, NGC~3115) were derived using the same
                        methodology. (5) Maximum galactocentric extent of SKiMS data, in arcseconds (see Section~\ref{sec:finalmaps} for definition).
			(6) Dimensionless proxy for stellar specific angular momentum within 1~\Reff\
                         (\citealp[see][]{Emsellem:2007hf}) with the majority of values from
                        \citet{Emsellem:2011br} and \citet[][NGC~720]{Cappellari:2007bn}.  For
                        galaxies without available IFU data (i.e., NGC~1400, NGC~1407 and NGC~3115), we
                        estimated $\lambda_{R_e}$ using SKiMS and literature kinematic measurements.
                        (7) Central rotator designation of fast (F) or slow (S) with values from \citet{Emsellem:2011br},
                        or derived from available kinematic data (for NGC~720, NGC~1400, NGC~1407, NGC~3115).
		       (8) Extinction corrected absolute $K$-band magnitudes derived
                        from the 2MASS extended source catalog \citep{Jarrett:2000fz}.  Values taken from
                        \citet{Emsellem:2011br} or measured using the same methodology for galaxies not in the
                        \atlas\ sample.  (9) Morphology from NED,	combining the RSA and RC3 classifications.
                        See \citet{Brodie14} for the Hubble stage parameters.
                        (10) Photometric position angles and (11) ellipticities.  Values taken from
                        \citet{Krajnovic:2011jj}, \citet[][NGC~720]{Cappellari:2007bn},
                        \citet[][NGC~1400, NGC~1407]{Spolaor:2008do}, and \citet[][NGC~3115]{Capaccioli:1987gi}.
                        (12) Systemic velocity in km~s$^{-1}$ from \atlas\ \citep{Cappellari:2011ej} where available, and otherwise from NED. 
                        (13) Distances measured in Mpc from \citet{Cappellari:2011ej}. For NGC~720 and NGC~3115 we 
			have used \citet{Tonry:2001ei} surface brightness fluctuation (SBF) distances with a $-0.06$~mag
                        correction. For NGC~1400 and NGC~1407 we assume that they are at the same distance and assign an average SBF distance of 26.8 Mpc. 
(14) Local environment type: field (F), group (G), or cluster (C).  For the environmental densities, see \citet{Brodie14},
		where we note also that there are some slight differences in the values adopted for \Reff, $M_K$, P.A., 
		$\epsilon$, and distance---particularly for the non-\atlas\ galaxies.
		\label{tab:galaxies}
		\end{table*}

	\section{Data}\label{sec:data}

		The SLUGGS observational strategy is designed around
                obtaining precise radial velocities of individual
                GCs around nearby ETGs (see \citealt{Pota:2013en}
	 	and \citealt{Brodie14} for details).  This is
                accomplished using custom-made multi-slit masks and the 
		DEIMOS spectrograph at the Nasmyth focus
                of the 10 meter Keck II telescope.  DEIMOS is wide-field
                ($16.7^\prime\times5^\prime$), highly multiplexing ($\sim100$
                slitlets per slitmask), and equipped with a
                flexure compensation system, making it well suited for
                precision kinematic studies.  We make optimal use of
                this instrument by extracting integrated
                galaxy-stellar-light spectra from the background
                regions of slits placed on GCs, thereby
                obtaining both types of spectra {\it simultaneously}.  
		This stellar-light aspect of the SLUGGS survey is called
		Spectroscopic Mapping of EArly-type Galaxies to their
		Outer Limits (SMEAGOL). The ensuing SKiMS measurements
		are discussed in Section~\ref{sec:kinematic_maps},
                while the GC measurements are presented
                in \citet{Pota:2013en}, in references therein,
		and in additional papers in preparation.

		\subsection{Observations}

			\begin{figure}
				\begin{center}
					\epsfxsize=8.4cm
					\epsfbox{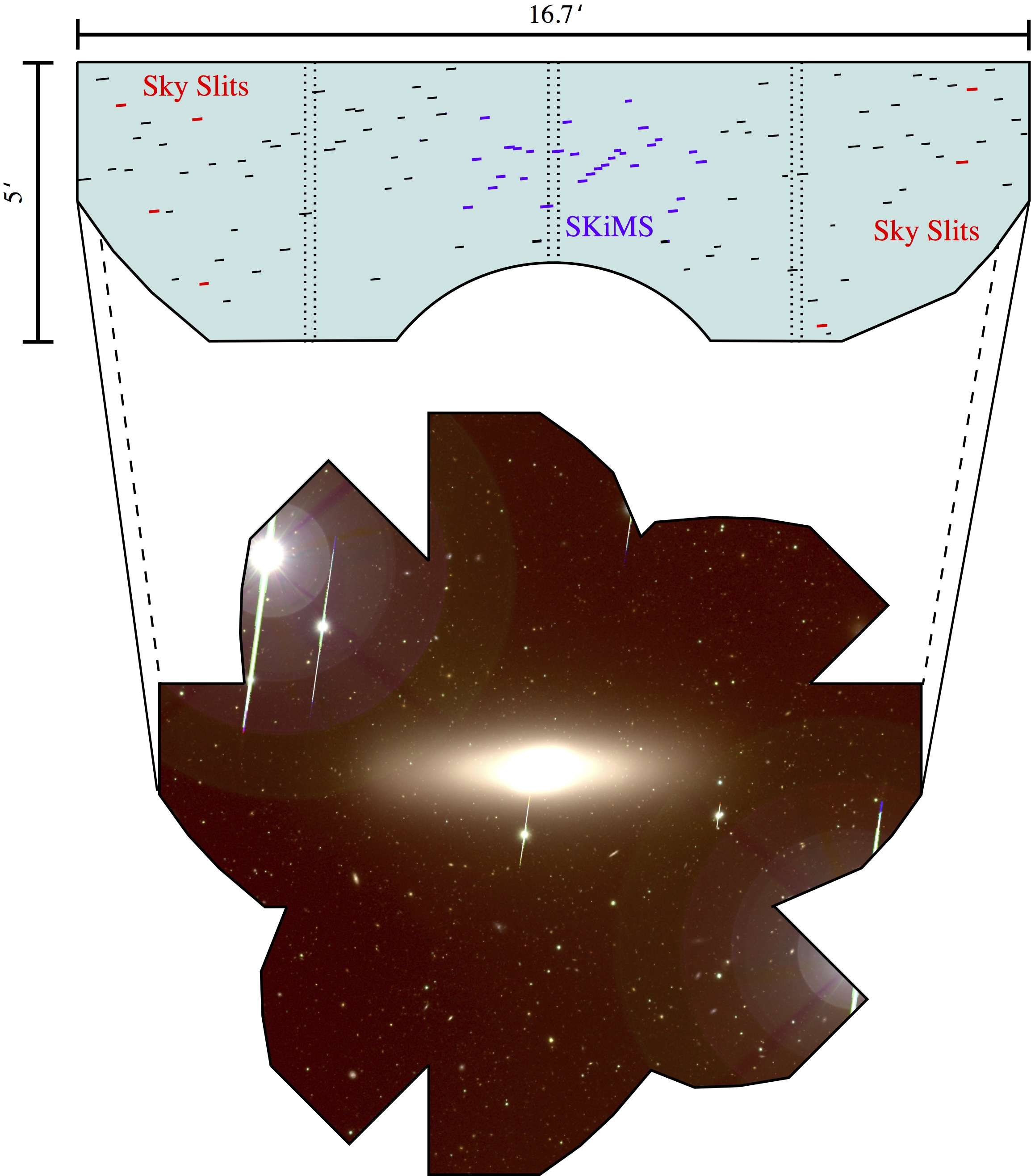}
					\caption{Observational footprint of multi-slit DEIMOS observations. 
					Top: the shape of a $16.7^\prime \times 5^\prime$ DEIMOS slitmask.
					Short lines denote milled slits, with selected sky slits shown in red, and selected SKiMS slits in purple. 
					Vertical dotted lines denote chip gaps.
					Bottom: a $gri$ composite Suprime-Cam
\citep{Miyazaki:2002wj} image of NGC 4526 with the areal coverage of four slitmasks arranged in a spoke pattern. 	}
					\label{fig:mask_layout}
				\end{center}
			\end{figure}

			DEIMOS has a roughly rectangular field-of-view with a few missing regions (see top of Figure~\ref{fig:mask_layout}).
			For comparison, the typical \Reff\ value
for galaxies in our sample is less than an arcminute.
			A single slitmask contains $\sim$100~slits, each with a minimum length of 4 arcsec.
			Additional filler slits are sometimes placed near the galaxy when there are no suitable GC candidates, and also at large radius to get sky spectra that are virtually uncontaminated by galaxy light.
			Normally, at least four slitmasks are observed per galaxy, and are arranged in a spoke pattern centered on the galaxy to ensure good azimuthal coverage (see Figure~\ref{fig:mask_layout} for an example).

			\begin{figure}
					\epsfxsize=8.0cm
					\epsfbox{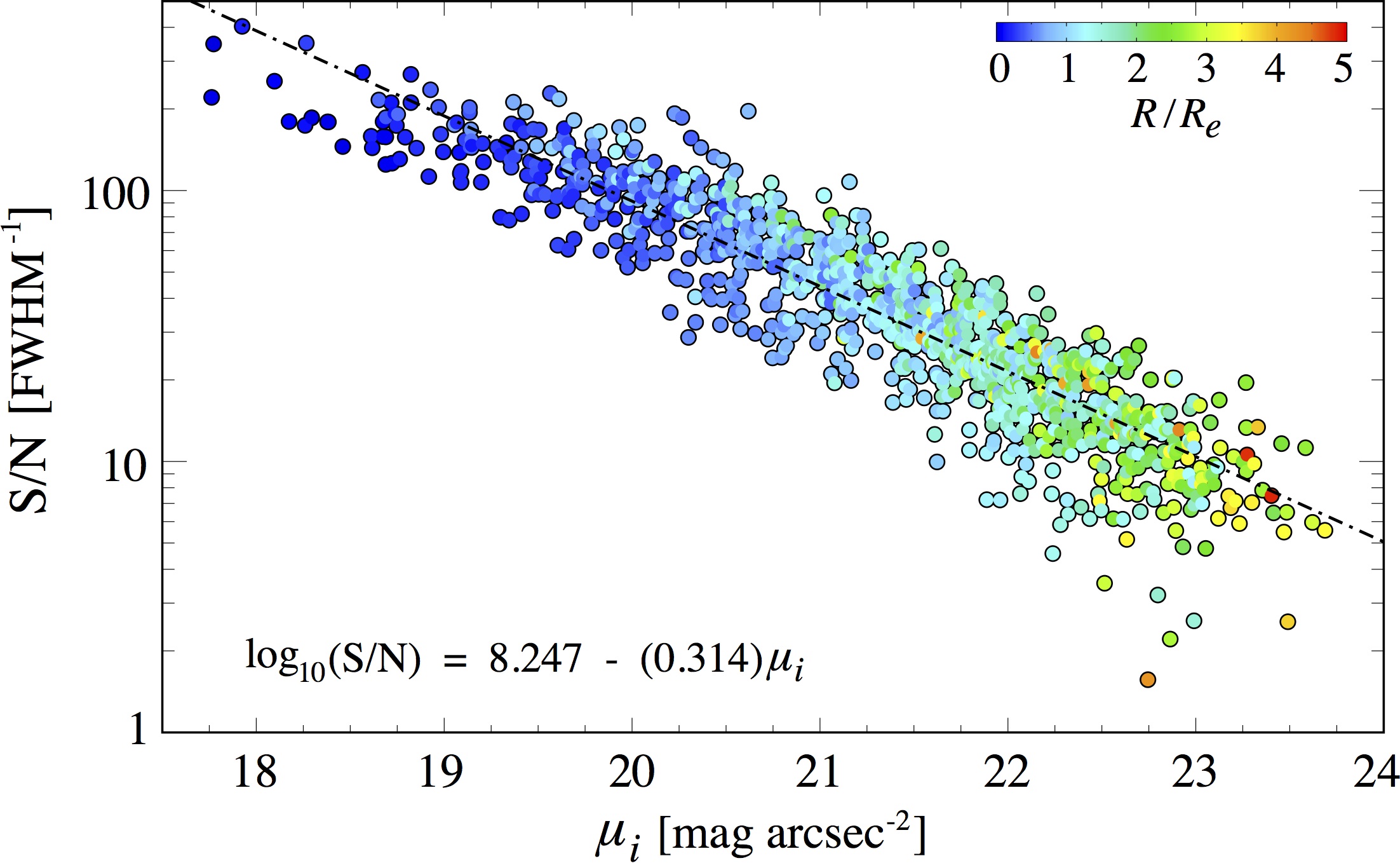}
					\caption{Photometric depth of the DEIMOS measurements.
					Spectral S/N is plotted vs.\ the local $i$-band surface brightness around each measurement location.
					Surface brightness values are measured from SDSS images, and only spectra from those galaxies within the SDSS footprint are shown.
					Points are colored according to the galactocentric radius normalized by \Reff\ (see the color bar in the top right corner).
					There is a roughly linear relation, shown as a dot-dashed line and given in the bottom left, between the base-10 logarithm of the spectral S/N and the local surface brightness.
					}
					\label{fig:snvsb}
			\end{figure}

			We collected over a thousand spectra for 22 galaxies over the course of 7 years and 19 observing runs.
			Our baseline observing strategy is to integrate on a mask for two hours in a series of four exposures, though conditions frequently cause deviations from this plan.
			The resulting spectra are very deep and permit kinematic information to be extracted from individual slits at surface brightnesses fainter than $\mu_i\!=\!23\,\rm{mag\,arcsec}^{-2}$ (see Figure~\ref{fig:snvsb}).
			
			\begin{figure}
					\epsfxsize=8.0cm
					\epsfbox{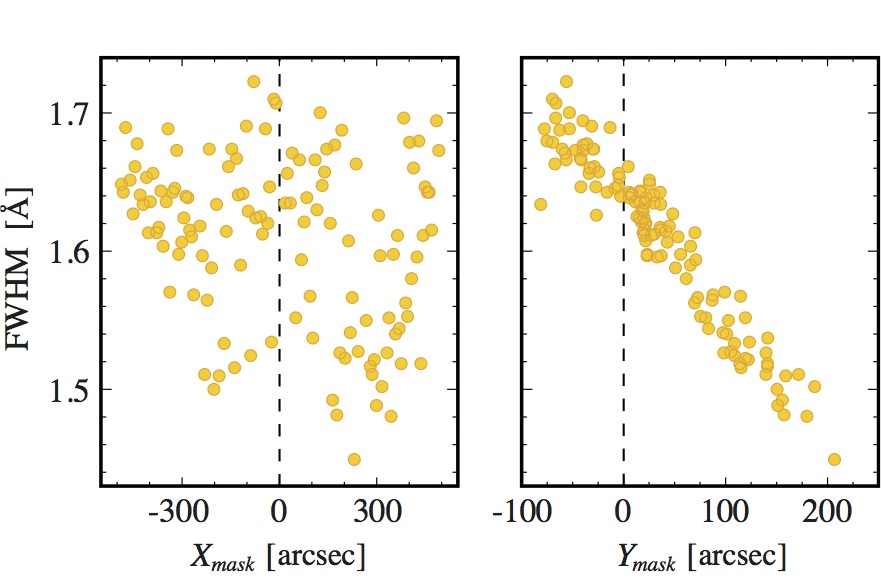}
					\caption{Spatial dependence of 
                                          DEIMOS spectral resolution.
                                          {\it Left:} measured resolution as a
                                          function of $x$-position
                                          and {\it right:} $y$-position
                                          across a single
                                          DEIMOS slit mask.  FWHM
                                          values are measured from a
                                          night sky emission line (at
                                          8415 $\rm{\AA}$) in the
                                          background spectrum of each
                                          slit.  The resolution
                                          appears uncorrelated with
                                          $x$-position on the mask (long
                                          axis), but has a 
                                          dependence on $y$-position (short axis).  }
					\label{fig:resolution}
			\end{figure}

			All of the spectra discussed here were
                        obtained using a 1200 line mm$^{-1}$ grating
                        centered at 7800 $\rm{\AA}$, and a 5500
                        $\rm{\AA}$ order-blocking filter.  This setup
                        is optimized to target the near-infrared (NIR)
                        Calcium~{\small II} triplet, with prominent lines at
                        8498, 8542, and $8662\,\rm{\AA}$.  In this
                        wavelength range the typical spectral
                        dispersion of $0.32\,\rm{\AA\,pixel^{-1}}$
                        translates to a velocity scale of
                        $11\;\rm{km\,s^{-1}\,pixel^{-1}}$.  With a
                        slit width of 1 arcsec, the average FWHM
                        resolution is $1.6 \rm{\,\AA}$ (see 
                        Figure~\ref{fig:resolution}), or
                        equivalently $\sim\!55$ km\,s$^{-1}$. This translates into 
an instrument velocity resolution of $\sigma_{\rm inst}$ $\sim$ 24 km\,s$^{-1}$. Although 
our full wavelength coverage of $\sim$6600--9100~\AA\ sometimes includes 
the H$\alpha$ line, it is not used in our kinematic analysis.

		\subsection{Data Reduction and Sky Subtraction}

			The DEIMOS spectra were all reduced using a modified version of the \texttt{spec2D} pipeline \citep{Cooper:2012uz}.
			The pipeline uses internal quartz-lamp flatfields to rectify each slit, measure a slit function, and make a fringing correction
			(that is minimal and straightforward to correct using the spectral flat-fields, owing to the accuracy of the
			flexure compensation system; \citealt{Wirth04}).
			Wavelength solutions are computed from NeArXeNe arclamp spectra taken through each mask,
			with residuals of 0.04\,\AA\ or better (equivalent to 1.4\,\kms) across the whole field.
			The background spectrum in each slit is modeled as a b-spline and the separate exposures are co-added together before a final one-dimensional spectrum is produced.
			We modified the pipeline to additionally return the associated pixel-by-pixel inverse variance array of the sky for use during kinematic fitting.

			Each background spectrum extracted from a slit is composed of two primary components: the integrated galaxy light that we are ultimately interested in, and foreground sky that must be subtracted out.
			An accurate measure of the latter component is crucial given the strong and variable emission of the night sky in the NIR.			
			Fortunately, the large field-of-view of DEIMOS relative to the luminous size of the galaxies under study means that there are numerous sky-slits toward the edges of each mask that are essentially free of galaxy light.
			We select a number of these slits in order to build up a suitable catalog of sky spectra for each mask.
			Specific care is taken to avoid slits with vignetting or other reduction issues that become more common toward the edges.
			Also, for reasons that are explained below, we select sky spectra from a variety of positions across the short axis of the mask.

			The next step would usually involve constructing a single, high S/N master sky spectrum by co-adding several of the selected sky spectra \citep[see e.g.,][]{Proctor:2009iy,Foster:2011hh}.
			However, the application of the conventional approach was complicated by the discovery of a spatially dependent resolution in DEIMOS. 
			Slits located at the bottom of a slitmask have slightly higher resolution (smaller FWHM) than slits placed near the top (see Figure~\ref{fig:resolution}).
			Consequently, the spectrum from an arbitrarily placed slit may have a slightly  different resolution than the constructed master sky spectrum.
			
		\begin{figure}
			\epsfxsize=8.6cm
			\epsfbox{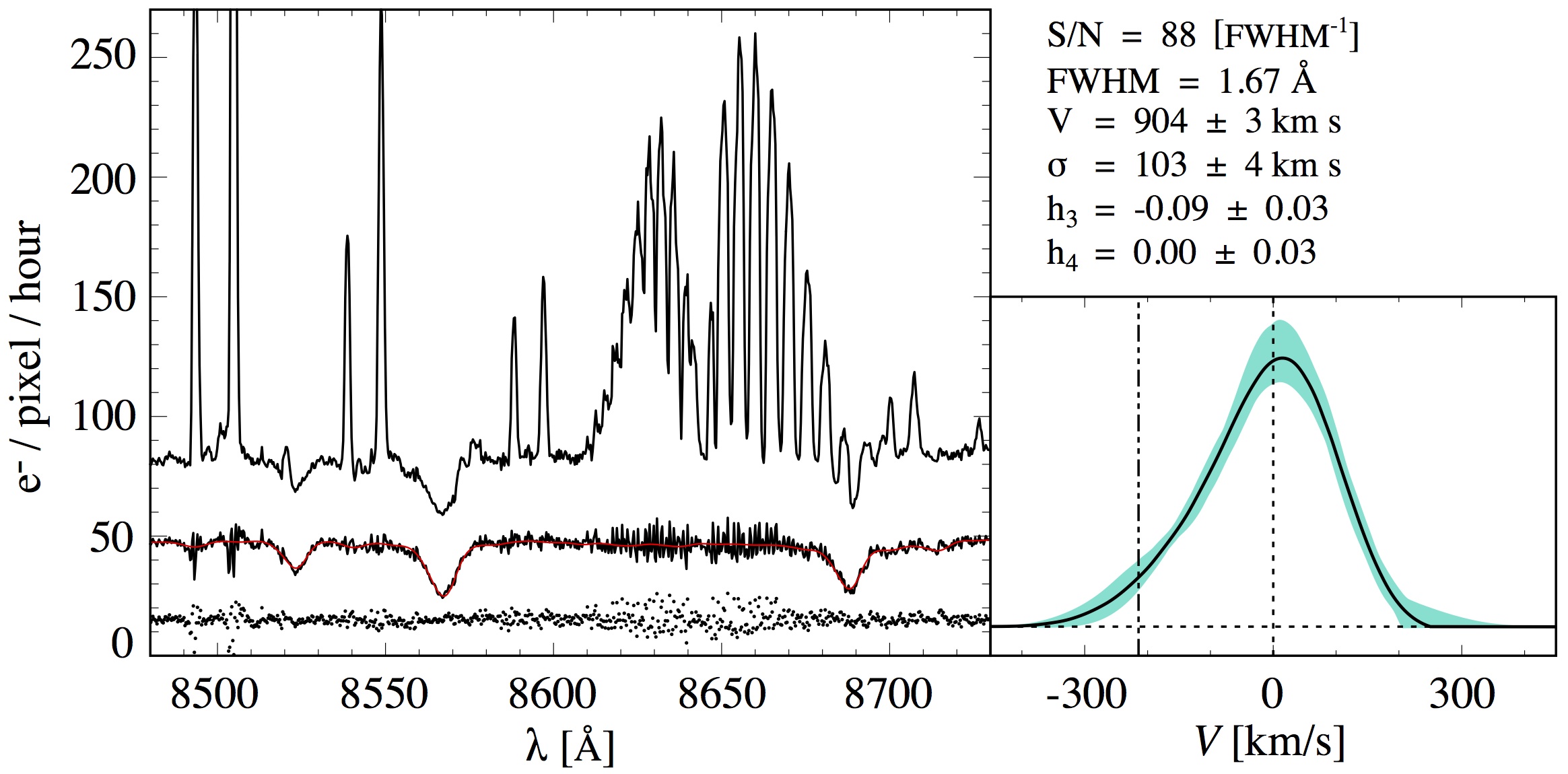}
			\caption{Example of an extracted stellar spectrum
			1.5~arcmin from the center of the galaxy NGC~3115.
			{\it Left:} a fiducial raw spectrum with galaxy stellar light and night sky background (top); 
			sky-subtracted spectrum with best fit model overlaid in red (middle); and fitting residuals (bottom).
			{\it Right:} best fitting parameterized LOSVD (solid black line).
			The envelope (solid cyan) shows the range of possible solutions as determined from 100 Monte Carlo realizations of the best fit model with added noise.
			The velocity axis has been shifted to the mean velocity of the measurement, and the vertical dot-dashed line marks the systemic velocity of the galaxy.
			}
			\label{fig:losvd}
		\end{figure}

			Ultimately, the effect is subtle, with the FWHM changing by less than a pixel between the top and bottom of the mask.
			However, we wanted to make every attempt at optimizing the sky subtraction given the potential for large residuals from strong near-infrared sky lines.
			We opted to use \texttt{pPXF} \citep[Penalized Pixel-Fitting method;][]{Cappellari:2004gm} to construct a unique sky spectrum for each slit, modeled as a weighted combination of the input sky spectra.
			This approach is suitable for sky-dominated spectra, which is the case for the vast majority of our data, and should result in a very accurate sky subtraction \citep[see e.g.,][]{Weijmans:2009hl}.			
			An example of a sky-subtracted science spectrum can be see in Figure~\ref{fig:losvd}.

	\section{Kinematic Measurements}\label{sec:kinmeas}

		The line-of-sight velocity distribution (LOSVD) of each
                spectrum is parameterized as a truncated Gauss-Hermite
                series \citep{1993ApJ...407..525V,Gerhard:1993tc};
                including $V$ (mean velocity), $\sigma$ (velocity
                dispersion), $h_3$ (a measure of skewness), and $h_4$
                (a measure of kurtosis).  These parameters are
                measured in pixel space using the \texttt{pPXF}
                software.  The best-fitting solution is determined by
                finding the weighted combination of template stars,
                which, when convolved by the parameterized LOSVD, most
                closely matches the input spectrum in a maximum
                likelihood sense.  We include a 4th order
                additive legendre polynomial to account for deviations
                in continuum shape between our spectra and the
                template stars.

		A potential pitfall in making kinematic measurements on low S/N spectra is fitting noise in the spectrum with anomalous $h_3$ and $h_4$ values.
		To help guard against this, \texttt{pPXF} uses regularization in the form of a penalty term that biases the fit toward a Gaussian when the S/N of a spectrum is insufficient to constrain the higher order moments of the LOSVD.
		From the Bayesian perspective, this is equivalent to imposing a prior on the $h_3$ and $h_4$ parameters that encapsulates the notion that higher order moments are less likely to be constrained as S/N decreases.
		A non-zero penalty requires a minimum decrease in the variance of the fit residuals in order to accept a larger amplitude $h_3$ or $h_4$ value \citep[see][]{Cappellari:2004gm}.
		While this penalization results in a bias toward Gaussian LOSVDs, it also reduces the likelihood of measuring spurious higher order moments.
		
		The appropriate penalty value is at the very least resolution and S/N-dependent, and can be determined via Monte Carlo simulation.
		Following the method outlined in \citet{Cappellari:2011ej}, we measured the requisite penalty at a range of S/N levels, and fit these results with a third order polynomial.
		The coefficients are $c_0 = 0.119$, $c_1 = 5.76\times10^{-3}$, $c_2 =-1.26\times10^{-5}$, and $c_3 = 9.37\times10^{-9}$.		
		The penalty value for a given science spectrum is determined using its measured S/N and the above polynomial.
		
		We restrict the \texttt{pPXF} fit to the spectral region immediately surrounding the NIR Calcium~{\small II} triplet between $8480\,\rm{\AA}$ and $8750\,\rm{\AA}$.
		The latter bound was increased on a galaxy by galaxy basis when a high systemic velocity redshifted the wings of the reddest CaT feature anywhere near the wavelength cutoff (e.g., NGC~4473).		
		No skylines are explicitly masked out since those pixels should, in principle, be appropriately weighted by the inverse variance arrays.
		However, we do allow for outlying pixel values to be clipped, which can occur when skylines are improperly subtracted.
		
		Template mismatch can be an issue with this method, so
                an appropriate suite of input template stars is
                crucial.  Previous SKiMS analyses (e.g., \citealt{Proctor:2009iy})
                used 13 template stars observed with DEIMOS
                that included both giant and main sequence stars at
                several different metallicities.  Here, we use an
                expanded set of 42 stars carefully selected from the
                \citet{Cenarro:2001go} library of 706 stars.  The
                primary differences in the new template catalog are
                the inclusion of main sequence early-type stars to
                allow for the presence of Paschen lines, and M-type
                dwarfs to account for any molecular absorption
                features.  The resolution of the selected template
                stars is $1.5\,\mathrm{\AA}$, which is well matched to
                the spectral resolution of our spectra
                (1.5\,--\,1.7$\;\mathrm{\AA}$).
		
		\texttt{pPXF} requires reasonable initial guesses of the $V$ and $\sigma$ fitting parameters in order to efficiently find the best fitting LOSVDs.
		This is especially important for lower S/N spectra where noise artifacts (e.g., large skyline residuals) can create likelihood barriers in parameter space.
		Thus, we devised the following automated method for finding reasonable initial parameter values.
		First, we create a grid of 120 ($V$,$\sigma$) pairs that span a realistic range of values for the galaxies under study.
		Second, we run \texttt{pPXF} in two moment mode (i.e., not fitting $h_3$ or $h_4$) using each of these pairs as initial guesses.
		Third, the resulting ensemble of solutions and reduced chi-squared ($\tilde{\chi}^2$) values are used to determine the best initial guesses of $V$ and $\sigma$.
		Finally, these values and the appropriate penalty terms are used to run \texttt{pPXF} in four moment mode to derive the $V$, $\sigma$, $h_3$, and $h_4$ values that maximize the likelihood function.

After determining the best fitting LOSVD for each spectrum, a Monte Carlo method is used to estimate the uncertainty of each kinematic parameter ($V$, $\sigma$, $h_3$, $h_4$).		
First, we use the best-fit model LOSVD to create a realization of each spectrum, and then add noise using the inverse variance array.
Each model spectrum is subsequently refitted using \texttt{pPXF} with the penalty set to 0 to derive conservative error estimates as suggested by \citet{Cappellari:2004gm}.
This process is repeated 100 times per spectrum, with the error on each kinematic parameter taken as the standard deviation of the returned values.
The best fitting solutions generally have $\tilde{\chi}^2\!\le1$, indicating that the inverse variance arrays are realistic estimators of the pixel by pixel uncertainties of the spectra. 
In those cases where $\tilde{\chi}^2>1$, we attempt to derive conservative Monte Carlo errors by multiplicatively scaling the inverse variance array such that $\tilde{\chi}^2=1$.
Finally, the measured velocities are corrected for heliocentric motion, and each individual fitted spectrum is inspected by eye for quality control.
The right panel of Figure~\ref{fig:losvd} shows an example of a measured LOSVD and its $99\%$ uncertainty envelope.

		\subsection{Measurement Repeatability}

\begin{figure}
\begin{center}
\epsfxsize=8.5cm
\epsfbox{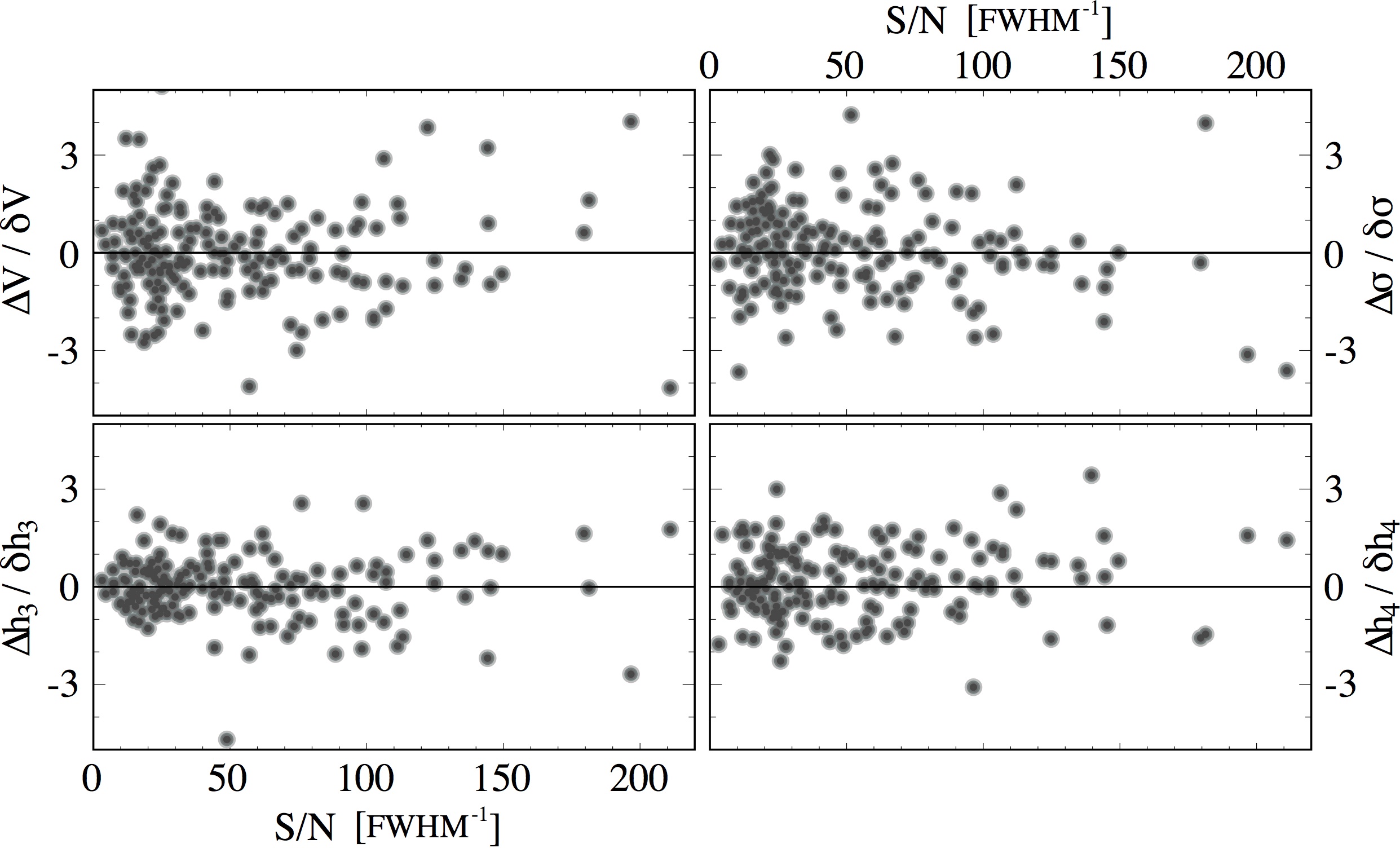}
\caption{Repeatability of measured velocity moments.
  The difference ($\Delta$) between measurements from
  spectra that are co-spatial to within 3 arcsec
  vs.\ S/N per resolution element.  Each $\Delta$
  value is normalized by the measurement errors added in
  quadrature.  If the derived errors, for velocity
  (top left), velocity dispersion (top right),
  $h_3$ (bottom left), and $h_4$ (bottom right), fully
  accounted for the true measurement error, then the
  standard deviation of the values on the $y$-axis would
  be exactly 1 in each panel.  However, at high S/N, both
  velocity and velocity dispersion show anomalously
  high values, suggestive of underestimated errors or a
  small systematic error term (see the text).
}
\label{fig:repeatability}
\end{center}
\end{figure}

			Repeated spectroscopic measurements can be useful in assessing the accuracy of Monte Carlo error estimates and in identifying unaccounted-for systematic error terms.
			We define two spectra to be co-spatial (i.e. repeated) if they are within 3 arcsec of each other.
			This search radius is small enough to ensure that only spectra from separate masks are paired.	
			Figure \ref{fig:repeatability} shows the difference between repeat measurements, normalized by their Monte Carlo errors added in quadrature, versus the average S/N per resolution element of each spectrum.			
			If the Monte Carlo errors capture all sources of uncertainty in the measurements, then the standard deviation of normalized differences should be nearly unity (i.e. $\tilde{\chi}^2=1$).
			Indeed, this is case for the vast majority of the measurements, reflecting the superb measurement precision of the DEIMOS spectrograph.
			
			However, it is also clear that the distribution widens with increasing S/N, indicating that the uncertainties are underestimated 
for the brightest spectra.
			This behavior is expected when there are small systematic error terms that only become non-negligible when the Monte Carlo error approaches zero.
			After some experimentation, we measured systematic errors in $V$ and $\sigma$ of $5\;\mathrm{km\,s^{-1}}$ and $6\;\mathrm{km\,s^{-1}}$, respectively.
			These values are added in quadrature to the previously derived Monte Carlo error estimates.
			The uncertainties on $h_3$ and $h_4$ were already conservative, leading to no apparent need for additional systematic error terms.

		\subsection{Literature Comparisons}\label{subsection:litComp}

\begin{figure}
				\begin{center}
					\epsfxsize=8.5cm
					\epsfbox{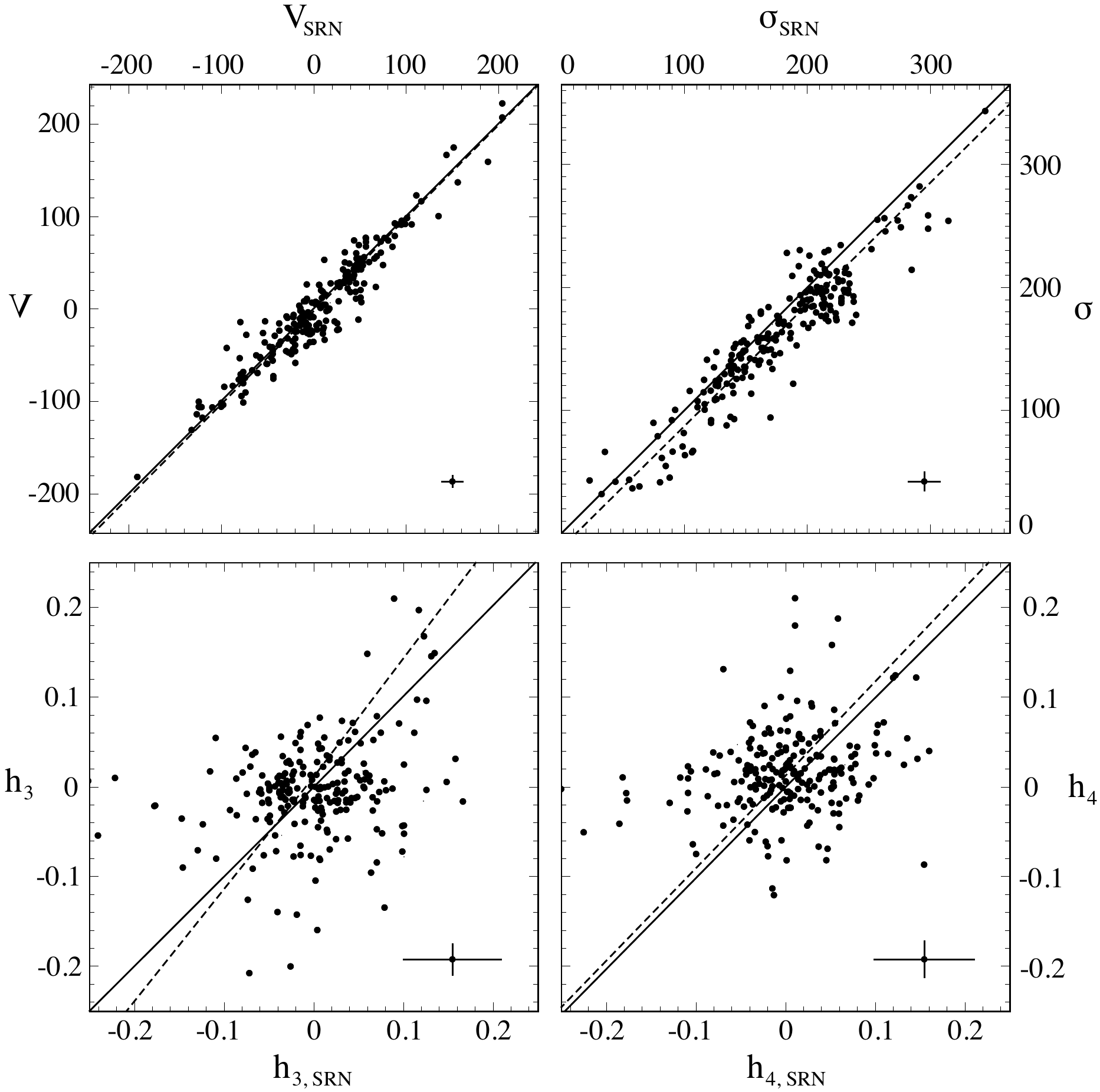}
					\caption{Literature comparison of kinematic measurements.
					Comparison between our measured values ($y$ axes) and literature values from 
					ATLAS$^{\rm 3D}$/SAURON (SRN; $x$ axes).
					Velocity (top left), velocity dispersion (top right), $h_3$ (bottom left), and $h_4$ (bottom right).
					The solid line is a 1:1 relation, while the dashed line denotes the best fitting linear relation to the data that takes
					into account the measurement uncertainties (where error bars in the figure show typical values).
					See the text for details.
					}
					\label{fig:srn_smg_compare}
				\end{center}
			\end{figure}

			While the majority of the DEIMOS spectra are at radii beyond 1~\Reff, there are still a fair number of spectra that are within the ATLAS$^{\rm 3D}$/SAURON observational footprint \citep{Cappellari:2011ej}.
			The galaxies in common between the two surveys span a range of systemic velocities, velocity dispersions, $h_3$, and $h_4$ values, providing a valuable opportunity to cross check measured kinematic quantities.
			Figure \ref{fig:srn_smg_compare} shows a comparison between measurements from each survey that are spatially coincident to within 3 arcsec.
			The dashed black lines in each panel are linear fits to the measurements that take $x$- and $y$-axis error bars into account while the solid black lines show the 1:1 relation.

			A few things are immediately obvious from this figure.  For instance, the general agreement
                        between the two surveys is very good, with the measurements showing a high
                        degree of correlation, and the median offsets for each parameter being very small. 
                        The exception is the velocity dispersion, where the SAURON values are
                        systematically higher than the DEIMOS values by $13\;\mathrm{km\,s^{-1}}$. 
                       We note that  \citet[][]{Murphy:2011hf} found higher velocity dispersions than SAURON for NGC~4486 (M87),
                       and that the mismatches were markedly reduced for a few galaxies after the \atlas\ improved reanalysis of
                       the original SAURON data \citep{Emsellem:2004kr}.

			Before discussing some specific galaxies, we should point out some important caveats regarding this comparison.			
			First of all, most of the matched spectra are near the outer edge of the SAURON field-of-view, and therefore represent some of the noisier measurements from that survey. Co-adding such data may lead to higher on average 
			velocity dispersions being measured. 
			Conversely, these data comprise the high S/N tail of our DEIMOS observations.
			Second, the spectral resolution and dispersions of each instrument are quite different.
			Operating in low resolution mode, SAURON has an instrumental velocity resolution of $\sigma_{\rm inst}=108\;\mathrm{km\,s^{-1}}$ and a dispersion of $66\;\mathrm{km\,s^{-1}\,pixel^{-1}}$.
			The corresponding values for our DEIMOS measurements are $24\;\mathrm{km\,s^{-1}}$ and $11\;\mathrm{km\,s^{-1}\,pixel^{-1}}$.
			Third, and perhaps most importantly, these two surveys utilize very different spectral regions, with SAURON probing shorter wavelengths (4800\,--\,5380$\;\mathrm{\AA}$) than our survey using the Calcium Triplet lines 
($\sim$8500$\;\mathrm{\AA}$).
			These spectral windows are quite different, each having benefits and drawbacks over the other.
			LOSVD fitting at bluer wavelengths benefits from more absorption features and fewer skylines than in the near-infrared. 
			However, this range is also more complicated, can have emission lines to contend with, and is sensitive to template mismatch issues that do not affect the near-infrared CaT \citep[][]{Barth:2002ba}.
			Given that the SAURON and DEIMOS spectral regions probe slightly different stellar populations, this could in principle be the cause of the systematic offset seen in velocity dispersion.
			This effect is explored in Appendix~\ref{sec:appC}, but we conclude that it is very unlikely to be responsible for the offsets we see. 
		
			In some cases, the discrepancy could be due to sampling issues in the SAURON spectra. 
			For example, in NGC~7457 the matched SAURON velocity dispersion values range between 30 and 70 $\mathrm{km\,s^{-1}}$ (i.e., at or below their spectral resolution) 
			while the corresponding DEIMOS values cluster tightly around 40 $\mathrm{km\,s^{-1}}$.  However, this cannot be the case for all galaxies. 

                            A possible contribution to the systematic offset may be the different values of the $h_4$ parameter in each survey. The $h_4$ parameter measures the kurtosis or how peaked the 
                            velocity profile is. The velocity profiles in the radial range of overlap of our data often have $h_4$ parameters close to zero, whereas those from the SAURON instrument are slightly 
                            positive. If the velocity profile is more peaked than a Gaussian, then a fit with $h_4$ = 0 will tend to underestimate the true width of the profile and hence derive a smaller velocity dispersion.    

						The primary focus of this paper is to present the large-scale kinematic structure of ETGs, and to highlight general trends with radius.
			To do this, we will be combining SAURON data and other literature data sets with our own in order to construct 2D kinematic maps (see the next section).
			Since we are concerned with general trends, we remove any mean systematic offset between the data sets so that the final maps do not have unphysical step functions in them.

		\section{Kinematic Mapping}\label{sec:kinematic_maps}
		
\begin{figure*}
\begin{center}
\epsfxsize=18.0cm
\epsfbox{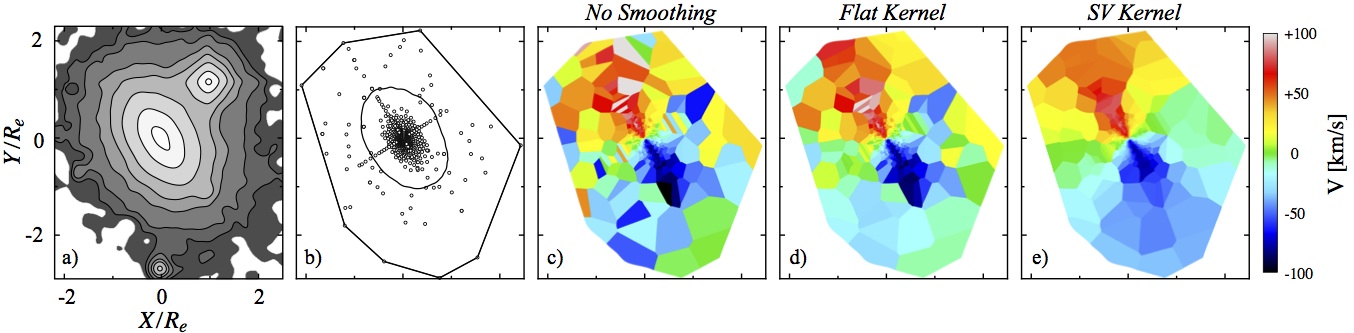}
\caption{Illustration of the interpolation-smoothing method for the velocity data of NGC 821.
A contour plot for the DSS image of NGC~821 is shown in (a), while the location of each stellar spectrum (SLUGGS and literature) is marked with a black circle in (b).
In the latter panel, the inner and outer solid black lines denote the 1~\Reff\ isophotal contour and the convex hull (i.e., bounding polygon) of the measurement locations, respectively.
As a point of comparison, a non-smoothed map of the velocity data is shown in (c).
Each solid color Voronoi bin corresponds to exactly one velocity measurement, and is colored according to the velocity scale at the far right of the plot.
Panels (d) and (e) contain maps that are smoothed using the 2D moving-window algorithm outlined in Section~\ref{smoothing_algorithm}.
The former uses a flat smoothing kernel, corresponding to a straight average, while the latter takes advantage of the spatial correlation information contained in the data by using the empirical semi-variogram (SV) as its smoothing kernel.
}
\label{fig:fig_explan}
\end{center}
\end{figure*}

In this section, we outline the methodology used to create 2D maps of $V$, $\sigma$, $h_3$, and $h_4$, for each galaxy in our sample.
These 2D kinematic maps are meant to provide an immediate sense of large-scale kinematic behavior and to highlight structures and general trends in the data.
While these types of maps are most commonly used to present the data products of IFU observations, they are also well suited for presenting other spatially resolved data sets, such as ours.
However, the spatial sampling density resulting from slitmask spectra is quite sparse compared to that of an IFU.
This results in large gaps in spatial coverage that require smoothing and/or interpolation to fill.

Previous approaches toward solving this problem include using a kinemetric model
\citep{Proctor:2009iy}, and kriging fitting \citep{Arnold:2011kp,Foster:2013}.
Both of these methods have their pros and cons,  e.g., the former requires a specific
 parameterization, while the latter is prone to over-smoothing or ringing.  
Here we introduce a
new methodology that requires no parameterization, preserves much of the
spatial information, and minimally smooths the data to suppress noisy measurements.

\subsection{Smoothing Algorithm}\label{smoothing_algorithm}
			
				We account for the empty regions between spectra by generating a Voronoi tessellation that assigns a spatial Voronoi bin to each stellar spectrum.				
				For example, the locations of the stellar spectra for NGC~821 are plotted in Figure~\ref{fig:fig_explan}(b), with the corresponding Voronoi bins visible as single colored polygons in Figure~\ref{fig:fig_explan}(c).
				We point out that the raw velocity map displayed in Figure~\ref{fig:fig_explan}(c) is noisy enough to obscure the general kinematic trends in the data, and is not very useful as is.			
				In principle, we could have partially ameliorated this problem by co-adding several noisy spectra together into a single higher S/N spectrum before fitting the LOSVD.
				However, while this approach is suitable for IFU measurements with ample spatial sampling, it would necessarily destroy valuable spatial information in our data.
				Additionally, many of our large-radius spectra are separated by rather large distances on the sky, and may be unsuitable for direct co-addition since they sample kinematically distinct regions of the galaxy.
			
				A better approach is to smooth the noisy data just enough to bring out the underlying trends, but not so much as to destroy spatial information.
				A common technique is to use a Gaussian smoothing kernel that de-weights more distant, and thus less correlated, points. 
				However, the varying spatial density of measurements makes a single-width Gaussian kernel unsuitable, since it would lead to maps with zones of under or over smoothing.
				The adaptive Gaussian kernel method used in \citet{Coccato:2009je} attempts to solve this problem, but still relies on an arbitrarily shaped Gaussian weighting function and an assumed kinematic model for the galaxy.
			
				\subsubsection{Weighting Function}
			
					Here, we use an empirical approach that uses the spatial correlation information contained in each data set to optimally smooth the associated kinematic maps.
					For each separate data set (e.g., $h_3$ values of NGC 821 or $\sigma$ values of NGC 1023), we calculate 
the semi-variance $\gamma$, as defined in
Equation~(\ref{eqn:semi_variance}) below.
This is measured by finding all pairs of
values that are separated by a particular distance~$\Delta$ to within
the specified tolerance~$\delta$, and then measuring half the variance
of their differences.

					\begin{eqnarray}\label{eqn:semi_variance}
						&&	\gamma(\Delta) = \frac{1}{2\,N_S} \sum_{(i,j)\in S}^{N_S} [Z(\boldsymbol{x}_i) - Z(\boldsymbol{x}_j)]^2	\\
						&&	S = \left\{(i,j) : -\delta < \left\lVert \boldsymbol{x}_i - \boldsymbol{x}_j \right\rVert - \Delta \le \delta; i > j \right\}					\nonumber
					\end{eqnarray}	
					Here $\boldsymbol{x}_k$ is the position of the $k$-th spectrum and $Z(\boldsymbol{x}_k)$ is the value of interest at that location.

                                        For each data set we  use the inverted form of the semi-variance as an
                                        empirical weighting function in the smoothing process outlined below.
			
				\subsubsection{Grouping Measurements}
			
					With the weighting function in hand, the next step is to combine the measurements together to suppress noise while not washing out legitimate large-scale structures.
					Our adopted technique is an extension of the one-dimensional moving-average algorithm, commonly used to filter time-series data, to two dimensions.
					The basic idea is to estimate the local mean of a data set for a series of spatial locations by taking the appropriately weighted average of nearby measurements.
					A key point here is that individual measurements can belong to many unique groupings of data, each centered at a different spatial location.
										
					The groupings (or clusters) of measurements are selected using a custom 2D search algorithm and the requirement that clusters meet certain connectedness criteria.
					Each group is assigned a unique identifier, based on the associated measurements.
					In each cluster, the semi-variance is calculated for each grouped measurement using (1) its distance to the centroid of the cluster and (2) the polynomial fit for the associated semi-variogram.
					These values are then combined with the Monte Carlo error estimates to derive a final weighting factor for each measurement.
					Next, these weights are used to compute a weighted sum of measurements for each cluster.
					At this stage, there are many unique clusters that overlap heavily with other groupings.
					The final step involves combining the overlapping clusters together, associating the results with the Voronoi bins associated with each individual measurement, and creating the final map.
												
					However, we have not yet specified how many points should go into a cluster---a crucial issue since this will control the degree of smoothness in the final map.
					Importantly, the number of points is not constant, and is determined separately for each grouping of measurements.
					We require that each cluster contains just enough points such that the resultant error of the weighted sum is within some tolerance (e.g., $\sim\!10$--$15\;\rm{km\,s^{-1}}$ for velocity maps).		
					This feature makes the smoothing method naturally adaptive because groupings in high spatial-measurement-density regions will be physically small and localized, and just the opposite in low-density regions.
					In basic terms, the smoothing is optimized so that the centers of maps are not over-smoothed, and the edges of maps are not under-smoothed.
					
					Figures~\ref{fig:fig_explan}(c) and (e) show maps for the raw and smoothed versions of the NGC~821 velocity data.
					The latter map reveals far more structure than the former, including a falling rotational amplitude along the major-axis.
					For comparison, Figure~\ref{fig:fig_explan}(d) shows the map that results from a flat weighting kernel, which is similar to taking a straight average in each cluster.

\subsection{Final Maps}\label{sec:finalmaps}

			Wide-field absorption line stellar kinematic
                        maps for the 22 ETGs in our sample are given in Appendix~\ref{sec:appA} and
                        displayed in Figure~\ref{fig:skims0}.  Each
                        map of $V$, $\sigma$, $h_3$, and $h_4$ is
                        optimally smoothed (see the previous section) to
                        highlight large-scale trends in the kinematics
                        while preserving the detailed structure
                        evident at the centers of those galaxies (from
                        long-slit and IFU observations).  Despite some residual
                        noise, it is clear that these galaxies exhibit
                        a wide variety of kinematic structures on
                        these large spatial scales and generally have
                        LOSVDs that are non-Gaussian at the few
                        percent level.

A brief description of each galaxy including its large radius
kinematics is given in Appendix~\ref{sec:appB}. We find a variety of 2D kinematic
structures. These include galaxies that rotate rapidly in their
centers but much more slowly farther out (e.g., NGC 3377), and rapid central rotators
that continue to rotate quickly to the limits of our starlight observations (e.g., NGC
2768). In the latter cases, the outer rotation is well aligned with the central regions.
All of the galaxies with slow rotation in the central regions
tend to remain slow rotators farther out (e.g., NGC 5846).
One particularly fascinating case is NGC~4365. The galaxy is well-known to host a kinematically
distinct core, where the very central regions show normal rotation along the
major axis, which twists to minor-axis rotation within the SAURON field-of-view.
Our data show that this misaligned rotation continues outward all the way
to the edge of our field of view at $\sim$~2~\Reff.

\section{Kinematical analysis and discussion}\label{sec:disc}

With the kinematics data and maps in hand, we carry out several
different analyses to characterize the 22 galaxies in our sample,
and to connect to the science questions raised in Section~\ref{sec:intro}.
We examine variations of angular momentum with radius in
Section~\ref{sec:angmom}, bulge--disk structure in Section~\ref{sec:BD},
and implications for galaxy formation in Section~\ref{sec:galform}.
						
                        \subsection{Angular Momentum}\label{sec:angmom}
			
The principal, novel parameter used in the \atlas\ survey is
the stellar specific angular momentum proxy, $\lambda_{R}$,
\begin{equation}
\lambda_R \equiv \frac{\langle R |V|\rangle}{\langle R\sqrt{V^2+\sigma^2}\rangle} ,
\label{eqn:lambda}
\end{equation}
calculated cumulatively inside either 0.5 or 1.0~\Reff.
This parameter is closely related to traditional $V/\sigma$ measurements,
and has limiting values of $\lambda_R=0$ to $\lambda_R=1$ for
minimum and maximum rotation support, respectively.
In \citet{Emsellem:2007hf}, a fixed value of $\lambda_{R}=0.1$ was
proposed as the dividing line between slow and fast rotators.
This divide was refined in \citet{Emsellem:2011br} by including a link with ellipticity:
\begin{equation}
\lambda_R = \lambda_0 \sqrt{\epsilon} \, ,
\label{eqn:fastslow}
\end{equation}
where $\lambda_0 \simeq 0.3$ and the exact value depends on whether 0.5 or 1.0~\Reff\ apertures are used.

The 0.5--1.0~\Reff\ apertures were motivated not only by the instrumentation
constraints but also by the finding that these observations reached
beyond most of the kinematically distinct components (KDCs) and into the region
where $\lambda_R$ apparently plateaued.
However, the asymptotic values of $\lambda_R$ including larger radius
data were not explored, and it should be noted that while 1\,\Reff\ is a
natural scale for the stellar luminosities and stellar masses of galaxies,
the natural scale for total masses and angular momenta is much farther out
(see the discussion in \citealt{Romanowsky:2012}).

Our goal is therefore to investigate further the scale dependence of
stellar specific angular momentum trends, and the robustness of the \atlas\
conclusions to the inclusion of larger-radius data.
Here we will take a first look at these issues, and defer a full
examination to another paper (C.~Foster et al., in preparation).
Also, in the interest of focusing on variations with radius, here we
will work with {\it differential} rather than {\it cumulative} values
(which require careful luminosity weighting).
This means that we will calculate a local $\lambda$ as in Equation~(\ref{eqn:lambda})
but in annuli rather than over filled apertures.

Our $\lambda$ calculations proceed as follows.  Each smoothed $V$ and $\sigma$
map  is subdivided into a fine 2D grid of pixels.
These pixels are binned into annuli with widths of 5~arcsec, and 
with fixed ellipticities from Table~\ref{tab:galaxies}.
The value of $\lambda$ within each annulus is calculated from
Equation~(\ref{eqn:lambda}), using the true galactocentric radius $R$
rather than the elliptical radius $R_m$, and assuming constant flux weighting per pixel.
Profiles of $\lambda$ with $R_m$ are generated using
an oversampled moving window approach, where each annulus is advanced
outward 1~arcsec relative to the previous one.
For stability, these profiles are computed only out to a radius $R_{\rm max}$ 
(see Section~\ref{sec:finalmaps} for definition).

\begin{figure}
\begin{center}
\epsfxsize=8.6cm
\epsfbox{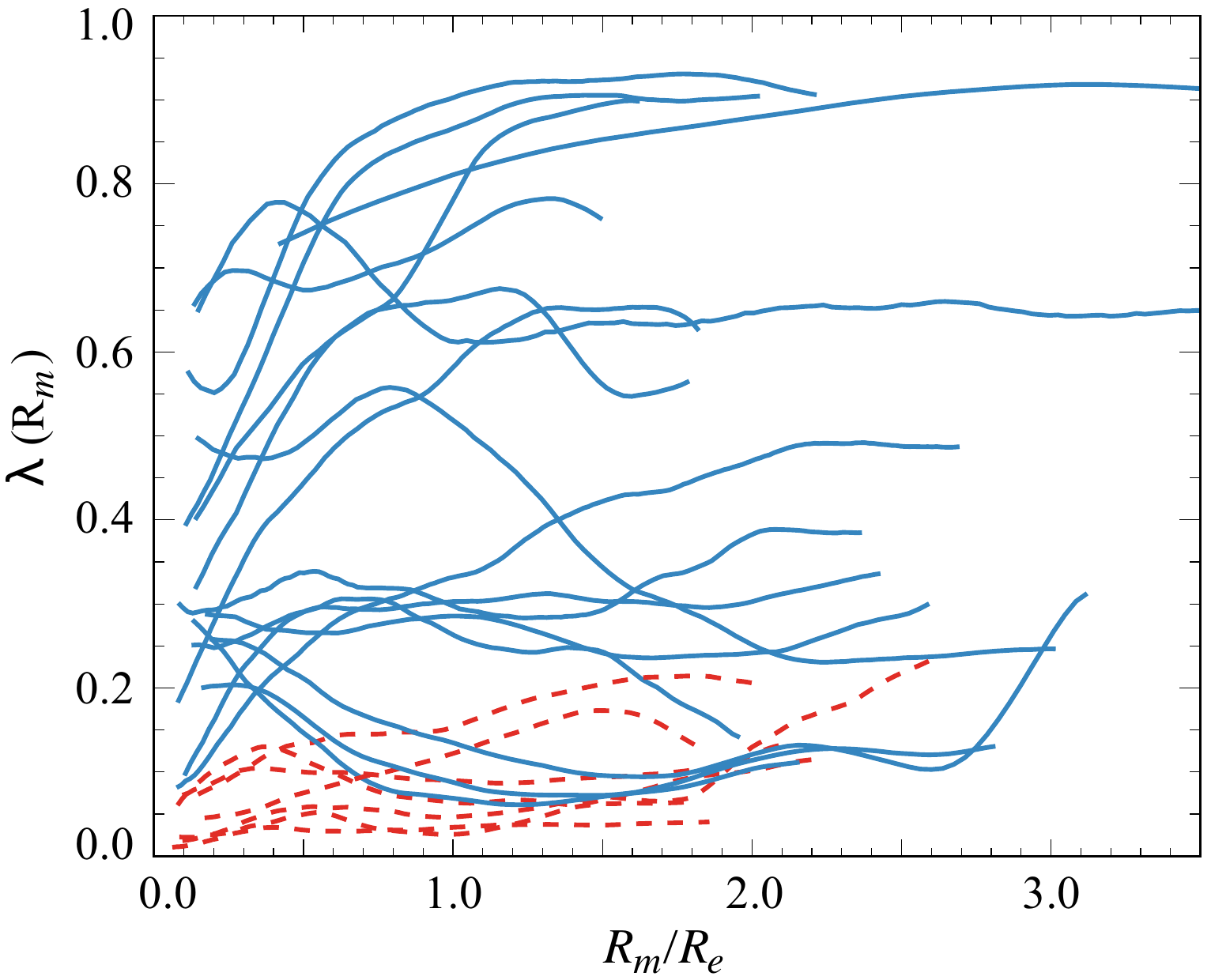}
\caption{Radial profiles of the local stellar specific angular momentum proxy, $\lambda$,
for our sample galaxies. Blue solid curves denote centrally 
fast rotators and red dashed curves denote centrally slow rotators.
A diversity of rotation profiles is seen beyond 1~\Reff, with centrally 
slow rotators remaining relatively slow, and centrally fast rotators ranging
from decreasing to constant to increasing $\lambda$.
}
\label{fig:lambda}
\end{center}
\end{figure}

In Figure~\ref{fig:lambda}, we show the resulting radial dependence of the
$\lambda$ parameter for our sample. Galaxies are color-coded by their
central fast/slow rotator designation. A diversity of $\lambda$ profiles is seen when the kinematics are probed beyond 1~\Reff.
Similar results for a sample of galaxies were found by \citet{Coccato:2009je} from PN kinematics,
and here we directly confirm that this complexity exists in the stellar kinematics
(as also seen by \citealt{Proctor:2009iy} in a much smaller galaxy sample;
variations in central kinematic profiles were also seen by \citealt{Emsellem:2011br} but are difficult to compare since these
were cumulative rather than differential $\lambda$ profiles).

The seven centrally slow rotators continue to have low angular momentum to large radii, 
indicating relatively slow rotation in the bulk of the stellar mass for these galaxies
(note that although $\lambda_R$ increases up to 0.2 for some of them, there is an accompanying increase of ellipticity,
which suggests a continued slow-rotator classification as in Equation~(\ref{eqn:fastslow})).
Slow rotating galaxies tend to be
the most massive galaxies and located at the high density cores of galaxy groups and clusters. They also reveal 
dynamically interesting features such as kinematically distinct cores (KDCs; e.g., NGC~3608, NGC~4365), 
minor-axis rotation (e.g., NGC~4365, NGC~4374, NGC~4486), and kinematic twisting (e.g., NGC~1407).
				
The centrally fast rotators, on the other hand, are more varied.
Half of the galaxies have $\lambda$ profiles that  continue to increase with radius  
(NGC 720, NGC 1023, NGC 2768, NGC 2974, NGC 4111, NGC 4526, NGC 4649, NGC 7457), while half have 
$\lambda$ nearly constant or decreasing 
(NGC~821, NGC~1400, NGC~3115, NGC~3377, NGC~4278, NGC~4473, NGC 4494, NGC 4697).
Some decreasing cases (NGC~821, NGC~4278, NGC~4473) even reach the domain of formally slow rotation.

\begin{figure}
\begin{center}
\epsfxsize=8.6cm
\epsfbox{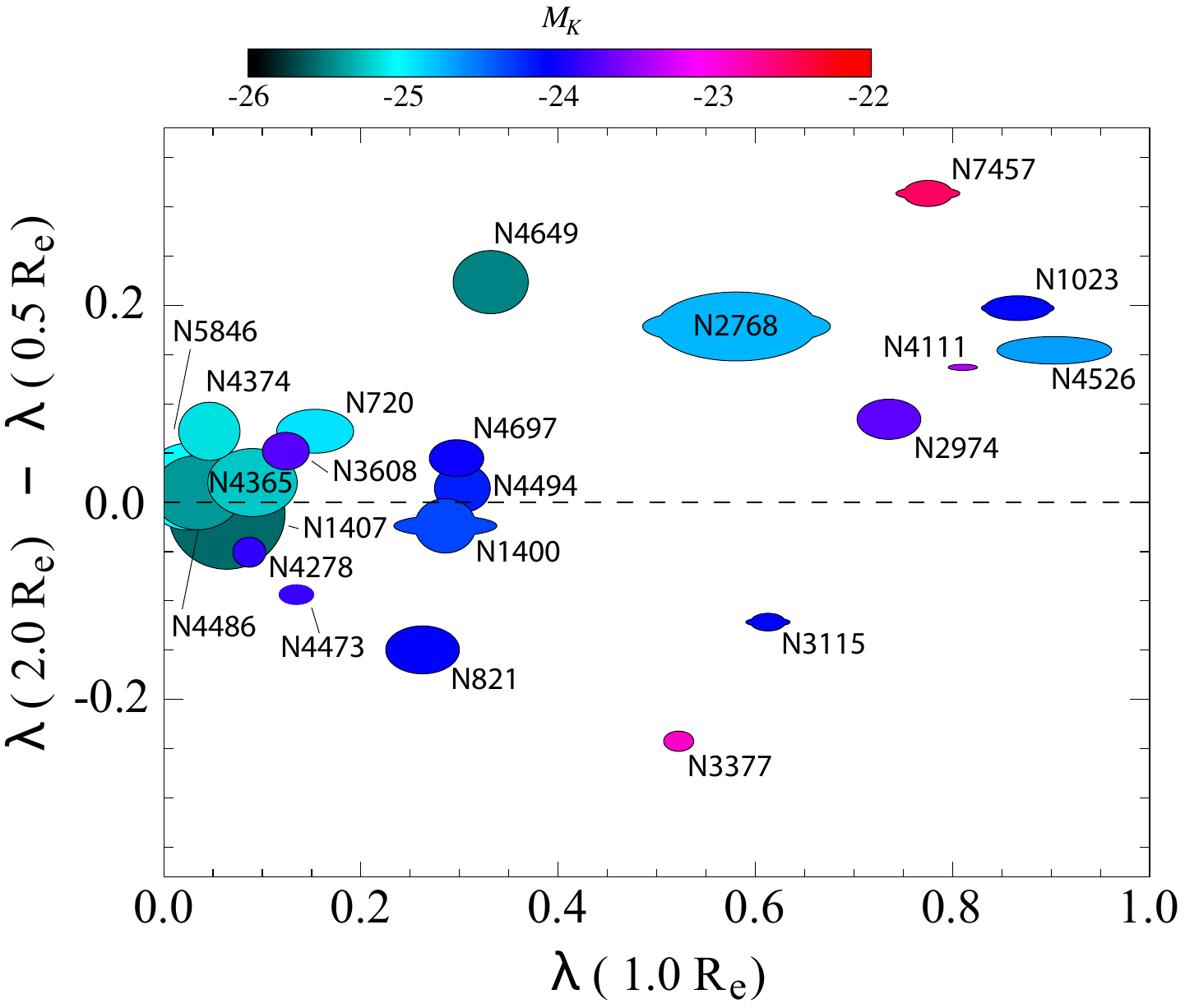}
\caption{Radial gradient of specific stellar angular momentum,
parameterized as the change in local $\lambda$ between 0.5 and 2.0\,\Reff\ scales (or $\Delta \lambda$)
vs.\ $\lambda$ at 1.0\,\Reff.
Each point is sized according to galaxy \Reff\ (in physical units),
while the shape reflects isophote flattening at 1\,\Reff\ and E or S0 classification, and the color reflects
$K$-band absolute magnitude (see the color bar at the top), respectively.
There appears to be a pattern for lenticulars to have $\Delta \lambda > 0$
(radially increasing profiles) and fast-rotator ellipticals to have
$\Delta \lambda < 0$ (radially decreasing).
}
\label{fig:fig_lambdaComp}
\end{center}
\end{figure}

We next consider a more simplified view of the $\lambda$ variations by introducing a $\lambda$-gradient, $\Delta \lambda$,
which is the change in local $\lambda$ between 0.5 and 2.0\,\Reff\
(or the farthest extent of the data, in a few cases where this is less than 2.0\,\Reff).
In Figure~\ref{fig:fig_lambdaComp}, we plot $\Delta \lambda$ versus the local $\lambda$ at 1.0\,\Reff\
for the 22 galaxies in our sample.  Galaxies with $\Delta \lambda > 0$ have rotation becoming more
dominant with radius, and less dominant for $\Delta \lambda < 0$.
The point styles reflect galaxy luminosities, ellipticities, physical sizes, and morphological types,
all of which could, in principle, relate to the gradients in specific angular momentum.

We do not discern any clear connecting patterns between $\Delta \lambda$
and the galaxy parameters including environment -- except for possibly the galaxy types.
The seven fast rotators with the most strongly increasing angular momentum
are all S0 or borderline E/S0 (if one considers that NGC~2974 was classified as an S0 or later
by all six observers in \citealt{Naim:1995}, and even exhibits spiral arms).
The remaining nine fast rotators, with more constant or declining $\lambda$,
include seven ellipticals, one E/S0, and one S0.

This intriguing segregation suggests that the standard morphological classifications may still be
relevant: although \atlas\ has established that the Es and S0s have similar kinematical properties inside 1\,\Reff, 
at larger radii their kinematics may diverge systematically.
Whether these galaxies are fundamentally the same or different will
require more detailed structural--dynamical analyses (including the links between radial variations in rotation and ellipticity), 
and ideally a larger sample of galaxies.
However, we will look at possible interpretations in the next subsections.

One notable case here is NGC~4649, which has the second highest luminosity in the entire \atlas\ sample of 260 ETGs.
It is somewhat peculiar in its central regions (at 0.35\,\Reff), where, unlike all of the other most massive ETGs, it is
(barely) a fast rotator ($\lambda_R=0.13$).  Its behavior is even more deviant in its outer regions, where
it becomes remarkably fast-rotating ($\lambda \sim 0.5$ at 2\,\Reff).
We will briefly consider the nature and origin of this galaxy in the following subsections.

In summary, we can now provide some insight into 
the radial dependence of specific angular momentum beyond one effective radius. 
We find that the central fast and slow rotator classifications remain valid
at larger radii in many, but not all, cases.
This suggests that centrally fast rotators are not necessarily
a homogeneous class of galaxies, and that the elliptical versus
lenticular galaxy morphological distinction may still be relevant.

\begin{figure*}
\begin{center}
\epsfxsize=3.3in
\epsffile{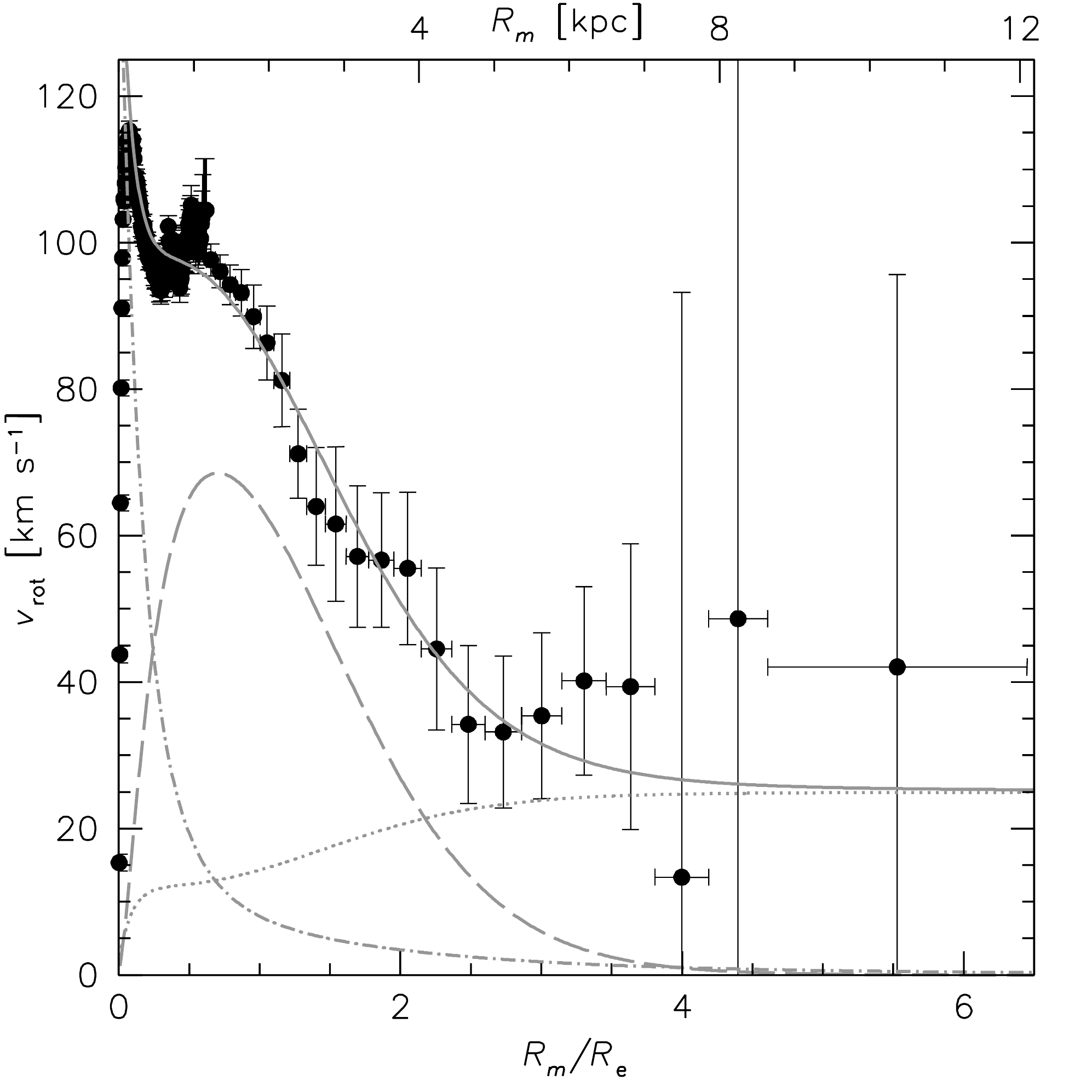}
\epsfxsize=3.3in
\epsffile{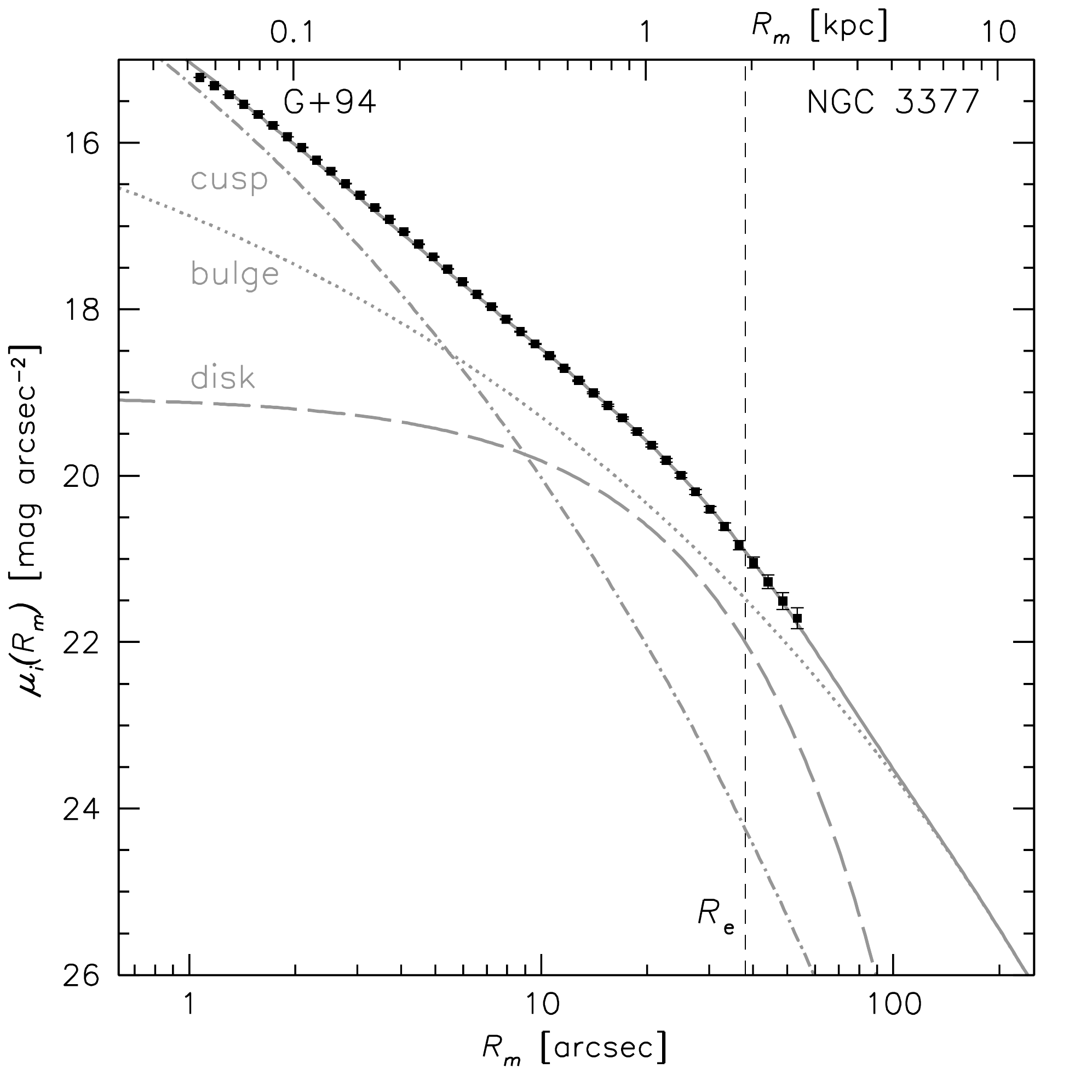}
\caption{Decomposition of the elliptical galaxy NGC 3377 into
cusp, disk, and bulge components (dot-dashed, dashed, and dotted curves,
respectively, with the sum shown as a solid curve).
{\it Left}: the rotation profile, with data as points with error bars plotted over the model profile.
{\it Right}: the azimuthally averaged $i$-band surface brightness profile,
with data plotted over the model fit. The vertical dashed line denotes one effective radius. 
We infer a fast-rotating disk-like component embedded in a slow-rotating boxy bulge.
\label{fig:decomp1}
}
\end{center}
\end{figure*}

Following the initial submission of this paper, there appeared a related survey of extended kinematics
in ETGs using the Mitchell Spectrograph (formerly VIRUS-P) integral field unit \citep{Raskutti14}.
Although detailed comparisons with our results are difficult owing to the low spatial resolution of the VIRUS-P survey at both small and
large radii, the qualitative conclusions appear the same: 
Raskutti et al.\ found that the slow and fast rotator classifications generally remain valid out to the limit of their data,
but with a few galaxies showing drops in rotation and extended kinematically distinct components.

We note that even our stellar kinematics data are still restricted
in radius to $\sim$\,2.5\,\Reff, and we cannot test how the results may change much farther out 
(where we recall that $\sim$\,5\,\Reff\ is the characteristic scale length for total stellar angular momentum).
Indeed, there are already two cases known from GC and PN kinematics
where fast rotation at $\sim$\,2.5\,\Reff\ dives toward slow rotation at larger radii (NGC~1023 and NGC~3115;
\citealt{Noordermeer:2008kx,Arnold:2011kp}).

\subsection{Embedded Disks}\label{sec:BD}

One key finding from the previous subsection
is the dramatic decrease of local angular momentum with radius in
a significant fraction of the galaxies.
\citet{Coccato:2009je} suggested a natural interpretation, that the
central regions host a significant stellar disk which fades toward
larger radii.  Hence the central kinematics may reflect a combination
of disk and bulge components, while the outer kinematics are ``pure bulge."

This interpretation leads us to
one of the recurring themes in extragalactic astronomy, of
decomposing galaxies into their constituent parts,
primarily bulges and disks.  This means determining the
relative fractions of these subcomponents in luminosity,
as well as their subcomponent density profiles and flattening.
There have been many techniques and software packages
developed to carry out bulge--disk decomposition based on photometry, 
but the results are notoriously model-dependent and uncertain.

In this context, kinematics offers a ray of hope, as it may be used in combination
with photometry to alleviate the degeneracies in deprojection and decomposition
\citep{Cinzano:1994tl,Scorza:1995ty,Romanowsky:1997,Magorrian:1999,Cappellari:2013b}.
The potential here is evident by the clear disk-like features
found in the \atlas/SAURON stellar kinematic maps,
when the surface brightness maps appeared much more ambiguous about the underlying galaxy structure.
In practice, however, photometric--kinematic decomposition has involved
fairly arduous modeling (e.g., based on detailed LOSVD shapes),
and the \atlas\ survey has so far applied only a conventional photometric approach \citep{Krajnovic:2013ho}.

We suggest instead a new approach where the availability of
wide-field stellar velocity maps may permit relatively painless and
well-constrained basic bulge--disk decompositions.
The starting assumption is that the disk and bulge components
each have fairly uniform rotational properties, and that the overall variation in rotation with radius
then reflects the transition between disk and bulge dominance.
The combined photometric and rotation profiles may then be fitted
with a disk--bulge model, as has been shown in recent pioneering
work using PN and GC data to decompose S0 galaxies
\citep{Cortesi:2011df,Cortesi:2013b,Forbes:2012bi}.
The same idea could also be used for fast-rotator ellipticals,
where kinematic transitions are found at smaller radii than in the S0s,
and our SKiMS data could be sufficient to constrain the decompositions.

As an illustration, we consider the case of NGC~3377, a low-luminosity
elliptical with a strong decline in $\lambda$ beyond $\sim$\,1~\Reff. 
The decline in rotation amplitude alone is even stronger, as shown
in the left panel of Figure~\ref{fig:decomp1}, which also
combines PN kinematics data in order to verify the behavior at larger radii \citep{Coccato:2009je}.
At a glance, we may suppose that the disk-to-bulge transition occurs at $\sim$\,1.5~\Reff.

\begin{figure*}
\begin{center}
\epsfxsize=3.55in
\epsffile{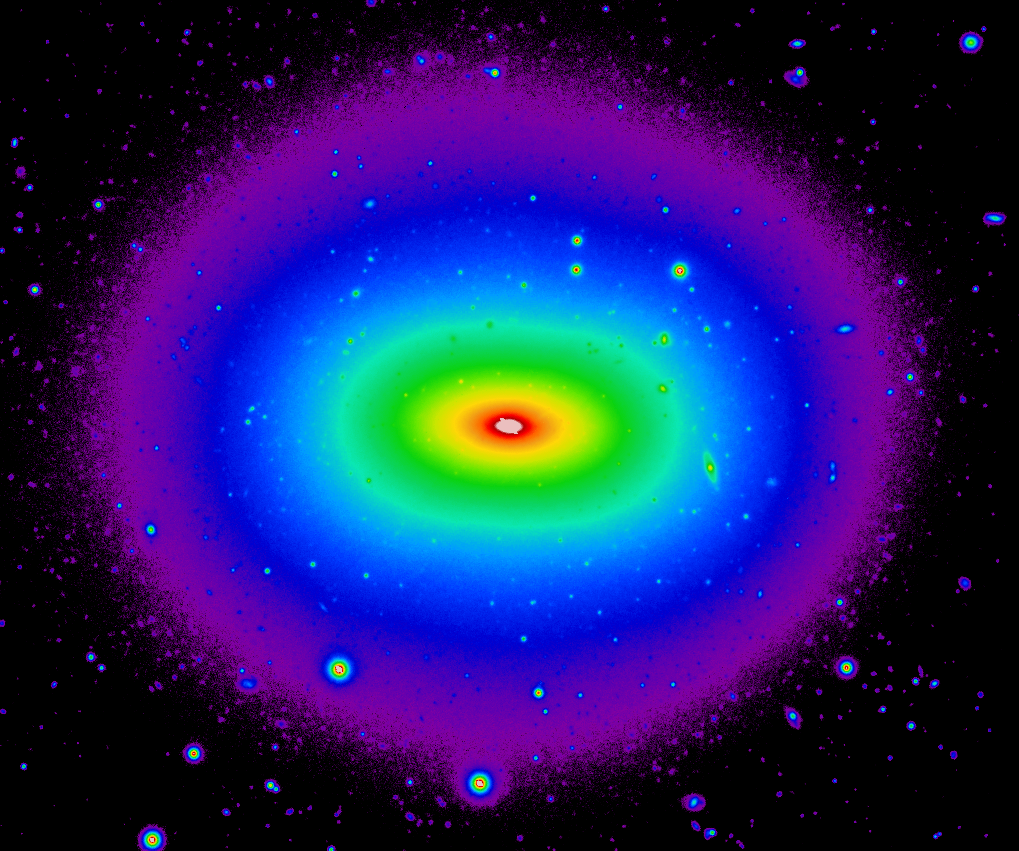}
\qquad
\epsfxsize=3.25in
\epsffile{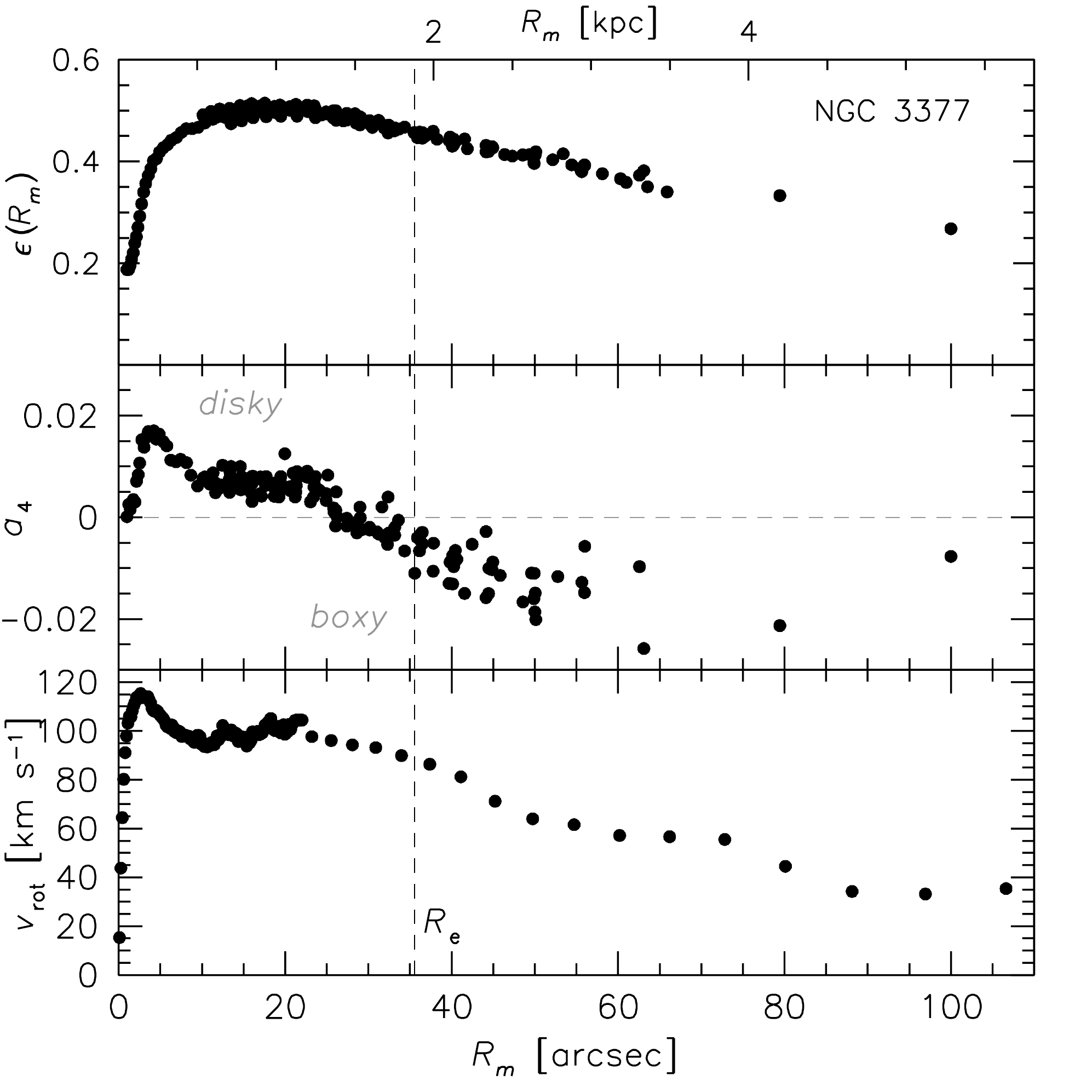}
\caption{Two-dimensional photometry of NGC~3377.
{\it Left:} $g$-band Subaru/Suprime-Cam image ($\sim$\,20\,kpc field of view),
rotated to align with the galaxy axes, and
with an arbitrary color scale used for contrast.
{\it Right:}
the top two panels show profiles of the isophote ellipticity and shape
parameter (fourth-order cosine term $a_4 > 0$ is disky, $< 0$ is boxy,
taken from literature studies in various optical filters;
\citealt{Jedrzejewski:1987un,Michard:1988,Scorza:1995ty,Peletier:1990dy,Goudfrooij:1994vj}).
Both expressions of the data suggest a disk-like component embedded in a
round and boxy bulge, which may be compared to the 1D rotation profile
(bottom right panel) and to the 2D velocity map in Figure~\ref{fig:skims0}.
\label{fig:decomp2}
}
\end{center}
\end{figure*}

The photometric data for NGC~3377 are shown in the right panel, as
a 1D surface brightness profile \citep{Goudfrooij:1994vj}.
We have carried out a three-component fit, where, in addition
to standard disk and bulge components (\citealt{Sersic:1968} indices $n=1$ and $n=4$),
we have included a cusp of ``extra light'' based on the decomposition of \citet{Hopkins:2009gc}.
This third central component is generic to many ellipticals
(where it may trace ancient starbursts) and here impacts the
overall decomposition results for the rest of the galaxy.
Our fitted parameters are \Reff$=3.5^{\prime\prime}$ for the cusp,
a disk scale radius of $R_{\rm d}=13.9^{\prime\prime}$, and
\Reff$=51.7^{\prime\prime}$ for the bulge.
The fractional luminosities in the different components are 19\%, 20\%, and 61\%, respectively.
These contrast with previous disk-fraction inferences of 5\% and 56\%--82\%
from \citet{Scorza:1995ty} and \citet{Krajnovic:2013ho}, respectively\footnote{NGC~821
is a similar case, where a cursory inspection of Figure~\ref{fig:skims0}
does not seem to support either
the 5\% or the 74\% disk-fraction estimates by \citet{Kormendy:2013} and \citet{Krajnovic:2013ho}.
The spread of these estimates, and the latter study's inference of zero disk fraction for some
S0s (e.g., NGC~2768: \citealt{Forbes:2012bi,Cortesi:2013b}), 
demonstrates the severe limitations of photometric-only decompositions.}.

After fitting the photometry, we assign
simple constant $v_{\rm rot}$ values to each component, constructing a total $v_{\rm rot}$ profile
that we compare to the data.
We find that values of 160~\kms, 170~\kms, and 25~\kms (cusp, disk, bulge)
fit the rotation profile remarkably well
(note that the maximum circular velocity was found from dynamical
modeling to be 212~\kms; \citealt{Cappellari:2013a}).
We also found it difficult to reproduce this profile
without a decent photometric fit that included all three components,
and we further note that the \atlas\ decomposition, with
\Reff$=7^{\prime\prime}$ for the bulge and $R_{\rm d}=22^{\prime\prime}$
for the disk, is not consistent with these kinematic data.

This simplistic 1D model fit is meant to be schematic only, and not
to fully explore the parameter uncertainties, nor to make use of all of the available data.
In particular, a robust fit should be carried out in two dimensions
for both the photometry and the kinematics, and would also incorporate stellar population differences
(cf.\ \citealt{Johnston:2012kl,Coccato:2013a,Fabricius14}).
Figure~\ref{fig:decomp2} illustrates some of the available 2D photometric information, 
where the galaxy appears flattened and disky in the inner regions, and round and boxy in the outer regions.
This morphology supports the notion of a distinct disk component
embedded in a more extended bulge, and the isophote shape transition
radius seems consistent with the kinematic transition
(for more discussion of how ellipticity correlates with rotational support at fixed anisotropy, see
\citealt{Emsellem:2011br}, Appendix B).
A similar morphology has also been found for another case from our
sample with a declining rotation profile (i.e., NGC~4697; \citealt{Scorza:1995ty,Huang:2013b}).

As another consistency check on the embedded-disk scenario, we consider
the velocity dispersion maps (Figure~\ref{fig:skims0}).
We expect that along the major axis, a cold disk will produce a lower
overall velocity dispersion, relative to the minor axis, which should
be dominated by the hotter bulge component.
Unambiguous instances of this behavior
are indeed observed in the maps for three of the declining rotators
(NGC~821, NGC~3115, NGC~3377).

Our final set of clues comes from the Gauss-Hermite $h_3$ moment.
An anti-correlation between $h_3$ and $V/\sigma$ implies an
extended tail of stars in the direction of the galactic systemic velocity, 
which in turn suggests a fast rotating disk superimposed on a more
slowly rotating bulge \citep{Rix:1990kp,Scorza:1995ty}.
Such an anti-correlation has been frequently observed in the central regions of ETGs
\citep[e.g.,][]{Bender:1994vo,Krajnovic:2008fi}, 
and now we have the opportunity to see what happens in their outer regions.

In our sample, all of the fast rotator galaxies with embedded disk-like structures
identified by their rapid rotation in the previous section have
anti-correlated $V/\sigma-h_3$ values in the same region.  The same is true
for the massive galaxy NGC~4649 as well as for the slow rotator NGC~1407,
which has a rotating structure within 1~\Reff\ that is misaligned with the
photometric major-axis and clearly exhibits anti-correlated
$V/\sigma-h_3$.  These observations support the embedded disk picture.

Intriguingly, the anti-correlation extends beyond the regions
of obvious fast rotating or disky embedded components.
This includes not only along the major axis at larger radii,
but also off-axis and well away from the disk.
The implication, along with the observation of weak kinematic twists,
may be that the central disk is embedded in a much puffier,
hot-disk component (cf.\ \citealt{Kormendy:2012})---which
seems consistent with the \atlas\ finding that these galaxies remain oblate--axisymmetric
out to $\sim$\,2.5--3\,\Reff, based on kinematic alignments \citep{Krajnovic:2011jj}.
These components could in turn be embedded in a more
extended slow-rotating spheroid that would produce a
second rotation drop, given observations at even larger radii \citep[see e.g.,][]{Arnold:2011kp}.
We thus speculate that ETGs may be kinematically multi-component systems,
with multiple layers of embedded structure.

In summary, our results reinforce the long-standing picture that
many ellipticals as well as lenticulars host embedded stellar disks.
These two morphological classes may simply differ in the relative
masses and length scales of the disks, and represent a continuous family
that may even connect smoothly to spirals
(e.g., \citealt{Scorza:1995ty,Cappellari:2011kh}).
Whether or not the galaxies appear kinematically different would then
depend on the spectroscopic field of view relative to the disk scale length,
which is nearly always larger than the SAURON aperture, but in many
cases smaller than our DEIMOS aperture.

If this simple picture is correct, it could validate the claims from \atlas\ that 
Es and S0s comprise a unified family of fast rotators.
However, just because two galaxies have disks and bulges does not necessarily
mean that they are closely related:
it could be a case of parallel evolution, as in the wings of birds and bats whose homology is
driven by common physical principles.
Resolution of this question will require detailed examination of the properties of the galaxy subcomponents
and of their scaling relations (see, e.g., Figure~34(b) in \citealt{Romanowsky:2012}).

\subsection{Two-phase Assembly}\label{sec:galform}

The existence of kinematical transitions and embedded disks discussed
in the preceding subsections may tie in with the picture of
two-phase galaxy assembly summarized in Section~\ref{sec:intro}.
Here the general idea is that a phase of rapid, dissipative processes
formed the central regions, followed by a prolonged period of dry minor mergers.
This pathway suggests that ETGs are
composed of an oblate rotating inner structure embedded within a more extended, spheroidal, and slowly rotating halo comprised of accreted stars
(e.g., \citealt{Vitvitska:2002ju,Cappellari:2013b}).
The transition radius between these two components will be different for each galaxy, but we can get a rough idea where it might occur from photometric observations between $z\,=0$ and $z\,=2$, which suggest significant mass growth in massive ETGs beyond a radius of 5~kpc or $\sim$\,1--2~\Reff\ \citep{vanDokkum:2010bn}.
This is indeed around the region where many of the galaxies in our sample show transitions in specific angular momentum
(\citealt{Raskutti14} also found several cases of kinematically distinct cores on 2--7\,kpc scales).

In more detail, we are only aware of one published prediction of large-scale
galaxy kinematics from a suite of hydrodynamical simulations of galaxy formation in 
a cosmological context, by \citet{Wu14}.
This study found a variety of rotation profiles that at first glance
resembles our observational findings.
However, the central regions of their galaxies did not reach anywhere
near the rapid rotation found in many \atlas\ galaxies.
This discrepancy is probably a combination of different selection effects
and of numerical resolution that was inadequate for reproducing cold disks.
				
The latter point is crucial, as our results suggest a rather
specific scenario for the first phase of assembly: for at least some of the galaxies,
the inner regions formed with a distinct stellar disk that has persisted until
the present day while being surrounded by a more extended,
spheroidal, and slowly rotating component.
This scenario would agree with observations of high-redshift galaxies (e.g., \citealt{vanderWel11,Chang13}),
and with the semi-analytic schematic of
\citet{Khochfar:2011gh}, where all galaxies evolve through
a continual interplay of cold gas cooling into disks, and mergers that erode or destroy the disks.
The most massive galaxies are expected to have the most active assembly
histories and thereby to have lost their initial disks.
For the lower-mass, fast rotators, the accretion was milder and more of the
initial diskiness was preserved.

We cannot rule out a reverse scenario for the embedded disks, where they
are grown {\it after} the bulge through later gas infall and star formation,
such as could occur during major mergers
(e.g., \citealt{Hoffman:2009ju}).
However, the rotational alignment between the disk and bulge
suggests a primordial rather than a major-merger origin
(see \citealt{Scorza:1995ty}), which
can produce more severe kinematic decoupling between the inner and outer
regions than we have generally observed \citep{Hoffman:2010dr}.
Very disky central regions are also difficult to produce
through major mergers \citep{Bois:2011kc}.

One potential candidate for a major merger remnant is NGC~4649,
which, as a very massive galaxy, is the most likely to have experienced a dry major merger (\citealt{Khochfar:2011gh};
see also the relatively high observed fraction of fast rotators among brightest cluster galaxies in \citealt{Jimmy13}).
It has remarkably high, apparently disk-like, outer rotation
(Section~\ref{sec:BD}), which suggests a massive S0 as a progenitor.
More detailed analysis of kinematics in this galaxy, along with others
in our sample, should be compared more directly to simulations in
order to better illuminate their formation mechanisms.

\section{Concluding Remarks}\label{sec:conclude}

We have presented wide-field (to $\sim$\,2.5~\Reff) kinematic maps of $V$, $\sigma$, $h_3$, and
$h_4$ for 22 nearby early-type galaxies.  Our sample consists of 16 centrally fast rotators and 6 centrally slow
rotators---close to the observed fractions measured in the volume limited sample from \atlas\
\citep[][]{Emsellem:2011br}.  These galaxies span a range of sizes
(\Reff{}~$=$~0.8--8.6\, kpc), ellipticities (0.04--0.58), luminosities ($-22.4\!>\!M_K\!>\!-25.5$),
morphologies (S0--E0), and environments (field, group, and cluster).  Given this
diversity of properties, it is perhaps unsurprising that these galaxies also exhibit
a range of large-scale kinematic behaviors (see Figure~\ref{fig:skims0}).

The wealth of structure evident in these large-scale kinematic maps will require
detailed dynamical modeling to fully understand and interpret within the context of
ETG assembly.  Here, our goal was to present the data and our analysis method.  We also
highlight some general trends. For example,
we find that the centrally slow rotator ETGs, which reside almost exclusively in the high
density cores of groups and clusters, remain slowly rotating in their halos. 
On the other hand, the centrally fast rotating galaxies exhibit
different behaviors at large radius with about half of our sample continuing to rotate rapidly at large
radii while the other half have declining angular momentum profiles so that they rotate
slowly in their halos. In other words, observations of central kinematics miss a
significant part of the overall structural and dynamical picture of ETGs.

The variations of rotation with radius appear to correlate with
the traditional galaxy morphological types (E and S0), and we
suggest that it would be premature to retire such classifications.
The rotational gradients are probably connected to the relative
dominance of embedded stellar disks, and we provide a conceptual
demonstration of how wide-field kinematics data might be
used to dramatically improve bulge--disk decompositions.

The dramatic cases of rapidly declining rotation profiles suggest
galaxies with distinct or decoupled inner and outer regions.
Such transitions may be fossil signatures of two-phase galaxy formation,
although they have yet to be produced in detail by cosmological simulations.

Future papers in this series will further characterize and explore the interesting structures shown here,
while the data itself will become public at the conclusion of the survey.

	\acknowledgements
We thank A.\ Burkert for useful discussions, E.\ Ramirez-Ruiz for computational support,
R.\ van den Bosch, M.\ Cappellari, E.\ Emsellem, D.\ Krajnovi{\' c}, and T.\ Lauer for helpful comments, and the referee for a positive report.
The data presented herein were obtained at the W. M. Keck Observatory, operated as a scientific partnership among the California Institute of Technology, the University of California and the National Aeronautics and Space Administration, and made possible by the generous financial support of the W. M. Keck Foundation. 
Based in part on data collected at Subaru Telescope, which is operated by the National Astronomical Observatory of Japan.
We thank the NSF for a Graduate Student Research Fellowship and Sigma-Xi for their financial support. We thank the National Science Foundation for funding via AST-0909237 and
AST-1211995. We thank the ARC for funding via DP130100388. 

\begin{figure*}
				\begin{center}
					\epsfxsize=17cm
					\epsfbox{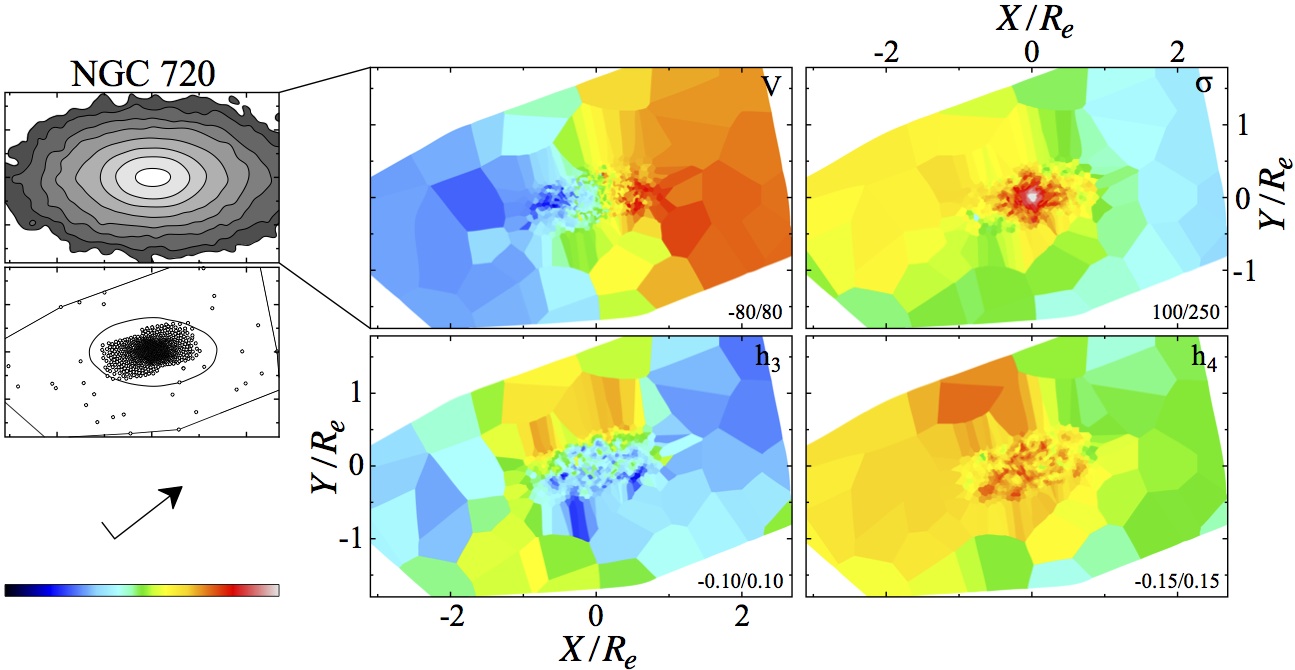}
					\epsfxsize=18cm
					\epsfbox{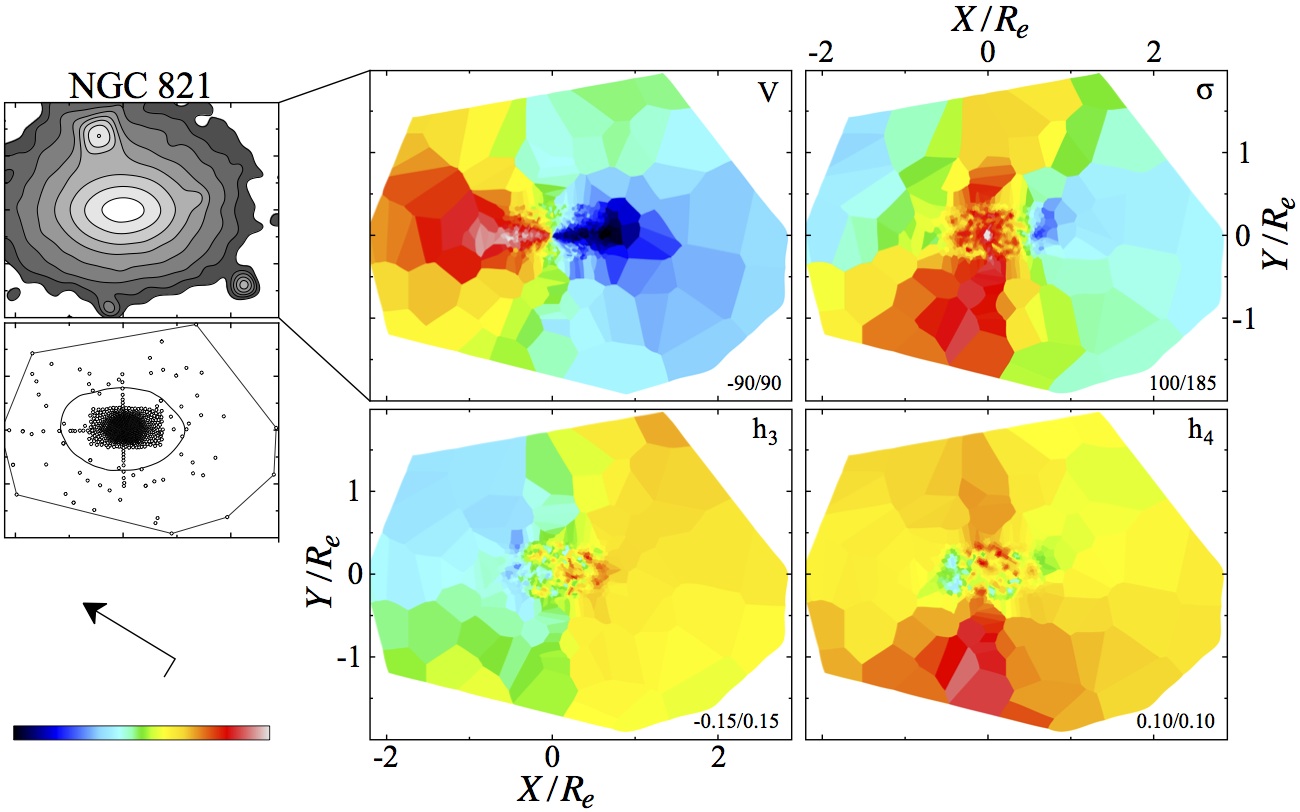}
					\caption{Smoothed kinematic maps of 22 early-type galaxies. 
					A separate subplot for each galaxy includes a DSS image (top left), a plot of stellar spectra (SLUGGS and literature) measurement locations (middle left), and an arrow indicating north and east.
					In the data location plot the 1~\Reff\ isophote is represented by a solid black ellipse, while each separate spectrum is denoted by a black dot surrounded by the convex hull of measurement locations (solid black polygon).
					Kinematic maps of $V$ (top middle), $\sigma$ (top right), $h_3$ (bottom middle), and $h_4$ (bottom right) are smoothed using the algorithm described in Section \ref{smoothing_algorithm}.
					The coloring scheme is set according to the color bar (bottom left), with the minimum/maximum of each range indicated in the bottom right corner of each panel; $V$ and $\sigma$ are in km\,s$^{-1}$, $h_3$ and $h_4$ are unitless. The color range is optimized for each galaxy. 
					The plotted spatial scale, which varies from galaxy to galaxy, is only displayed for the $\sigma$ and $h_3$ maps; with $X$- and $Y$-axes normalized by the \Reff\ values listed in Table~\ref{info}.
					See Section~\ref{sec:finalmaps} for further details.
					}
					\label{fig:skims0}
				\end{center}
			\end{figure*}

			\addtocounter{figure}{-1}
			\begin{figure*}
				\begin{center}
					\epsfxsize=18cm
					\epsfbox{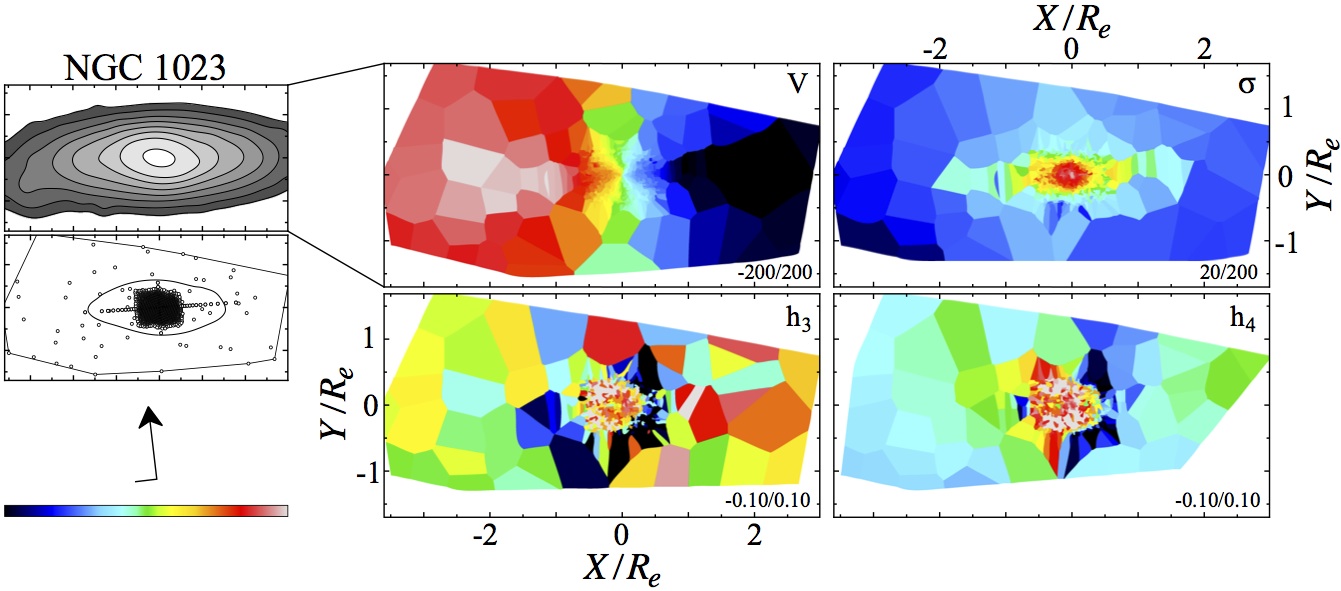}

					\epsfxsize=18cm
					\epsfbox{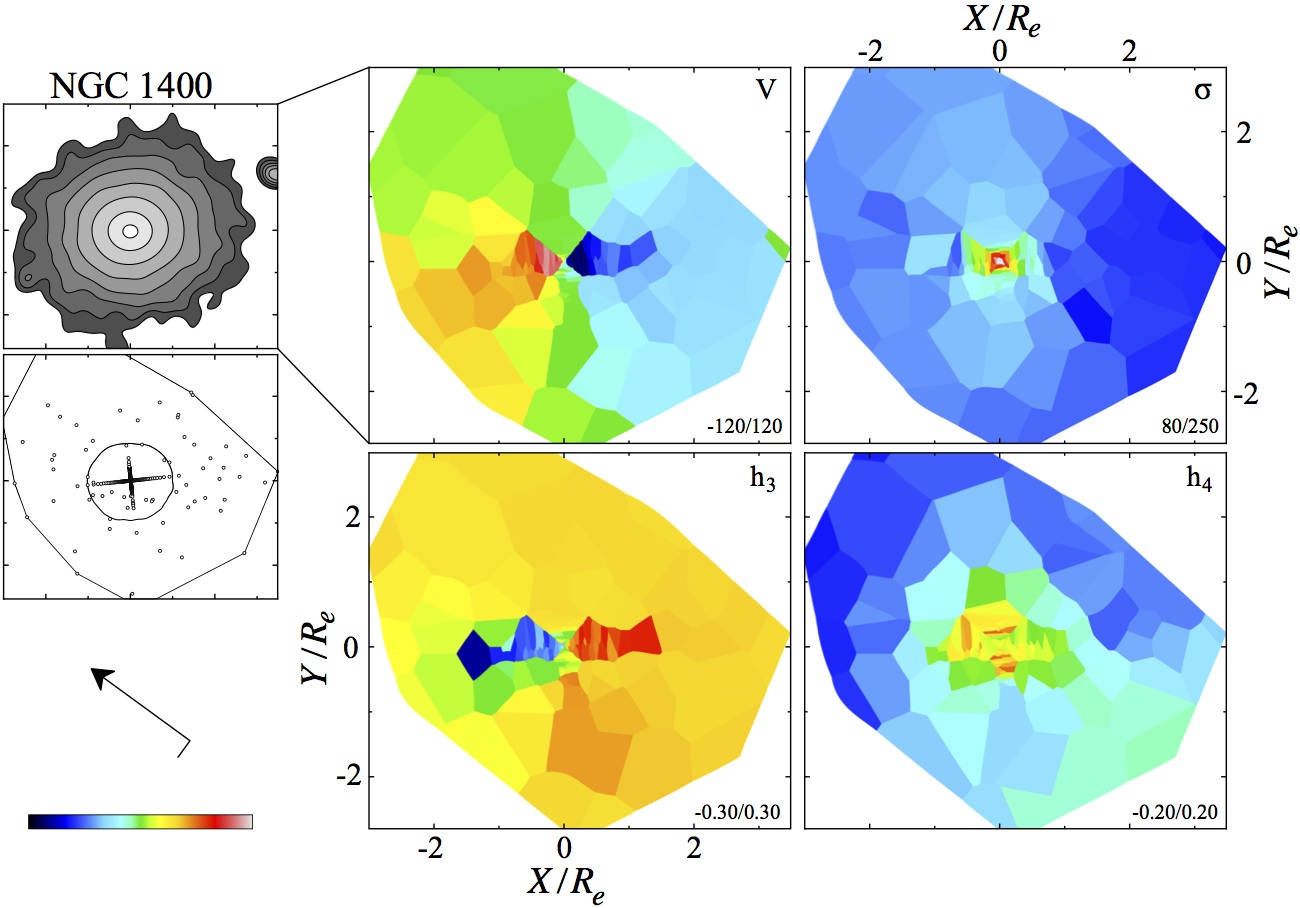}
					\caption{Continued.
					}
				\end{center}
			\end{figure*}			

			\addtocounter{figure}{-1}
			\begin{figure*}
				\begin{center}
					\epsfxsize=18cm
					\epsfbox{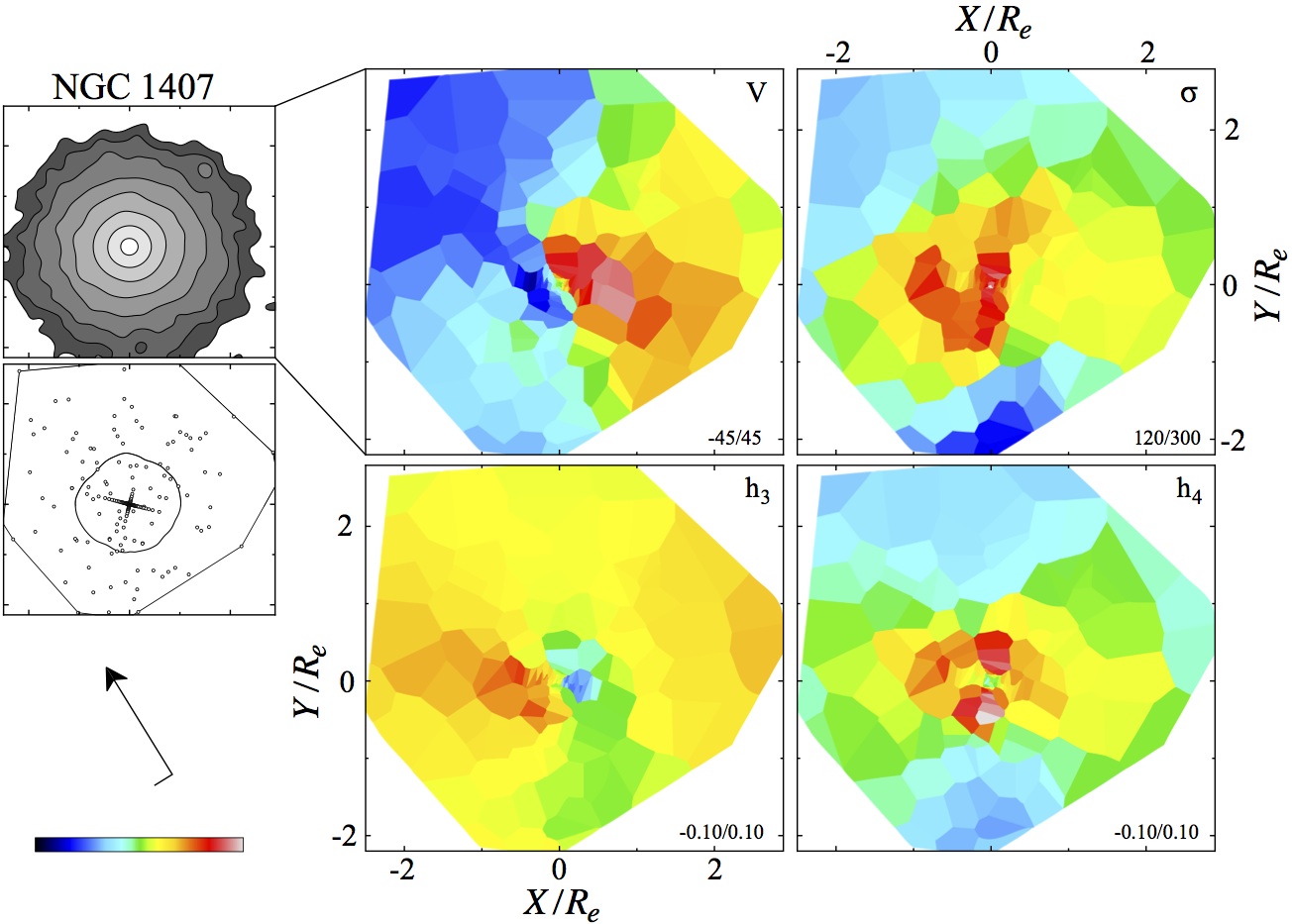}
%
					\epsfxsize=18cm
					\epsfbox{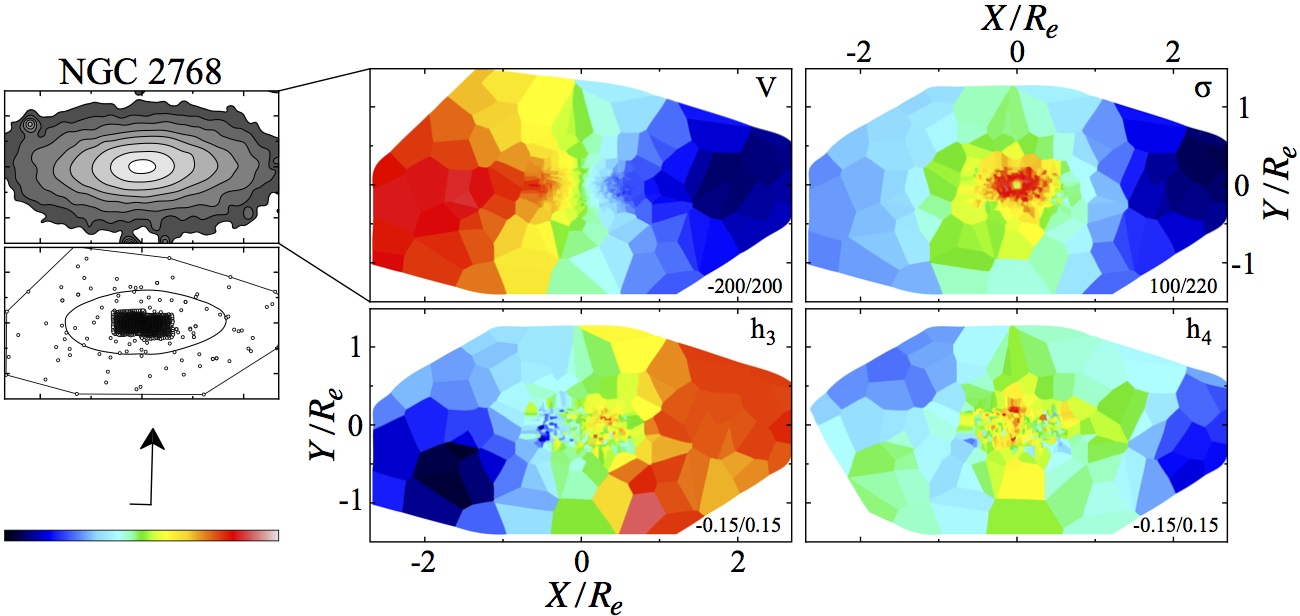}
					\caption{Continued.
					}
				\end{center}
			\end{figure*}

			\addtocounter{figure}{-1}
			\begin{figure*}
				\begin{center}
					\epsfxsize=18cm
					\epsfbox{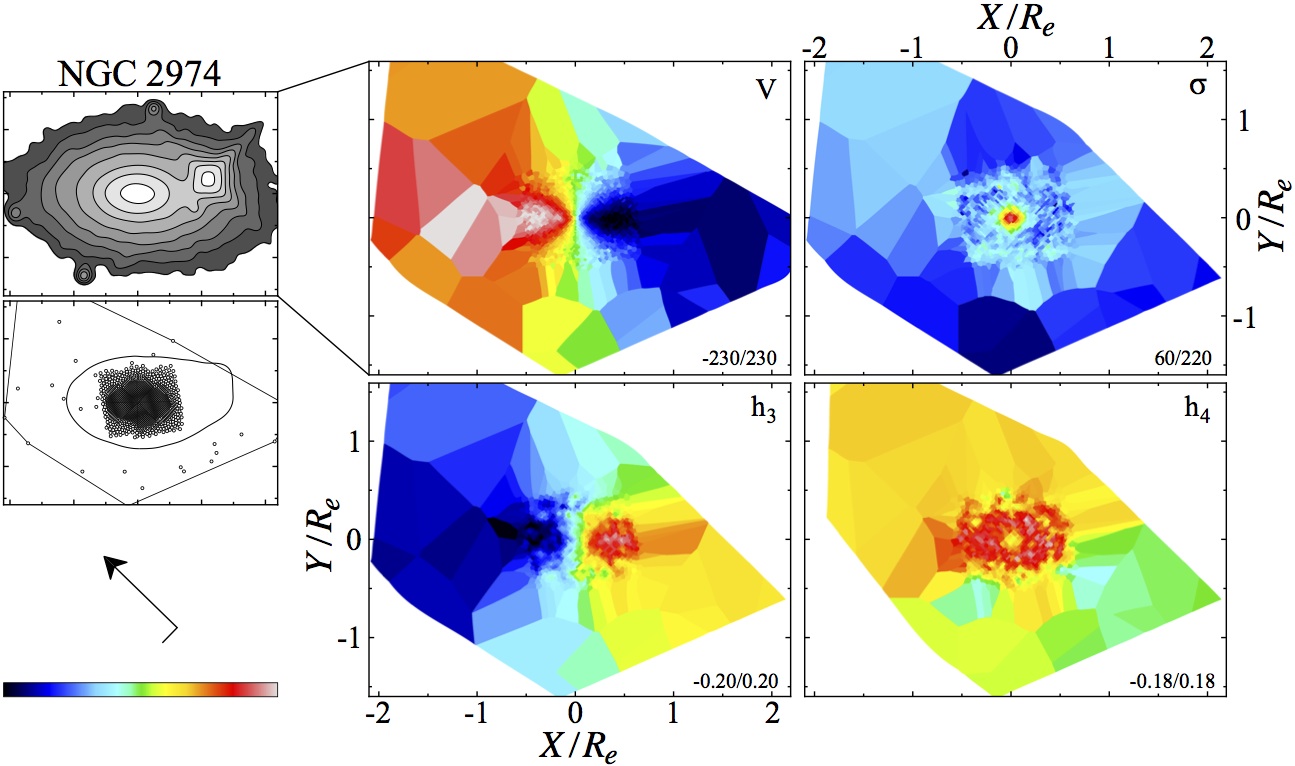}
%
					\epsfxsize=18cm
					\epsfbox{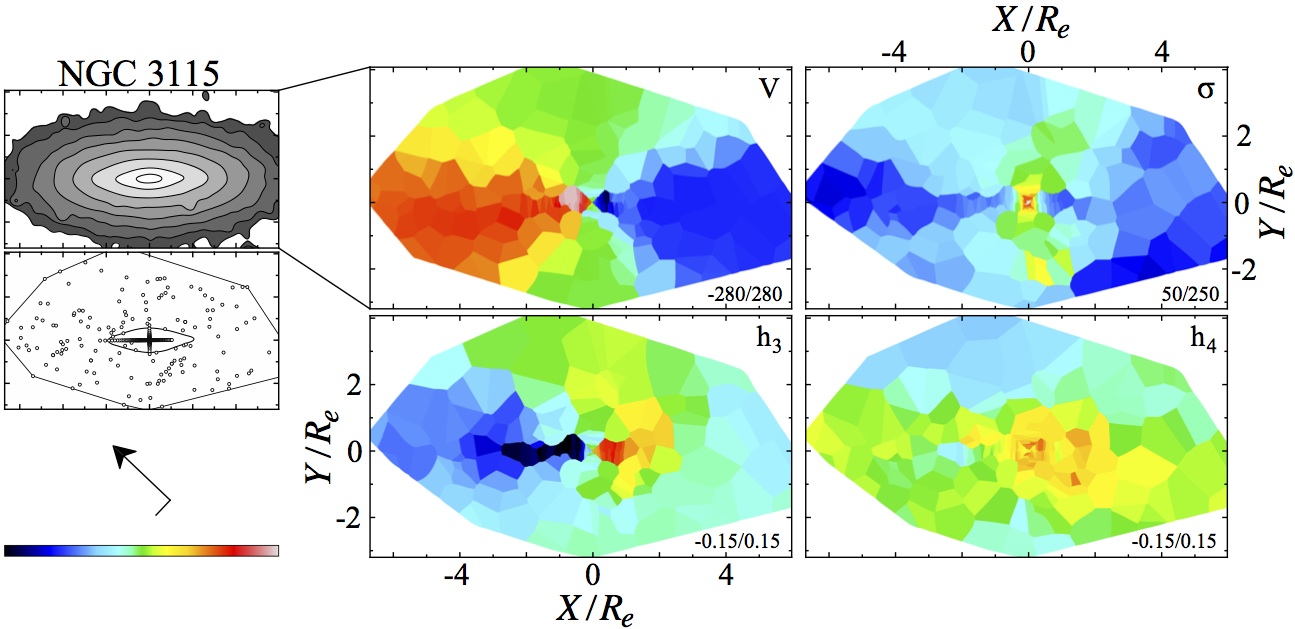}
					\caption{Continued.
					}
				\end{center}
			\end{figure*}

			\addtocounter{figure}{-1}
			\begin{figure*}
				\begin{center}
					\epsfxsize=18cm
					\epsfbox{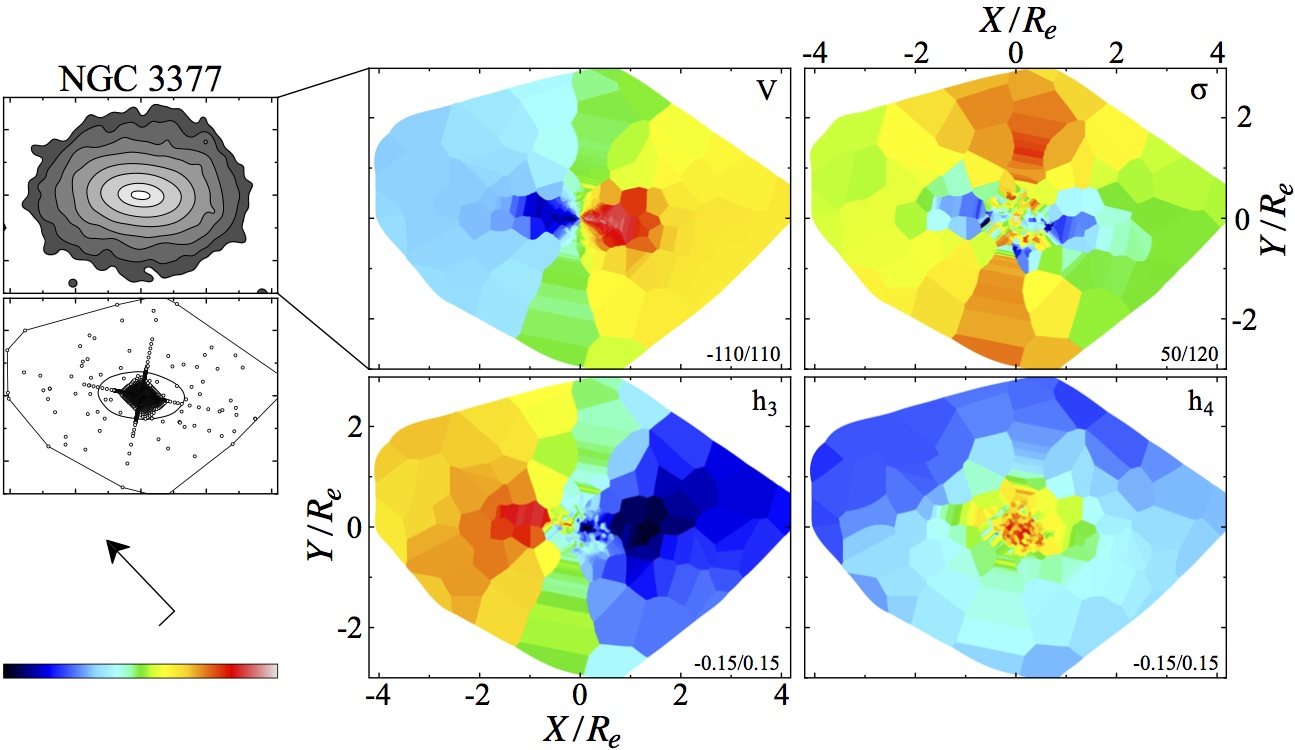}

					\epsfxsize=18cm
					\epsfbox{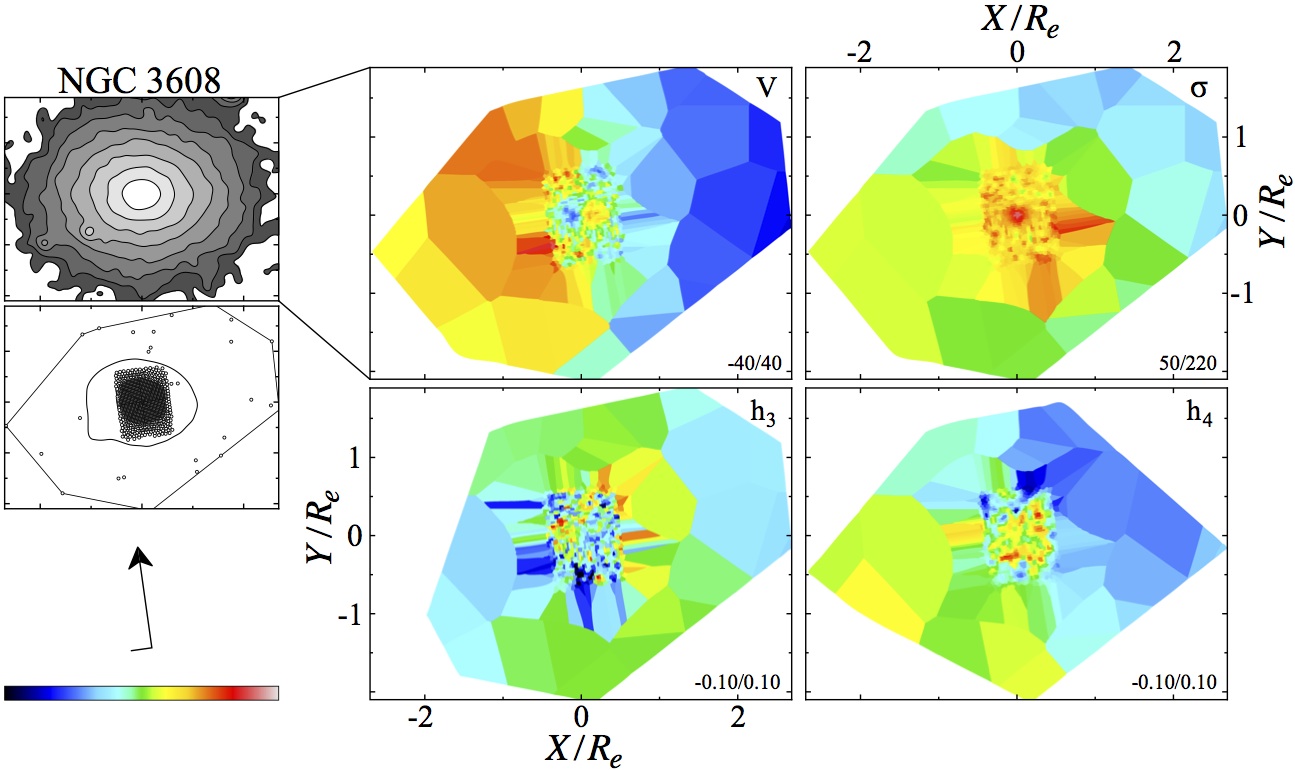}
					\caption{Continued.
					}
				\end{center}
			\end{figure*}

			\addtocounter{figure}{-1}
			\begin{figure*}
				\begin{center}
					\epsfxsize=18cm
					\epsfbox{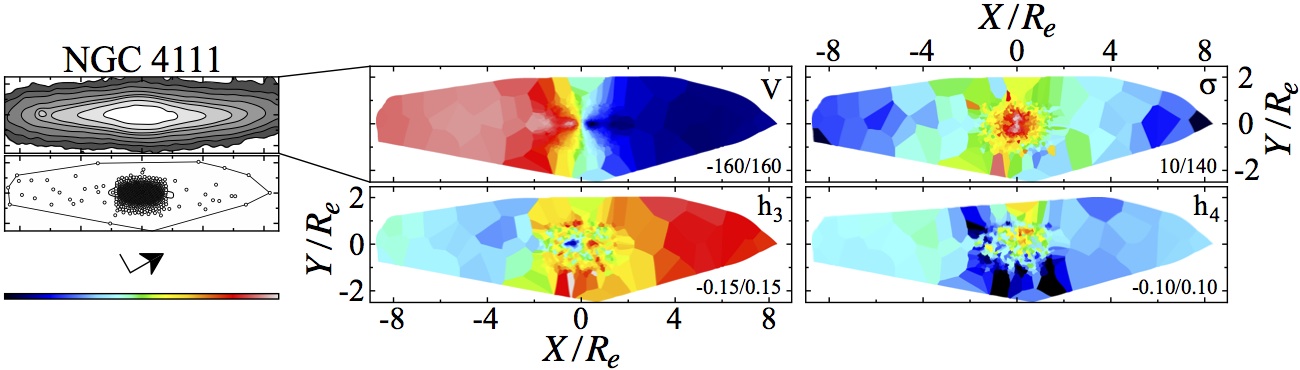}

					\epsfxsize=18cm
					\epsfbox{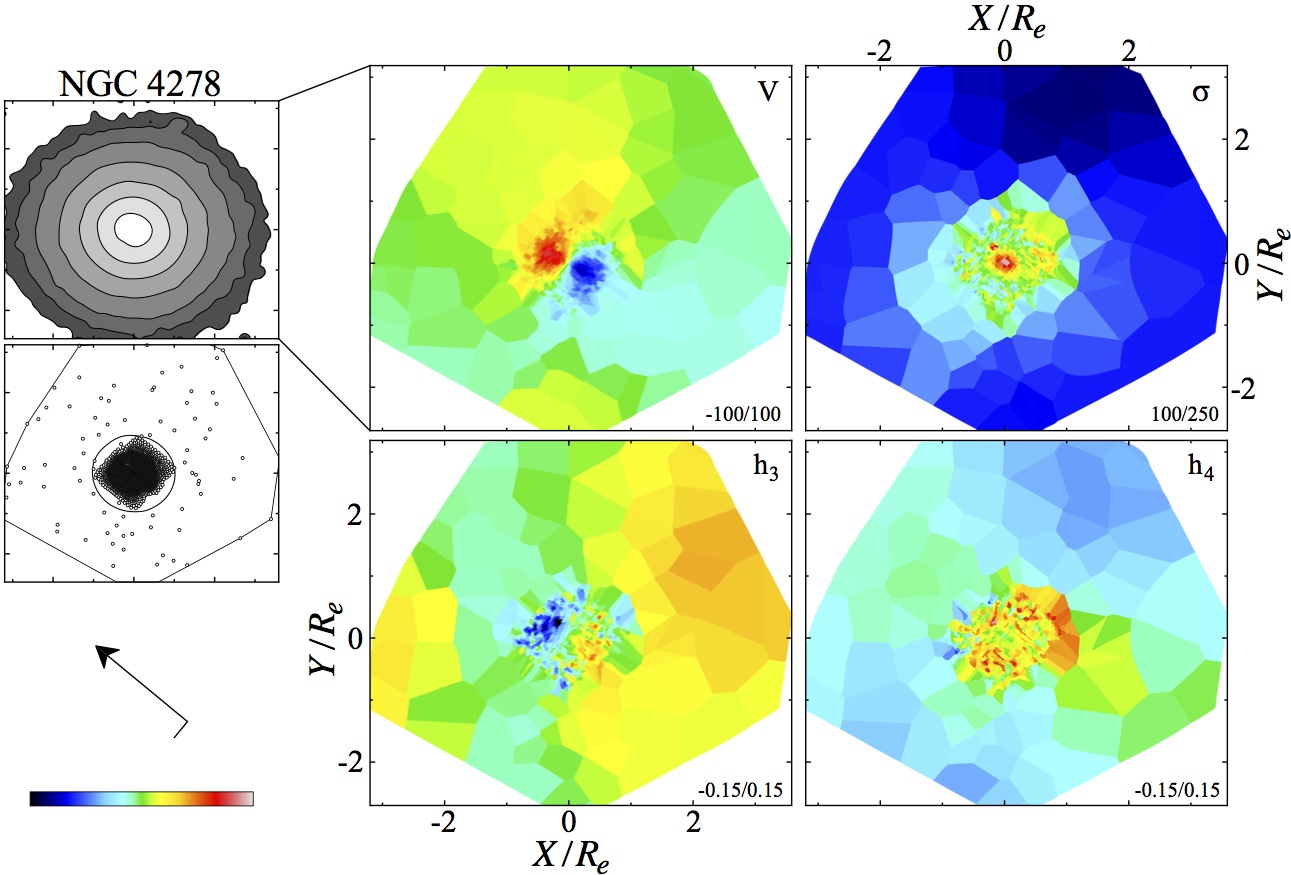}
					\caption{Continued.
					}
				\end{center}
			\end{figure*}

			\addtocounter{figure}{-1}
			\begin{figure*}
				\begin{center}
					\epsfxsize=18cm
					\epsfbox{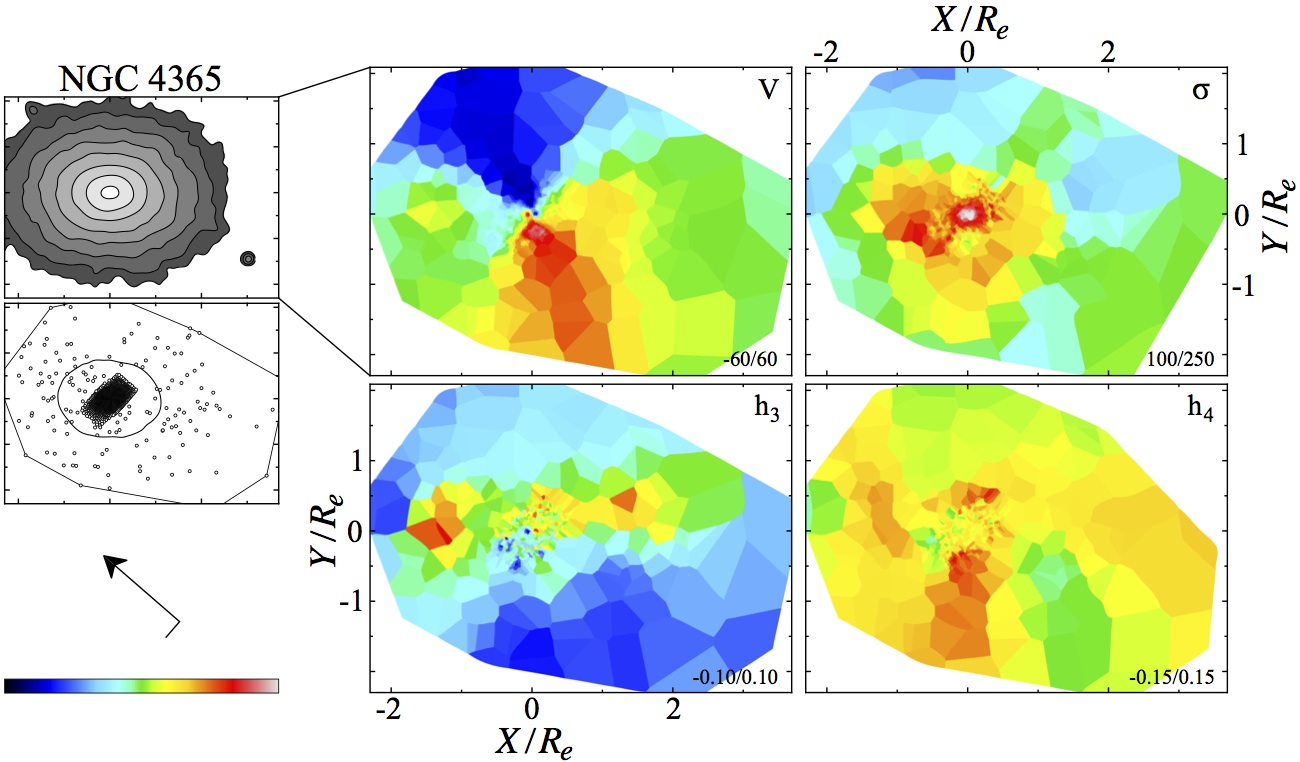}

					\epsfxsize=18cm
					\epsfbox{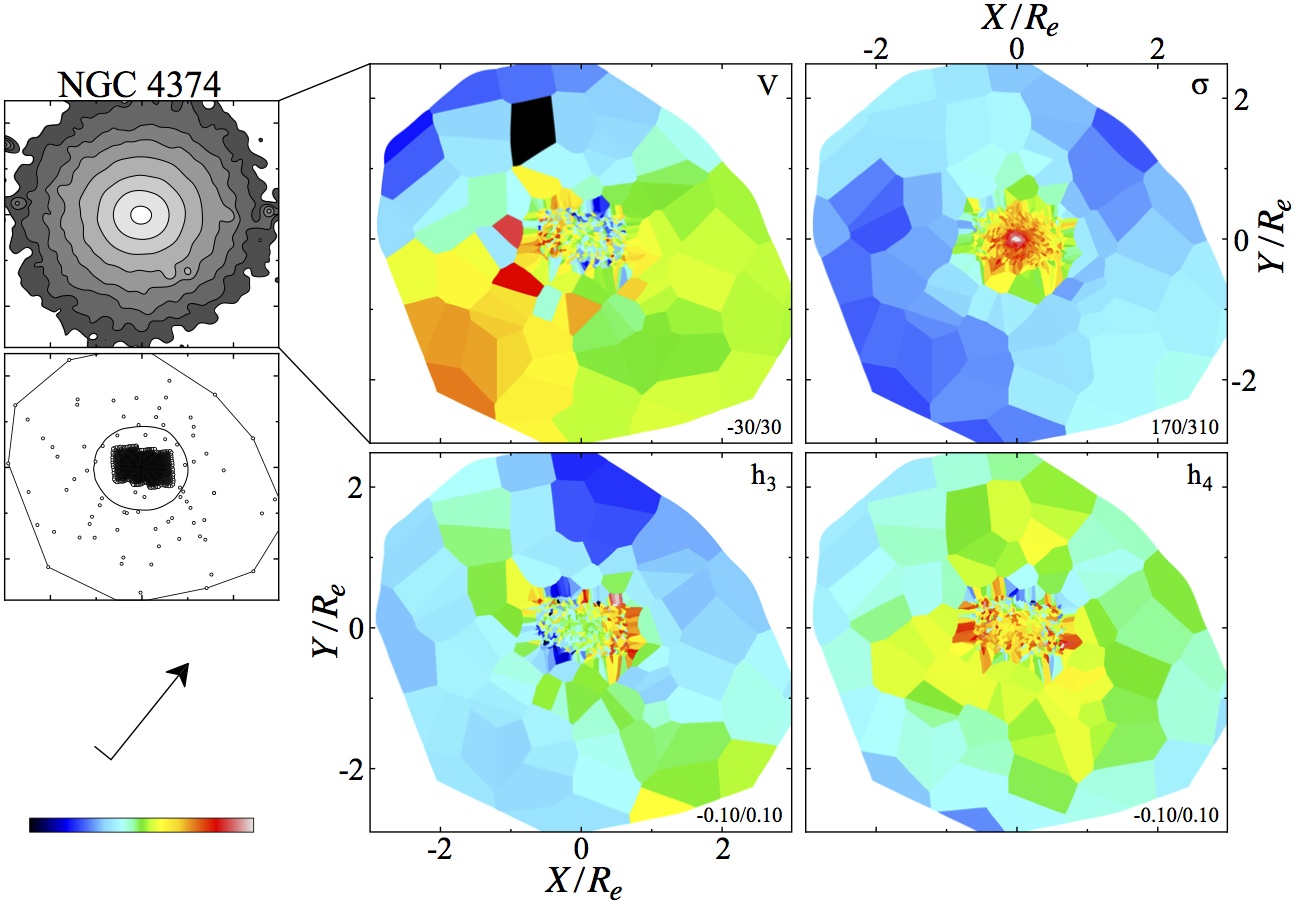}
					\caption{Continued.
					}
				\end{center}
			\end{figure*}
	
			\addtocounter{figure}{-1}
			\begin{figure*}
				\begin{center}
					\epsfxsize=18cm
					\epsfbox{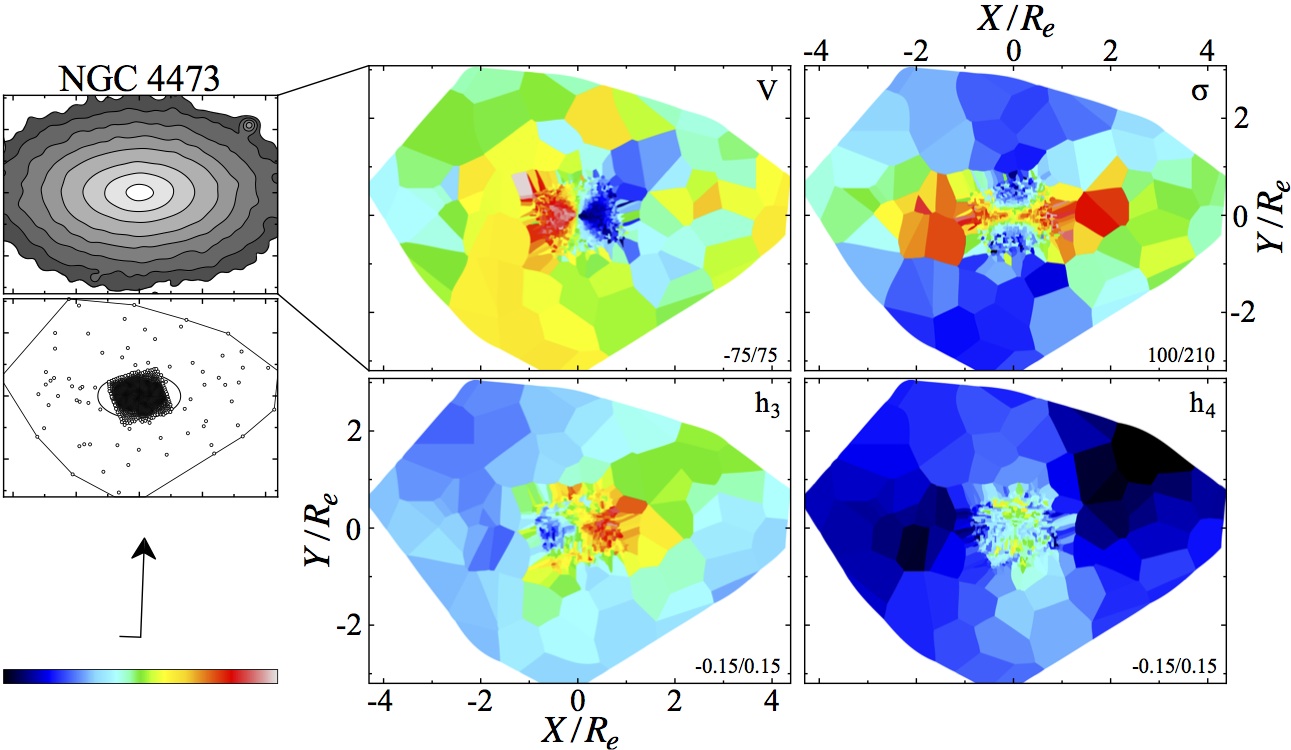}
		
					\epsfxsize=18cm
					\epsfbox{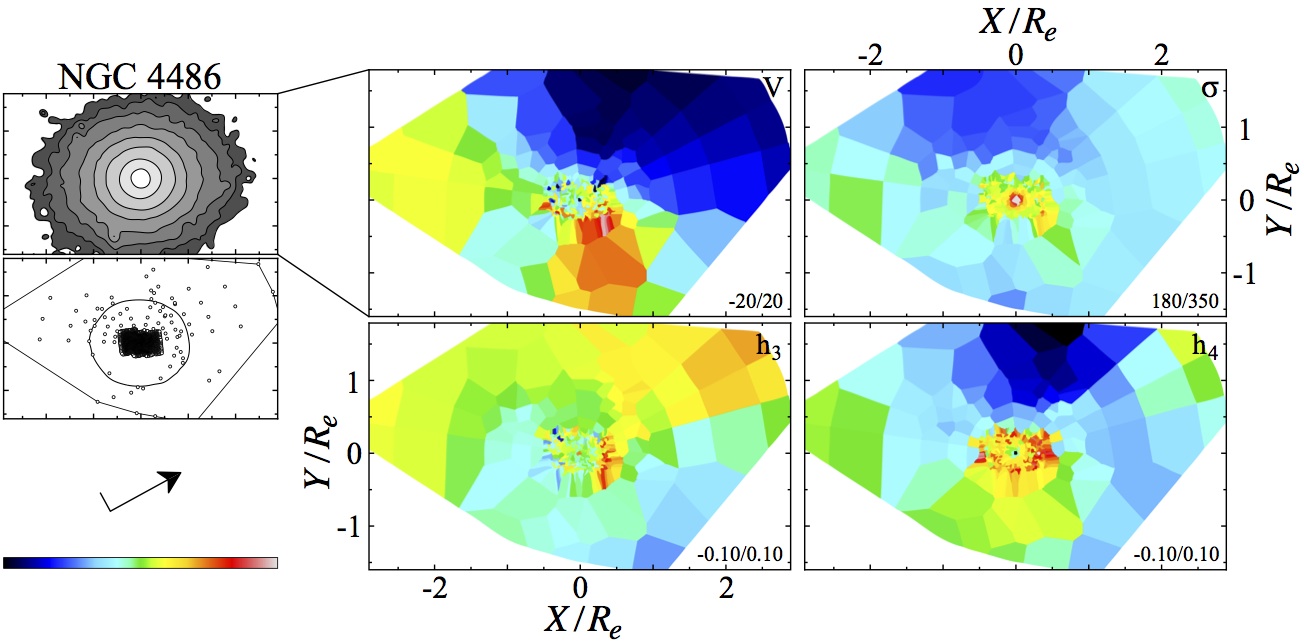}
					\caption{Continued.
					}
				\end{center}
			\end{figure*}

			\addtocounter{figure}{-1}
			\begin{figure*}
				\begin{center}
					\epsfxsize=18cm
					\epsfbox{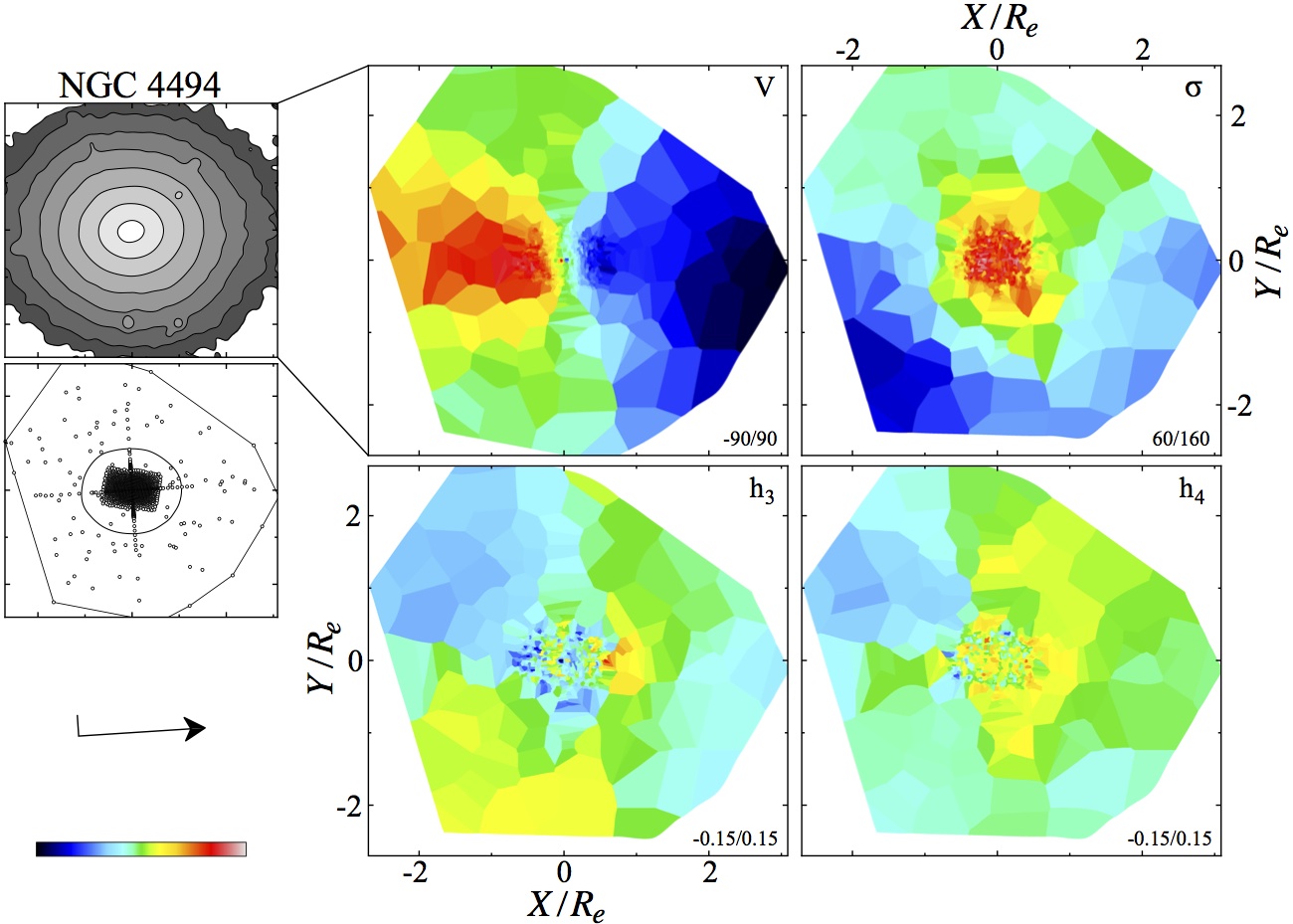}

					\epsfxsize=18cm
					\epsfbox{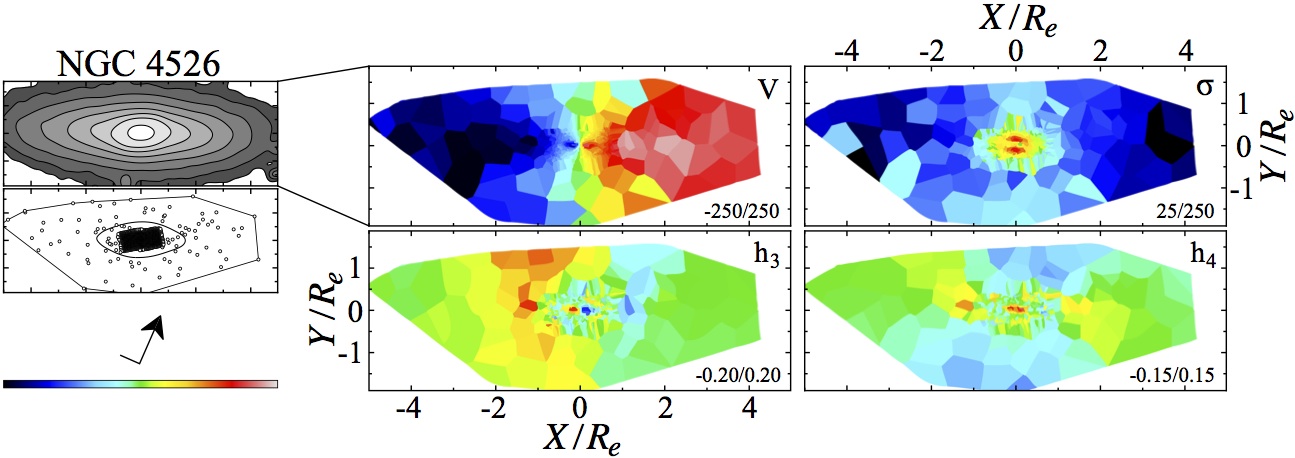}
					\caption{Continued.
					}
				\end{center}
			\end{figure*}

			\addtocounter{figure}{-1}
			\begin{figure*}
				\begin{center}
					\epsfxsize=18cm
					\epsfbox{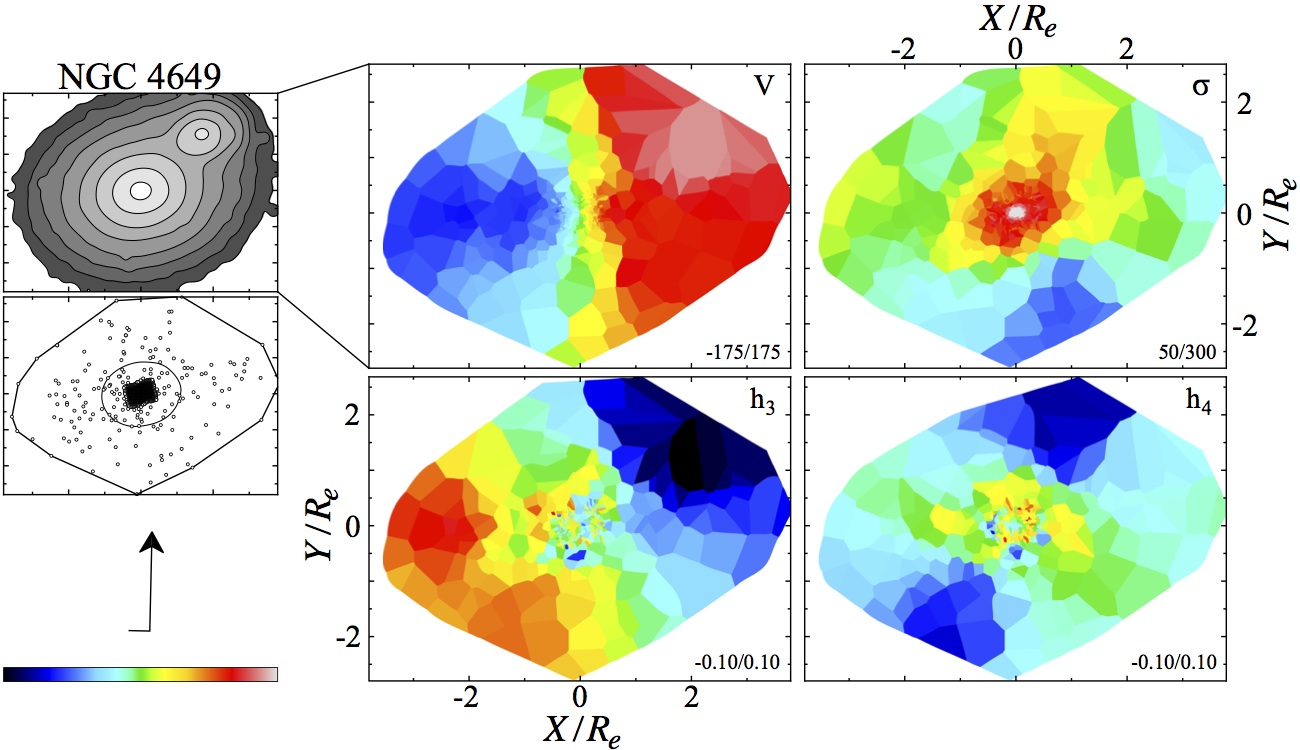} 

					\epsfxsize=18cm
					\epsfbox{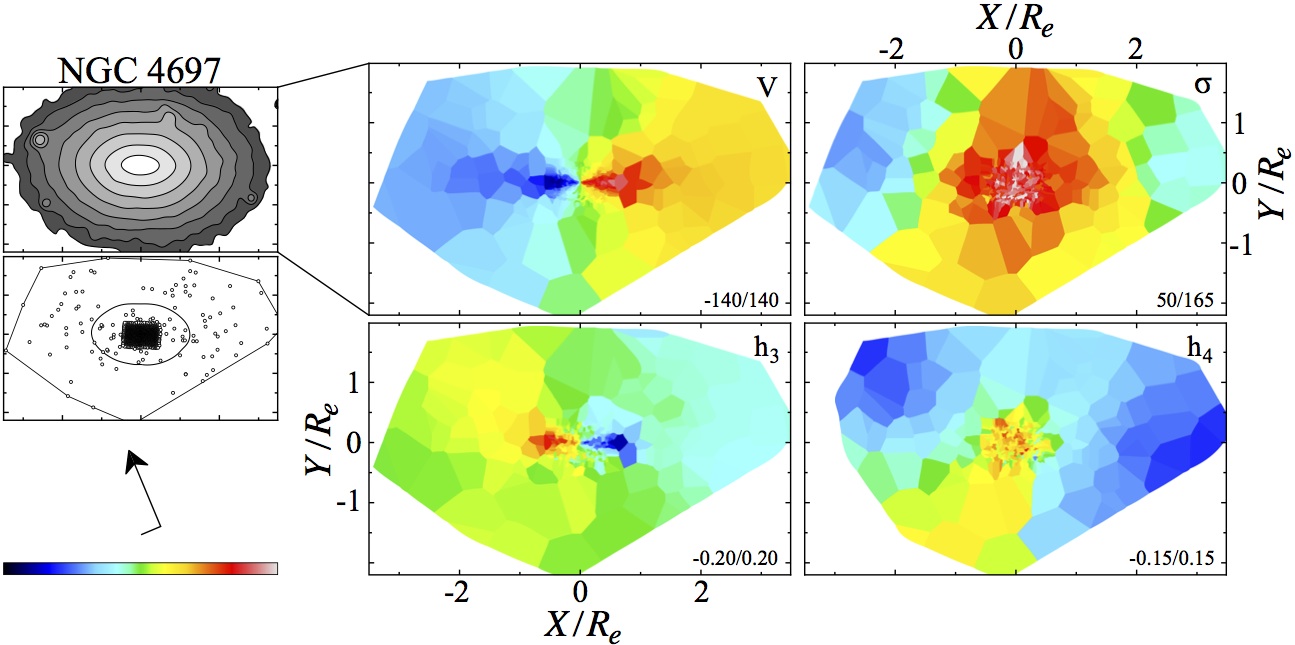}
					\caption{Continued.
					}
				\end{center}
			\end{figure*}
			
			\addtocounter{figure}{-1}
			\begin{figure*}
				\begin{center}
					\epsfxsize=18cm
					\epsfbox{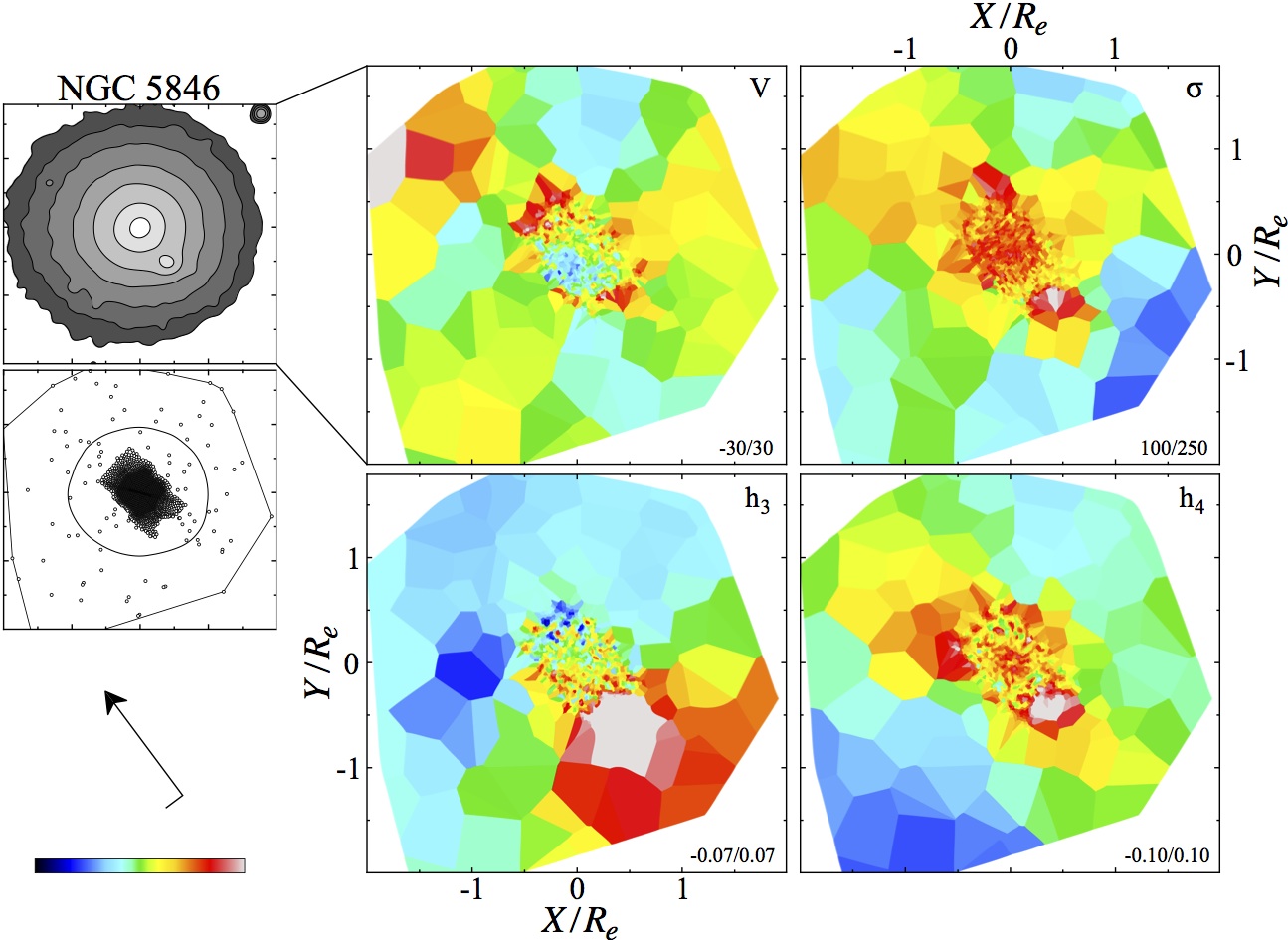}

					\epsfxsize=18cm
					\epsfbox{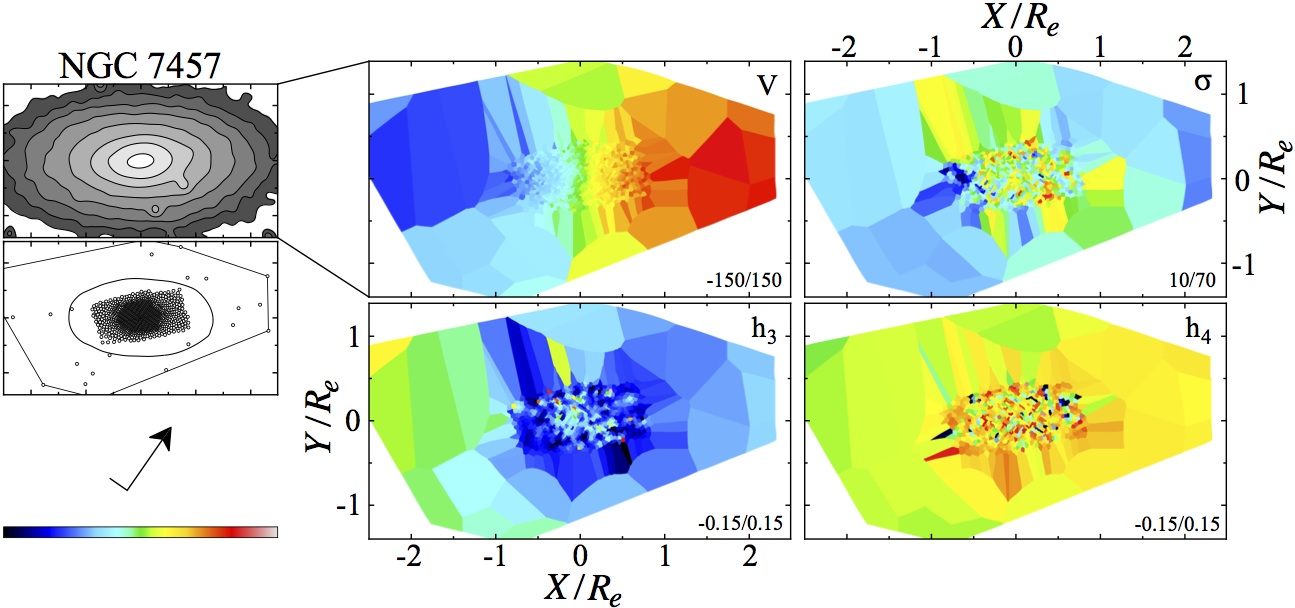}
					\caption{Continued.
					}
				\end{center}
			\end{figure*}

		
	\appendix
	
	\section{Kinematic Maps}\label{sec:appA}

In Figure~\ref{fig:skims0}, the data for each galaxy are laid out as follows.
At left, the galaxy's name is listed above the isophotal contour map of the respective $I$-band DSS image---oriented so that the photometric major-axis is horizontal\footnote{ The adopted major-axis position angles generally correspond to the {\it outer}
isophotes, which are, in many cases, twisted relative to the inner isophotes and kinematics---hence the apparent tilt of many of the
maps.}.
The associated compass, which points north by east, illustrates the rotation of the data (see Table~\ref{tab:galaxies} for adopted position angles).
The panel below the contour map shows plots of the locations of each measurement used to construct the kinematic maps and includes the 1~\Reff\ contour and the convex hull (bounding polygon) of data locations.
This hull is slightly larger than $R_{\rm max}$, the maximum radial extent of the data,
which we define as the largest intermediate-axis radius at which at least 75\% of the area within an elliptical annulus lies inside the outer measurement hull.

In the smoothed kinematic maps shown at right, each measurement is assigned a Voronoi bin (see the previous section) and plotted as a function of $X$ and $Y$ position in the galaxy, normalized by \Reff.
Panels are labeled according to measurement type ($V$, $\sigma$, $h_3$, or $h_4$), and Voronoi bins are colored according to value using the color bar shown at the bottom left.
The minimum/maximum of the range is indicated in the bottom right corner of each panel.

\section{Notes on Individual Galaxies}\label{sec:appB}

In this Appendix, we list a (non-exhaustive) selection of salient
properties for our sample galaxies along with a qualitative description of
their respective combined stellar and GC (when available) kinematics. In
all cases, morphological classifications are from the RC3 as reported in NED.

The following literature kinematic data (IFU and long slit) are included along with our Keck/DEIMOS measurements in order to provide coverage of the galaxy centers and improve spatial sampling:
\textbf{NGC 720} (\citealt{Cappellari:2007bn}, private communication);
\textbf{NGC 821} \citep{Weijmans:2009hl,Pinkney:2003jb,Proctor:2005hj,Forestell:2010bm,Cappellari:2011ej}; 
\textbf{NGC 1023} \citep{Cappellari:2011ej,Fabricius:2012ea};
\textbf{NGC 1400} \citep{Spolaor:2008do};
\textbf{NGC 1407} \citep{Spolaor:2008do};
\textbf{NGC 2768} \citep{Cappellari:2011ej};
\textbf{NGC 2974} \citep{Cappellari:2011ej};
\textbf{NGC 3115} \citep{Norris:2006if};
\textbf{NGC 3377} \citep{Coccato:2009je,Cappellari:2011ej};
\textbf{NGC 3608} \citep{Cappellari:2011ej};
\textbf{NGC 4111} \citep{Cappellari:2011ej};
\textbf{NGC 4278} \citep{Cappellari:2011ej};
\textbf{NGC 4365} \citep{Cappellari:2011ej};
\textbf{NGC 4374} \citep{Cappellari:2011ej};
\textbf{NGC 4473} \citep{Cappellari:2011ej};
\textbf{NGC 4486} \citep{Cappellari:2011ej,Murphy:2011hf};
\textbf{NGC 4494} \citep{Coccato:2009je,Cappellari:2011ej};
\textbf{NGC 4526} \citep{Cappellari:2011ej};
\textbf{NGC 4649} \citep{Pinkney:2003jb,Cappellari:2011ej};
\textbf{NGC 4697} \citep{Cappellari:2011ej};
\textbf{NGC 5846} \citep{Cappellari:2011ej};
\textbf{NGC 7457} \citep{Cappellari:2011ej}.

\textbf{NGC~720.}
This is a highly elongated E5 galaxy possibly residing in a triaxial halo
\citep{Buote:2002a}. \citet{Cappellari:2007bn} found clear
but moderate central rotation within 1~\Reff, and similarities to both
fast and slow rotators. Our kinematic data show that this slow but significant rotation
continues essentially constant out to the edge of the SLUGGS data around
2~\Reff.

\textbf{NGC~821.}
This E6 galaxy is a classic example of a disky elliptical
\citep{Lauer:1985tk}, although \citet{Kormendy:2013} argued for a reclassification as an S0.
Its characteristic disky isophotes disappear beyond 1~\Reff\ (about 30 arcsec;
\citealt{Bender:1988vj,Nieto:1991vh,Goudfrooij:1994vj}). Rotation is prominent inside of this radius
\citep{Dressler:1983dv,Bender:1994vo,Pinkney:2003jb,Emsellem:2004kr}, but
diminishes farther out (see Figure
\ref{fig:skims0}; \citealt{Proctor:2009iy,Coccato:2009je,Weijmans:2010bn}).
Our data show that both the velocity dispersion and the $h_4$ maps exhibit
large-scale structure with distinctly different amplitudes along major and
minor axes. Only the metal-poor GCs show significant rotation and that is
along the photometric minor axis in the inner parts \citep{Pota:2013en},
consistently with the PN data \citep{Coccato:2009je}.

\textbf{NGC~1023.}
This is a barred lenticular galaxy (SB0). Our measured velocity dispersion
profile declines strongly with radius, with $\sigma$ values eventually
falling below the resolution limit. Over the same radial range, the
rotational amplitude gradually increases with radius before plateauing
near 1~\Reff\ (to $\sim$\,200~km~s$^{-1}$) and eventually dropping beyond 3~\Reff\
(see also, e.g., \citealt{Noordermeer:2008kx,Coccato:2009je,Cortesi:2011df}).

\textbf{NGC~1400:} This galaxy (E1/S0) is the second brightest galaxy
of the NGC 1407 group. The combined velocity map indicates a moderately
sized embedded disk on the scales of the effective radius within an
aligned and moderately rotating bulge. Additional evidence for this
interpretation comes from the clear anti-correlation between $V/\sigma$
and $h_3$ that is present only in this region. Its GC system kinematics
are consistent with the stellar kinematics \citep{Pota:2013en}.

\textbf{NGC~1407.} This is a giant elliptical galaxy (E0) at the center of
the Eridanus A group. Weak rotation at all radii with anti-correlated
$h_3$ within a half-light radius are evident. Overall, the velocity
dispersion falls with radius, though a notable upward ``annular'' spike is
apparent around 1~\Reff\ \citep[first detected by][]{Proctor:2009iy}.
There is a hint of a spatially corresponding structure in $h_4$.
\citet{Pota:2013en} found that the metal-rich GCs rotate along the major
axis while metal-poor GCs only show significant rotation in the inner
parts along the minor axis. This rotation declines and progressively
aligns with the major axis in the outskirts. The metal-rich GC velocity
dispersion is consistent with that of the galaxy stars. Both GC
subpopulations have tangentially biased orbits \citep{Romanowsky:2009gq}.

\textbf{NGC~2768.}
This is a well studied flattened (E6/S0) galaxy. Its cylindrical
rotational field within $\sim$\,40~arcsec \citep{Emsellem:2004kr} and
sustained overall rapid rotation are obvious in our combined maps and have
already been well documented \citep{Fried:1994dd,Proctor:2009iy}. $V$ and
$h_3$ are globally anti-correlated as expected from an S0 with bulge and
disk components of similar scale size and relative brightness
($B/T=0.7$; \citealt{Forbes:2012bi}). According to \citet{Pota:2013en}, the
metal-rich GC subpopulation rotates along with the stars, while the metal-poor
GCs do not show significant rotation.

\textbf{NGC~2974.} Though classified as an E4, this galaxy exhibits some
spiral structure in addition to more subtle substructure in the form of
shells \citep{Tal:2009ht}. It rotates rapidly within 1~\Reff\ with
corresponding anti-correlated $h_3$ values \citep{Cinzano:1994tl,Emsellem:2004kr}. 
At larger radius the rotation amplitude
plateaus out to 3~\Reff\ along the major-axis, and also remains
anti-correlated with $h_3$.

\textbf{NGC~3115.}
The kinematics for this nearby S0 is described in \citet{Arnold:2011kp},
who also studied the GC kinematics. A
rapidly rotating thin disk, apparent out to about 2~\Reff, is embedded
within a bulge that also exhibits fairly rapid rotation. Anti-correlated
velocity and $h_3$ is apparent out to about 2~\Reff\ in the bulge. As
evidenced by the dispersion map, these two structures are both embedded in
a more spheroidal and slowly rotating structure that only becomes evident
at larger radius. \citet{Arnold:2011kp} also found rapid metal-rich GC
rotation marginally consistent with the stellar rotation in the inner part
that decreases steeply farther out. The metal-poor GCs rotate with slightly
lower amplitude and also exhibit an outer decline beyond 4~\Reff.

\textbf{NGC~3377.}
The combined stellar kinematic map of this elongated elliptical (E5-6)
galaxy presents several key characteristics of an embedded disk within a
more spheroidal component. The inner regions are dusty, have disky
isophotes (e.g. \citealt{Bender:1988vj}), 
rotate rapidly, and have depressed velocity dispersions compared to
larger radii. The outer regions are boxier and rotate more slowly but
remain aligned with the inner regions within about 4~\Reff\ (see also
\citealt{Coccato:2009je}). \citet{Pota:2013en} reported significant rotation
along the major axis with outer twist of the metal-rich GCs in agreement
with stars and long-slit data. The metal-poor GCs do not show significant
rotation. The velocity dispersion of the metal-rich GCs is consistent
with that of the stars while that of the metal-poor GCs is higher.

\textbf{NGC~3608.}
This slightly flattened low luminosity elliptical galaxy (E1--2) shows only
mild outer rotation along the major axis. It is one of the first
known galaxies with a central counter-rotating kinematically decoupled
core \citep{Jed:1988a}. As such, it is one
of the few non-regular rotators in the \atlas\ sample \citep{Krajnovic:2011jj}.

\textbf{NGC~4111.}
This is the most disky galaxy (S0) in our sample, and has the smallest effective
radius (i.e., \Reff~$<1$ kpc). As a result, our kinematic data extend out
to 4~\Reff. In the inner parts, long slit data from \citet{Loyer:1998a}  show a bump in the major axis rotation at a
galactocentric radius of roughly 5 arcsec with otherwise constant
amplitude farther out around 150 km s$^{-1}$. Our map shows that this
rotation is sustained up to 4~\Reff.

\textbf{NGC~4278.}
This well-studied small round elliptical (E1--2) contains a distorted H~{\small I} disk
\citep{Knapp:1978a}. Our combined kinematic maps
indicate that the stars are rotating only within 0.5~\Reff, with very weak to
no rotation farther out. The velocity dispersion falls rapidly with
radius. \citet{Pota:2013en} did not find evidence that the metal-rich GCs
rotate but concluded that the metal-poor GCs rotate mildly in the outer
regions along a direction between the major and minor axes. The velocity
dispersion of both GC subpopulations agrees broadly with the stellar
kinematics.

\textbf{NGC~4365.}
This giant elliptical (E3) galaxy is well known for its
kinematically decoupled core rotating along the major axis and otherwise
minor axis ``rolling'' rotation \citep{Davies:2001bf}, which 
can be explained as a signature of long-axis tube orbits in a triaxial potential
\citep{Statler:2004,vandenBosch:2008}.
Such a configuration could arise through a major merger involving a spiral
galaxy, with the central ``normal'' rotating component originating from newly
formed stars, and the outer rolling component from the progenitor's cold disk
\citep{Hoffman:2010dr}. Our large-scale
combined kinematic maps reveal that the minor-axis rolling continues to
large radii and hence involves the bulk of the stellar mass. 
The galaxy also appears to be in the process of further halo growth via 
an interaction with NGC~4342 in the same group \citep{Blom:2014}. 
\citet{Pota:2013en} found that the two `normal' GC
subpopulations have identical velocity dispersion radial profiles. 
\citet{Blom:2012} reported that the unusual intermediate metallicity subpopulation
(green) of GCs rotates like the stars along the minor axis while both
metal-rich and metal-poor GCs rotate along the major axis.

\textbf{NGC~4374.}
This Virgo round elliptical (E1) galaxy harbors a rapidly rotating
nuclear gas disk \citep{Bower:1998a}. 
\citet{Napo:2011a} used bright PNe as tracers to
measure the stellar kinematics out to $\sim 350$ arcsec. They found a
steeply declining dispersion profile in the inner $\sim 30$ arcsec beyond
which the profile flattens out at a value of $\sim230$ km s$^{-1}$. As might be
expected for such a slow rotator \citep{Emsellem:2011br}, we detect only
mild to no rotation in the outer parts of the combined kinematic map, in
agreement with \citet{Coccato:2009je}.

\textbf{NGC~4473.}
This elongated elliptical (E5) galaxy presents disky isophotes \citep{Bender:1988vj}. It is an interesting case of a
so-called ``$2\sigma$'' galaxy \citep{Krajnovic:2011jj}
which has off-center velocity dispersion peaks. The favored explanation
for this feature is the presence of two counter-rotating disks \citep{Cappellari:2007bn}. 
The inner rotation quickly declines with radius, eventually giving way to 
two-component retrograde major-axis and minor-axis rotation in the outskirts.
This outer kinematic configuration is a tell-tale sign of triaxiality,
and the galaxy overall may be considered a case of a ``kinematically
distinct halo'' \citep{Foster:2013}.

\textbf{NGC~4486.}
Also known as M87, this well known bright central galaxy of Virgo is
morphologically classified as an E0--1 and cD galaxy. Our kinematic map shows that the
velocity dispersion falls with radius, though the values remain high at
all radii probed ($\gtrsim200$ km s$^{-1}$). Some weak amplitude
$\sim20$ km s$^{-1}$ off-axis rotation is apparent beyond 0.5~\Reff\ (see
also \citealt{Murphy:2011hf}). Its GC kinematics are well-studied (e.g.,
\citealt{Cote:2001a,Strader:2011z}) and only weak
rotation along the major axis is detected in both subpopulations. The
overall GC velocity dispersion is only mildly declining at all probed
radii. Helped by the sheer number of measured GC velocities, \citet{Roman:2012z}  
found significant substructures in phase space
indicating that this otherwise apparently quiet galaxy is actively
accreting other systems.

\textbf{NGC~4494.} This ``ordinary'' elliptical (E1--2) galaxy contains an
inner dust ring \citep{Lauer:2005bk}.  The central regions contain a
kinematically distinct core \citep{Bender:1994vo} and double
rotation amplitude maxima \citep{Krajnovic:2011jj}. Our map shows that
rotation is sustained in the outskirts, along with a 10 deg twist in the
kinematic position angle around 1.5 \Reff\ \citep{Proctor:2009iy}. In
agreement with our velocity dispersion map, \citet{Napo:2009a} found a radially mild decrease in velocity dispersion
using combined PN and long-slit data. The GC kinematics were
studied in \citet{Foster:2011hh}, who found significant
rotation for the metal-poor GCs consistent with the stellar motions, but
little to no rotation in the metal-rich GCs. On the other hand, both GC
subpopulations have a velocity dispersion profile consistent with that of
the stars.

\textbf{NGC~4526.}
This is an edge-on barred lenticular (SB0) galaxy with indications of possible
spiral structure \citep{Michard:1994tl}. Our
combined velocity map exhibits a steep decrease to very low values below
our resolution. The central regions of both the $\sigma$ and $h_4$ maps
suggest embedded components along the major and minor axes. There is a cold
central disk \citep{Bureau:2002a} in the inner parts that transitions into a hotter co-rotating bulge.

\textbf{NGC~4649.}
Also known as M60, this massive Virgo E2/S0 galaxy is of comparable size to NGC 4486. The
velocity maps show a clear contribution from the nearby spiral galaxy NGC~4647 to the
northwest although there is no obvious sign of interaction. In contrast
to M87, it exhibits rapid rotation with radially declining $\sigma$ (also
see \citealt{Teo:2011a} for a similar result using
PNe). Kinematic maps for higher order moments show clearly
anti-correlated $V-h_3$ and radially dropping $h_4$. 
\citet{Das:2011a} used long-slit and PN tracers to model the
mass of the system and concluded that the stellar orbits are centrally
isotropic to mildly radial. \citet{Bridges:2006a} found
no significant rotation of the GC system around NGC~4649, and a
relatively flat velocity dispersion radial profile out to 3.5 \Reff.

\textbf{NGC~4697.}
This elongated elliptical (E6) galaxy exhibits fast rotation
in the inner parts \citep{Emsellem:2011br}. 
Our extended kinematic map shows that this fast rotation is due to an
embedded disk with a clear rotation amplitude decline with radius. The inner
parts show a clear $V-\sigma$ anti-correlation that vanishes in the
outskirts. The velocity dispersion map is radially declining. The
declining outer rotation and velocity dispersion radial profile was
detected using PN tracers in \citet{Mendez:2009a}. The PN data also 
imply mild radial anisotropy \citep{Lor:2008a}  and evidence has been found
that they may in fact belong to two separate kinematic populations
\citep{Sam:2006a}.

\textbf{NGC~5846.}
This massive early-type (E0--1/S0) galaxy has quite circular isophotes
\citep{Michard:1994tl}. Our combined kinematic map shows that there is
no obvious rotation in the central regions nor in the outskirts.
There are clear contributions from the nearby galaxy NGC~5846A to the south in
all four maps. \citet{Pota:2013en} demonstrated that neither GC
subpopulation shows significant rotation apart from a marginal detection in
the metal-poor GCs around 150--300 arcsec. They concluded that this may be due
to ``contaminating'' GCs associated with NGC~5846A. The velocity dispersion
radial profile of the metal-rich GCs is relatively flat, in contrast to
that of the declining PNe \citep{Das:2008a}.

			\textbf{NGC~7457.}
			This small S0 hosts a small bar \citep{Michard:1994tl} 
and a pseudo-bulge \citep{Pinkney:2003jb}. As with other lenticular galaxies, it
is classified as a fast rotator based on its inner kinematics 
\citep{Emsellem:2011br}. Our combined kinematic map shows global rotation (also
see \citealt{Cortesi:2013b}) and anti-correlated $h_3$. The velocity
dispersion values are universally low. The kinematics of this poor GC
system have been presented by \citet{Pota:2013en}, who found that the GCs
are globally rapidly rotating, in agreement with the stars.			

	\section{Stellar Populations and Wavelength}\label{sec:appC}
	
		The conventional wisdom is that ETGs are primarily composed of old stars.
		However, accretion can bring in small amounts of gas that reinvigorate star formation for short periods of time, thus contaminating the predominantly old stellar population with some number of young stars.
		Given a dissipational origin, this young stellar component may then exhibit its own distinct kinematic structure; e.g., it could be both colder and more rapidly rotating than the preexisting stars.
		Since young stellar populations emit more of their light in the blue region of the spectrum, we were interested to know whether adding a young stellar component led to measured LOSVD shapes that were systematically different between blue and red spectral regions.
		Specifically, will a small population of young stars yield different kinematic measurements at blue wavelengths (4800--5380~\AA) than at red wavelengths (8480--8750~\AA)?
		
		To test for this differential sensitivity to young stars, we created a set of composite spectra using the MILES single stellar population model spectra \citep{Vazdekis:2010en}.
		Each composite spectrum contains a particular fraction of younger (3~Gyr) stars ranging between 0\% and $100\%$, with the balance made up of old (10~Gyr) stars.
		To keep the test simple, we kept both sub-populations at solar metallicity.
		To simulate the effect of a moderately rotating component, we convolved the 3~Gyr old spectrum with a Gaussian LOSVD with $V = 100$~ km s$^{-1}$ and $\sigma$ = 75 km s$^{-1}$.
		Conversely, the 10~Gyr population is meant to represent a dynamically hotter and slowly rotating spheroid, and was thus convolved with a Gaussian LOSVD with $V = 0$~km s$^{-1}$ 
		and $\sigma$ = 150 km s$^{-1}$.

\begin{figure*}
			\begin{center}
				\epsfxsize=17cm
				\epsfbox{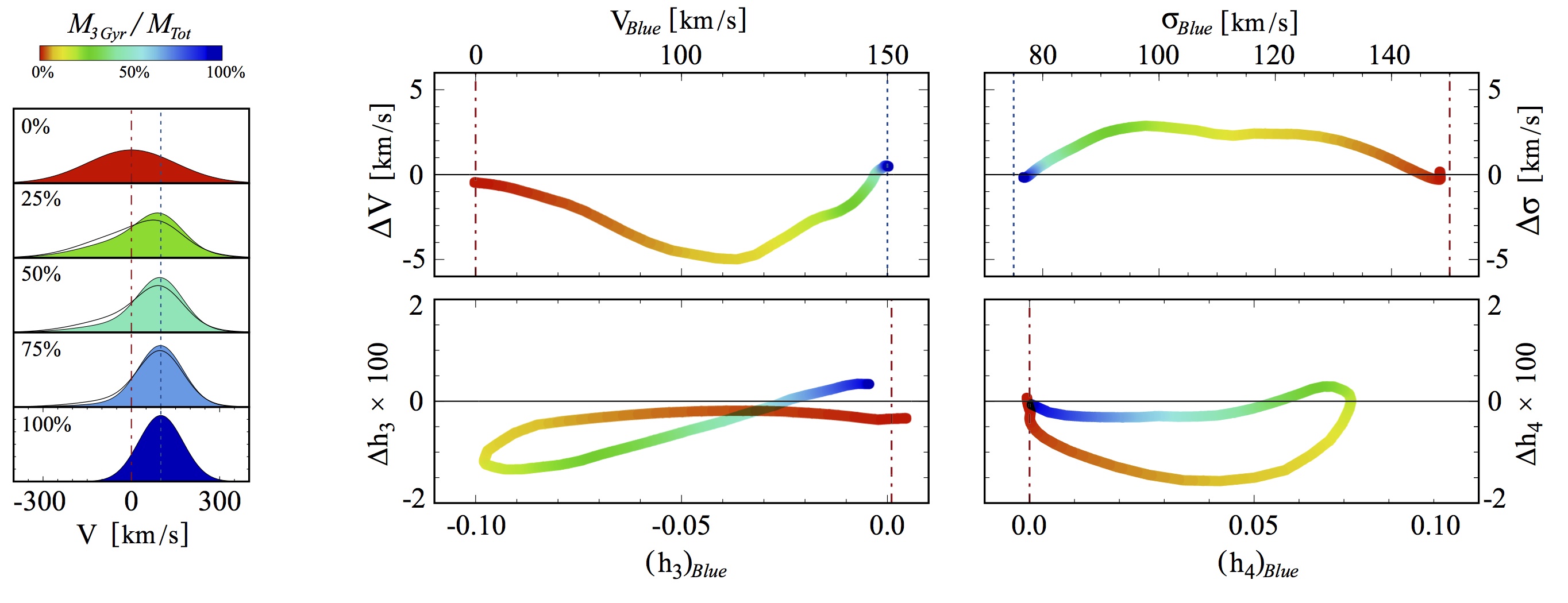}
				\caption{Differential kinematic effects of a young stellar component on blue and red spectral regions.
				Composite spectra of single stellar populations are created by combining spectra together from a 3~Gyr and a 10~Gyr old population.
				The input spectra are first convolved with Gaussian LOSVDs (with  
				$V = 100$~km s$^{-1}$ and $\sigma = 75$ km s$^{-1}$).
				Fiducial composite LOSVDs are shown (left) and colored according to young star fraction.
				Each panel (middle and right) shows the difference between the measured LOSVD parameters for red spectra centered on the calcium triplet (8480--8750~\AA) and blue spectra (4800--5380~\AA).
				If both spectral regions were equally sensitive to young stars, the curves would all be horizontal indicating exactly equal measurements.
				While there is some obvious differential sensitivity apparent, the amplitude of the difference is small enough to be ignored, given the typical uncertainties on these parameters for our spectra.
				}
			\end{center}\label{fig:redvblue}
		\end{figure*}

		If the young stellar population preferentially affects the blue spectral measurements, then we should detect a systematic offset in the derived kinematic parameters between the red and blue spectra.
		Specifically, we expect that the LOSVD parameters derived from the blue spectra will be biased toward the kinematics of the young stellar population.
		We used \texttt{pPXF} to measure the kinematic parameters for each composite spectrum separately in the blue and red wavelength ranges.
		The results are shown in Figure~\ref{fig:redvblue}15.
		Filled histograms (at left) show the LOSVDs for composite spectra with a young star fraction of 0, 25, 50, 75, and $100\%$.
		We note that young stellar populations are brighter than older populations, which results in a natural overrepresentation of younger stars in the LOSVD.
		The overplotted open histograms show the resulting LOSVDs if this luminosity difference were unaccounted for.
		Importantly, the relative brightnesses of young and old stellar populations is similar in the chosen blue and red spectral ranges.
		This means that red and blue spectra are roughly equally effected by young (3~Gyr) stellar populations, and thus no significant differential sensitivity to young stars is expected.
		
		The right panels illustrate this point a bit more clearly.
		In each panel we plot the difference in the measured kinematic quantity between the red and blue spectral regions versus its value as measured in just the blue spectral region.
		The curve follows our ensemble of composite spectra from a young-star fraction of $0\%$ to $100\%$ as shown by the color bar at the top left.
		When the young star fraction is 0, the spectrum contains only old stars and has a purely Gaussian LOSVD with zero mean velocity and a velocity dispersion of 150 km s$^{-1}$.
		As the young star fraction increases, the composite LOSVD becomes non-Gaussian (non-zero $h_3$ and $h_4$) until the population is purely made of young stars and the LOSVD is again Gaussian.
		Note that this effect is very non-linear in that a small amount of young stars is able to appreciably shift all four kinematic 
		parameters---a consequence of the extremely bright intrinsic luminosity of young stellar-populations.

		If the red and blue spectra were equally affected by the young stellar population, then each curve would be exactly horizontal.
		While this is not strictly the case, the deviations are small enough to be considered negligible given the typical uncertainties of our stellar kinematic measurements.
		In other words, there is relatively little differential sensitivity to stellar population effects between these wavelength regions.
		This simple test suggests that the discrepancies in measured kinematic values between the SAURON and SLUGGS data sets (see Section~\ref{subsection:litComp}) are not caused by a differential sensitivity to young stellar populations in the different wavelength ranges used in each survey.

	\bibliography{skims-bib}

\end{document}